\documentclass[12pt]{scrartcl}
\deffootnote[1em]{1em}{1em}{\textsuperscript{\thefootnotemark}}
\DeclareOldFontCommand{\rm}{\normalfont\rmfamily}{\mathrm}
\DeclareOldFontCommand{\sf}{\normalfont\sffamily}{\mathsf}
\DeclareOldFontCommand{\tt}{\normalfont\ttfamily}{\mathtt}
\DeclareOldFontCommand{\bf}{\normalfont\bfseries}{\mathbf}
\DeclareOldFontCommand{\it}{\normalfont\itshape}{\mathit}
\DeclareOldFontCommand{\sl}{\normalfont\slshape}{\@nomath\sl}
\DeclareOldFontCommand{\sc}{\normalfont\scshape}{\@nomath\sc}
\DeclareRobustCommand*\cal{\@fontswitch\relax\mathcal}
\DeclareRobustCommand*\mit{\@fontswitch\relax\mathnormal}
\usepackage{color}
\usepackage{type1cm}
\usepackage{graphicx}
\usepackage[usenames,svgnames]{xcolor}
\usepackage[hyphens]{url}
\usepackage[colorlinks=true,linkcolor=blue,citecolor=blue,breaklinks=true]{hyperref}
\usepackage[numbers,sort&compress]{natbib}

\usepackage{amsmath,amssymb,bm,amsthm,cases}

\def\beq{\begin{equation}}
\def\eeq{\end{equation}}
\def\nbeq{\begin{equation*}}
\def\neeq{\end{equation*}}

\def\<{\langle}
\def\>{\rangle}
\def\kket{\rangle\!\rangle}

\renewcommand{\d}{\partial}

\begin{document}
\title{Thermalization and prethermalization in isolated quantum systems: \\ a theoretical overview}
\author{\large Takashi Mori$^1$, Tatsuhiko N. Ikeda$^2$, Eriko Kaminishi$^1$, and Masahito Ueda$^{1,3}$
}
\date{\normalsize{$^1$\textit{
Department of Physics, Graduate School of Science, The University of Tokyo, Bunkyo-ku, Tokyo 113-0033, Japan
}\\
$^2$\textit{
Institute for Solid State Physics, The University of Tokyo, Kashiwa, Chiba 277-8581, Japan
}\\
$^3$\textit{
RIKEN Center for Emergent Matter Science (CEMS), Wako, Saitama 351-0198, Japan
}}}
\maketitle

\begin{abstract}
The approach to thermal equilibrium, or thermalization, in isolated quantum systems is among the most fundamental problems in statistical physics.
Recent theoretical studies have revealed that thermalization in isolated quantum systems has several remarkable features, which emerge from quantum entanglement and are quite distinct from those in classical systems.
Experimentally, well isolated and highly controllable ultracold quantum gases offer an ideal system to study the nonequilibrium dynamics in isolated quantum systems, triggering intensive recent theoretical endeavors on this fundamental subject.
Besides thermalization, many isolated quantum systems show intriguing behavior in relaxation processes, especially prethermalization.
Prethermalization occurs when there is a clear separation in relevant time scales and has several different physical origins depending on individual systems.
In this review, we overview theoretical approaches to the problems of thermalization and prethermalization.
\end{abstract}
\newpage
\tableofcontents
\newpage

\section{Introduction}
\label{sec:introduction}

It is a longstanding yet not completely resolved problem to reconcile reversible microscopic laws and irreversible macroscopic phenomena.
The second law of thermodynamics dictates that the entropy in an isolated system be a non-decreasing function of time.
In 1872, Boltzmann derived the Boltzmann equation for classical gases and used it to prove the renowned $H$-theorem, $dH(t)/dt\leq 0$ with $H(t)=\int d\bm{r}\int d\bm{v}\;f(\bm{r},\bm{v},t)\ln f(\bm{r},\bm{v},t)$, where $f(\bm{r},\bm{v},t)$ is a single-particle distribution function of position $\bm{r}$ and velocity $\bm{v}$ at time $t$~\cite{Boltzmann1872}.
The $H$-theorem proves the second law of thermodynamics by identifying the entropy with $-H(t)$ (in this review, the Boltzmann constant is set to unity).
However, the derivation of the Boltzmann equation was not entirely mechanical, as pointed out by Loschmidt~\cite{Loschmidt1876}.
According to the reversibility of the microscopic dynamics such as Newton's equations of motion in classical systems and the Scr\"odinger equation in quantum systems, if we had a solution in accordance with the second law, we would also have its time-reversed solution which shows a decrease in entropy.
Boltzmann had thought that he succeeded in proving that $H(t)$ either decreases or stays constant, but Loschmidt pointed out that the second law of thermodynamics should not be considered an absolute law. 
One may then wonder why one direction of time accompanied by the entropy increase is always observed and the other has never been.

Boltzmann acknowledged the importance of Loschmidt's criticism, and recognized that in his derivation of the Boltzmann equation, a nontrivial property of probabilistic nature (the assumption of ``molecular chaos'') is implicitly assumed.
Boltzmann then developed the crucial idea that macroscopic irreversibility is, as a matter of principle, a probabilistic notion~\cite{Boltzmann1877}.
He realized the importance of the distinction between microscopic and macroscopic variables, and pointed out that it is not all the microscopic states for a given set of macroscopic variables but an overwhelming majority of them that will evolve in accordance with the second law.
In other words, although the microscopic mechanical law allows time evolutions compatible with the entropy decrease, those initial states that violate the second law are rarely realized.
The approach to thermal equilibrium in isolated systems, which is called \textit{thermalization}, should be understood as a \textit{typical} behavior of macroscopic systems.
In this connection, it is worth pointing out that when Maxwell introduced a statistical method in deriving the Maxwell velocity distribution~\cite{Maxwell1860}, he already noticed the probabilistic nature of the second law of thermodynamics~\cite{Maxwell1871}.

Thermal equilibrium is also understood on the basis of a probabilistic consideration.
Boltzmann argued that the equilibrium state corresponds to the state with the maximum probability~\cite{Boltzmann1877}.
Here, all the allowed microscopic states under given constraints (e.g. a fixed value of the energy and the number of particles) are assumed to be equally probable, which is referred to as the \textit{principle of equal probability}.
Boltzmann succeeded in deriving the Maxwell velocity distribution just by counting the number of states, which he called ``Komplexion'', without solving any dynamical equation.
This probabilistic consideration led Gibbs to the ensemble theory of statistical mechanics~\cite{Gibbs1902}.

Einstein~\cite{Einstein1902,Einstein1903,Einstein1904} developed statistical mechanics independently of Gibbs, and arrived at results which are essentially equivalent to those obtained by Gibbs.
However, there is a difference in attitude between Einstein and Gibbs.
Gibbs aimed at establishing a ``rational foundation of thermodynamics'' and stressed that fluctuations of thermodynamic quantities are negligible in macroscopic systems.
In contrast, Einstein aimed at using fluctuations as a tool for investigating unknown microscopic physics~\cite{Einstein1904}.
By making ingenious use of statistical fluctuations, Einstein reached several remarkable results such as the determination of Avogadro's number using the Brownian motion~\cite{Einstein1905} and the demonstration of the wave-particle duality of light through the investigation of energy fluctuations~\cite{Einstein1909}.
Einstein's theory of fluctuations played seminal roles in the development of the linear-response theory~\cite{Kubo1957}.
We now know that statistical fluctuations obey fluctuation theorems, which are valid far from equilibrium beyond linear response and are particularly important for small systems~\cite{Evans1993, Gallavotti1995, Jarzynski1997}.

The principle of equal probability provides a definition of the probability of each macrostate; it is proportional to the phase-space volume of a set of microstates belonging to a given macrostate.
On the other hand, in 1868, Boltzmann introduced another definition of the probability based on a dynamical consideration~\cite{Boltzmann1868}.
Suppose that we observe a system during a long time period $T$, and that the system stays in a macrostate during a time $\tau$.
Then, the probability of this macrostate is defined by $\lim_{T\rightarrow\infty}\tau/T$.
This definition has a clear operational meaning, and Einstein~\cite{Einstein1903} preferred the latter definition to the former.

Encouraged by Maxwell's paper~\cite{Maxwell1879}, Boltzmann tried to establish a mechanical foundation for the principle of equal probability by attempting to show that the above two definitions of the probability are equivalent.
Through this attempt, Boltzmann proposed the \textit{ergodic hypothesis}, which states that the equal-energy surface in the classical phase space consists of a single trajectory obeying the equations of motion~\cite{Boltzmann1887}.
This original statement by Boltzmann turned out to be mathematically wrong, but the ergodic hypothesis was properly formulated later on the basis of the notion of metric transitivity by Birkhoff and Smith~\cite{Birkhoff1928} and Birkhoff~\cite{Birkhoff1931}.

If we assume the ergodic hypothesis, we can prove the equivalence of the two definitions of the probability.
In particular, the ergodic hypothesis ensures the equality between the long-time average and the microcanonical ensemble average of a physical quantity, which we shall call \textit{ergodicity}.
Khinchin~\cite{Khinchin_text} argued that the ergodic hypothesis is unnecessary and not essential to derive the ergodicity relevant to macroscopic systems.
He emphasized the importance of properly restricting the class of physical observables, which he called ``sum functions'', instead of considering arbitrary functions in phase space.
The fact that a thermodynamic system consists of large degrees of freedom plays a crucial role in his formulation.

Fermi, Pasta, and Ulam performed a numerical test of the validity of ergodicity in an anharmonic chain~\cite{Fermi1955}.
Surprisingly, they found that the ergodicity is violated, and the recurrence in the energy distribution to harmonic modes occurs (this work also shows that an early work by Fermi himself on the quasi-ergodic theorem~\cite{Fermi1923} is flawed).
This numerical finding led to the discovery of solitons~\cite{Zabusky1965} and played a crucial role in the chaos theory in connection with the KAM theorem developed by Kolmogorov, Arnold, and Moser~\cite{Kolmogorov1954, Moser1962, Arnold1963}.

In quantum systems, Pauli~\cite{Pauli1928} discussed the connection between reversible microscopic dynamics and irreversible relaxation to thermal equilibrium.
Pauli derived the master equation and proved the $H$-theorem by repeatedly using the random-phase assumption, which he considered to be a quantum counterpart of the assumption of molecular chaos in deriving the Boltzmann equation.
Later, van Hove~\cite{van_Hove1955} showed that Pauli's derivation is justified in some cases.
However, Pauli's treatment is rather restrictive.

In 1929, von Neumann published his work~\cite{Neumann1929} on the quantum ergodic theorem under quite a general setup.
By clearly distinguishing macroscopic variables from microscopic ones, he proved that thermalization is a typical behavior among macroscopic quantum systems.
His result follows from the fact that individual energy eigenstates already share an essential character of thermal equilibrium, which is now known as the \textit{eigenstate thermalization hypothesis}~\cite{Deutsch1991, Srednicki1994}.
Von Neumann's work may be regarded as a quantum extension of Boltzmann's idea on the derivation of the second law of thermodynamics from reversible microscopic dynamics, and it may be regarded as a milestone in the theory of thermalization in isolated quantum systems.
This is, however, not the end of the story.

Firstly, it turned out that there is a conceptual difficulty in von Neumann's result.
The quantum ergodic theorem is a statement about a vast majority of systems, but it does not prove anything about individual systems.
More precisely, von Neumann introduced an ensemble of Hamiltonians (or equivalently, an ensemble of ``macro observers'') that are generated by performing random unitary transformations to a reference Hamiltonian.
Then, it is found that a vast majority of Hamiltonians constructed in this way are highly unphysical since they typically involve nonlocal many-body interactions.
Thus von Neumann's quantum ergodic theorem cannot be considered as a complete theory on thermalization in isolated quantum systems.

Secondly, there is a new feature in quantum mechanics which is not considered in von Neumann's work; even \textit{microscopic} quantities can relax to its equilibrium value at the level of expectation values.
We call it the \textit{microscopic thermalization} in this review.
It is known that almost every pure state with a small energy fluctuation is locally indistinguishable from the microcanonical ensemble if the density of states (the number of energy eigenstates within a unit energy interval) is huge.
This property is called the canonical typicality~\cite{Goldstein2006, Popescu2006, Sugita2006}.
It results from quantum entanglement between subsystems, and has no classical counterpart.
It is noted that the density of states grows exponentially with the system size and is already very large in a relatively small system, e.g. a system involving 10 spins.
Therefore, thermalization in this sense can occur even in such small systems, which is also an interesting feature of the problem.

Thirdly, recent experimental advances in ultra-cold atomic systems have provided a strong impetus to theoretical studies of nonequilibrium dynamics in isolated quantum systems.
Microscopic thermalization has been observed in several experiments~\cite{Trotzky2012, Kaufman2016, Neill2016, Clos2016}.
In Ref.~\cite{Kaufman2016}, not only the expectation values of local quantities, but also the purity of a quantum state was measured, and this work provides a clear evidence that an isolated system shows microscopic thermalization even though its quantum state remains pure. 
The absence of thermalization has also been reported in integrable systems~\cite{Kinoshita2006,Gring2012} and many-body localized systems~\cite{Schreiber2015, Kondov2015, Choi2016, Smith2016}.

It has been experimentally reported that some isolated quantum systems exhibit \textit{prethermalization}~\cite{Gring2012, Kuhnert2013, Langen2013, Langen2015,Tang_arXiv2017}; the system first relaxes to a quasi-stationary nonequilibrium state before reaching thermal equilibrium.
It is also a subject of great current interest to understand the mechanism of prethermalization.

In this review, we provide a theoretical overview on the problems of thermalization and prethermalization in isolated quantum systems.
We discuss thermal equilibrium in classical systems and explain Boltzmann's idea in Sec.~\ref{sec:classical_equilibrium}, and
thermal equilibrium in quantum systems in Sec.~\ref{sec:quantum_equilibrium}.
We review general theory of thermalization in isolated quantum systems in Sec.~\ref{sec:thermalization}.
In Sec.~\ref{sec:prethermalization}, we discuss prethermalization in several systems.
Finally, we make concluding remarks in Sec.~\ref{sec:conclusion}.

%
%
%

\section{Thermal equilibrium in classical systems}
\label{sec:classical_equilibrium}

Before discussing quantum systems, we shall describe the notion of thermal equilibrium in classical systems, which dates back to Boltzmann's idea~\cite{Lebowitz1999_review,Gallavotti_text}.
In Sec.~\ref{sec:equilibrium}, we explain that thermal equilibrium is characterized by its typicality among microscopic states in an energy shell.
In Sec.~\ref{sec:entropy}, we introduce the Boltzmann entropy for each macrostate.

\subsection{Typicality of thermal equilibrium}
\label{sec:equilibrium}

Let us consider a classical system of $N$ particles contained in the region $\Lambda\subset\mathbb{R}^d$ of volume $V$ on $d$-dimensional space, whose microstate is specified by a set of positions $\bm{q}^N=(\bm{q}_1,\bm{q}_2,\dots,\bm{q}_N)\in\Lambda^N$ and momenta $\bm{p}^N=(\bm{p}_1,\bm{p}_2,\dots,\bm{p}_N)\in\mathbb{R}^{dN}$.
Here, $\bm{q}_i\in\Lambda$ and $\bm{p}_i\in\mathbb{R}^d$ denote the position and momentum of the $i$th particle, respectively.
A microstate is represented by a point on the phase space $\Gamma=(\bm{q}^N,\bm{p}^N)\in\Lambda^N\times\mathbb{R}^{dN}$, and the time evolution of the system is represented by a trajectory on the phase space $\Gamma_t$ as a function of time $t$.
We denote the Hamiltonian of the system by $H(\Gamma)$, and the classical Hamiltonian dynamics conserves the energy of the system, $H(\Gamma_t)=E$.
We can therefore restrict the phase space to an energy shell
\beq
\Omega_{E,N,\Lambda}:=\left\{\Gamma\in\Lambda^N\times\mathbb{R}^{dN}: H(\Gamma)\in[E-\Delta E,E]\right\}.
\eeq

Next, we fix a set of \textit{macrovariables}
\beq
\mathcal{M}=\{M_1(\Gamma),M_2(\Gamma),\dots,M_K(\Gamma)\},
\eeq
which are macroscopic quantities of interest and specifies a \textit{macrostate} of the system.
Natural choices of macrovariables are extensive quantities of the total system such as the Hamiltonian $H(\Gamma)$, the total number of particles $N(\Gamma)$, the total momentum, and the total magnetization.
We may also introduce extensive quantities for macroscopic subsystems consisting of the system, which are proper macrovariables. 

A macrostate of the system is a set of microstates which are grouped according to the values of macrovariables.
We divide these macrovariables into small intervals $M_i\in((\nu_i-1)\Delta M_i,\nu_i\Delta M_i]$ characterized by an integer $\nu_i$ and width $\Delta M_i$ which is much smaller than the typical magnitude of $M_i$ but large enough to contain many microstates in each interval.
The set of $\nu_i$, denoted by $\nu=(\nu_1,\nu_2,\dots,\nu_K)$, specifies the values of the macrovariables within precision $\Delta M_i$, and each macrostate is labeled by $\nu$.
The energy shell $\Omega_{E,N,\Lambda}$ is decomposed into macrostates $\{\nu\}$ as
\beq
\Omega_{E,N,\Lambda}=\bigcup_{\nu}\Omega_{\nu},
\eeq
where $\Omega_{\nu}$ is defined as
\begin{align}
\Omega_{\nu}:=\left\{\Lambda=(\bm{q}^N,\bm{p}^N)\in\Lambda^N\times\mathbb{R}^{dN}: M_i(\Gamma)\in((\nu_i-1)\Delta M_i, \nu_i\Delta M_i]\right.
\nonumber \\
\left. \text{ for all } i=1,2,\dots,K\right\}.
\end{align}
Note that $\Omega_{\nu}\cap\Omega_{\nu'}=\emptyset$ if $\nu\neq\nu'$.
We say that a microstate $\Gamma$ belongs to a macrostate $\nu$ if $\Gamma\in\Omega_{\nu}$.
We call $\Omega_{\nu}$ the subspace corresponding to the macrostate $\nu$.
The two microstates $\Gamma_1$ and $\Gamma_2$ are regarded as being macroscopically identical if they belong to the same macrostate.
When the system is in a macrostate $\nu=(\nu_1,\nu_2,\dots,\nu_K)$, the value of a macrovariable $M_i(\Gamma)$ is almost the same as $\nu_i\Delta M_i=:M_i^{(\nu)}$.

\begin{figure}
\centering
\includegraphics[width=10cm]{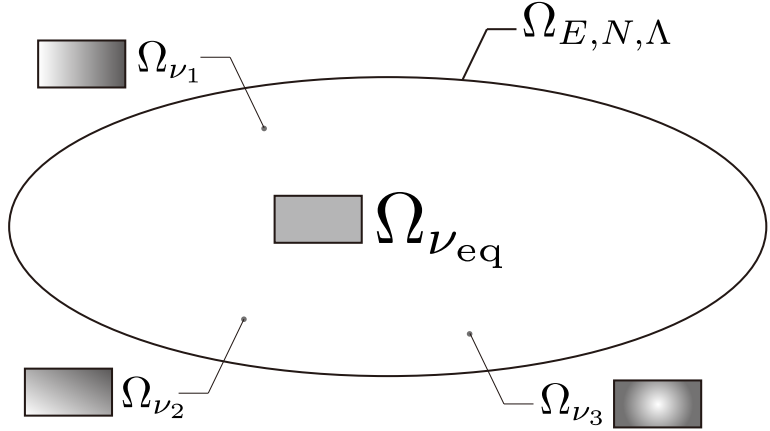}
\caption{Schematic illustration of the typicality of thermal equilibrium.
The elliptic region represents the energy shell $\Omega_{E,N,\Lambda}$,
which is constituted from macrostates labeled by $\nu$.
The typicality of thermal equilibrium means that
a single macrostate corresponding to thermal equilibrium $\Omega_{\nu_\mathrm{eq}}$ dominates the energy shell
and out-of-equilibrium states $\{\Omega_{\nu}\}_{\nu\neq \nu_\mathrm{eq}}$ are extremely rare.
Here each shaded rectangle schematically illustrates how particles distribute inside the box.
For a gas system in a box, $\{\Omega_{\nu}\}_{\nu\neq \nu_\mathrm{eq}}$ represent, for example,
nonuniform states, whereas $\Omega_{\nu_\mathrm{eq}}$ represents a uniform equilibrium state.
We here follow similar figures in Refs.~\cite{Tasaki2016,Tasaki_text}.
}
\label{fig:typicality}
\end{figure}

It is expected to be generally true and can be shown in many cases that when the system is macroscopic, there is a special macrostate $\nu_{\mathrm{eq}}$ that satisfies
\beq
\frac{\left|\Omega_{\nu_{\mathrm{eq}}}\right|}{\left|\Omega_{E,N,\Lambda}\right|}\approx 1,
\label{eq:therm_typicality1}
\eeq
where $|\Omega|$ is the phase space volume of the region $\Omega\subset\Lambda^N\times\mathbb{R}^{dN}$.
Moreover, we have
\beq
1-\frac{\left|\Omega_{\nu_{\mathrm{eq}}}\right|}{\left|\Omega_{E,N,\Lambda}\right|}=e^{-\mathcal{O}(V)},
\label{eq:therm_typicality2}
\eeq
which means that an overwhelming majority of microstates belong to $\nu_{\mathrm{eq}}$.
This macrostate $\nu_{\mathrm{eq}}$ corresponds to thermal equilibrium, and Eq.~(\ref{eq:therm_typicality2}) expresses the \textit{typicality of thermal equilibrium}; thermal equilibrium is characterized as a common property shared by an overwhelming majority of microstates in the energy shell.
See Fig.~\ref{fig:typicality} for a schematic picture.
In particular, we shall call the property of Eq.~(\ref{eq:therm_typicality2}) the \textit{thermodynamic typicality}.
The system in a microstate $\Gamma$ is said to be in thermal equilibrium if $\Gamma$ belongs to the equilibrium subspace $\Omega_{\nu_{\mathrm{eq}}}$.

The thermodynamic typicality explains why equilibrium statistical mechanics works so well.
The value of a macrovariable $M_i$ in thermal equilibrium is given by $M_i^{\mathrm{eq}}:=M_i^{(\nu_{\mathrm{eq}})}$, and it is also expressed as the average of $M_i(\Gamma)$ over the equilibrium subspace $\Omega_{\nu_{\mathrm{eq}}}$ (within a precision $\Delta M_i$):
\beq
M_i^{\mathrm{eq}}=\frac{1}{|\Omega_{\nu_{\mathrm{eq}}}|}\int_{\Omega_{\nu_{\mathrm{eq}}}}d\Gamma\, M_i(\Gamma),
\eeq
where $d\Gamma=d\bm{q}^Nd\bm{p}^N=d\bm{q}_1d\bm{q}_2\dots d\bm{q}_Nd\bm{p}_1d\bm{p}_2\dots d\bm{p}_N$ and the integration is taken over the equilibrium subspace $\Omega_{\nu_{\mathrm{eq}}}$.
Because of Eq.~(\ref{eq:therm_typicality1}) (or Eq.~(\ref{eq:therm_typicality2})), this average can be extended to an exponentially good approximation to the entire energy shell,
\beq
\frac{1}{|\Omega_{\nu_{\mathrm{eq}}}|}\int_{\Omega_{\nu_{\mathrm{eq}}}}d\Gamma\, M_i(\Gamma)
\approx\frac{1}{|\Omega_{E,N,\Lambda}|}\int_{\Omega_{E,N,\Lambda}}d\Gamma\, M_i(\Gamma)
=:\langle M_i\rangle_{\mathrm{mc}},
\eeq
where $\langle M_i\rangle_{\mathrm{mc}}$ is the average of $M_i$ over the microcanonical ensemble.
Thus we obtain
\beq
M_i^{\mathrm{eq}}\approx\langle M_i\rangle_{\mathrm{mc}}.
\eeq
The equilibrium value of a macrovariable is calculated by using the microcanonical ensemble.
This is the cornerstone of equilibrium statistical mechanics.

It is noted that $\langle M_i\rangle_{\mathrm{mc}}$ depends only on a few macroscopic parameters\footnote
{Strictly speaking, the energy shell depends on $E$, $N$, and $\Lambda$,
where $\Lambda$ is specified by its volume $V$ and its shape.
Therefore, $\langle M_i\rangle_{\mathrm{mc}}$ depends on $E$, $N$, $V$, and the shape of the material.
It is known that the shape dependence can be ignored for a macroscopic system in many cases.
Here, we do not consider the shape dependence, but it is noted that in long-range interacting systems, the shape dependence is important due to the lack of additivity.
}
such as the total energy $E$, the total number of particles $N$, and the volume $V$.
All the macrovariables in thermal equilibrium are determined by $(E,N,V)$, which justifies the validity of the thermodynamic description of macroscopic systems from a microscopic point of view.

\subsection{Boltzmann entropy}
\label{sec:entropy}

We can associate the Boltzmann entropy $S_{\nu}$ to each macrostate $\nu$ as follows.
The entropy in thermodynamics quantitatively characterizes which transitions are possible by thermodynamic operations.
In thermodynamic operations, a macroscopic system of our interest $A$ may be in contact with another macroscopic system $B$, and hence we must discuss a composite system consisting of the two macroscopic systems $A$ and $B$ in order to define the entropy of the system $A$.

Suppose that the two systems can exchange energy and particles but the interaction energy between them is negligible.
We shall calculate the phase-space volume of the subspace of a macrostate $\nu=(\nu_A,\nu_B)$, where each subsystem $X=A,B$ is in the macrostate $\nu_X$ with the total energy $E_X$ and the total number of particles $N_X$.
We assume that particles are macroscopically identical but microscopically distinguishable.
Then the phase-space volume of $\Omega_{(\nu_A,\nu_B)}$ is calculated as
\begin{align}
|\Omega_{(\nu_A,\nu_B)}|&=\frac{N!}{N_A!N_B!}\int_{\Omega_{\nu_A}^A}d\Gamma_A1\cdot\int_{\Omega_{\nu_B}^B}d\Gamma_B 1
\nonumber \\
&=N!\cdot\frac{|\Omega_{\nu_A}^A|}{N_A!}\cdot\frac{|\Omega_{\nu_B}^B|}{N_B!},
\label{eq:composite}
\end{align}
where $\Omega_{\nu_X}^X$ ($X=A,B$) is the subspace of a macrostate $\nu_X$ of $X$, $d\Gamma_X$ is its volume element, and $N=N_A+N_B$ is the total number of particles in the entire system.
The factor $N!/(N_A!N_B!)$ in the first line of Eq.~(\ref{eq:composite}) gives the number of ways of partitioning $N$ particles into the subsystems $A$ and $B$.
This factor is necessary because $N$ particles are indistinguishable at the level of macrostates.

The value of $|\Omega_{(\nu_A,\nu_B)}|$ is proportional to the probability of the macrostate $(\nu_A,\nu_B)$ of the composite system if we admit the principle of equal probability (all the microstates in the energy shell are equally probable in the sense of the Lebesgue measure in phase space).
Here, $N!$ in Eq.~(\ref{eq:composite}) is a constant since the composite system is considered to be isolated, and hence this probability is essentially proportional to $(|\Omega_{\nu_A}^A|/N_A!)\cdot(|\Omega_{\nu_B}^B|/N_B!)$.
Then, the factor $|\Omega_{\nu_X}^X|/N_X!$ can be interpreted as the statistical weight of a macrostate $\nu_X$ of the system $X$.
This weight is universal in the sense that the weight of a macrostate $\nu_A$ is independent of any macrostate $\nu_B$ of the subsystem $B$ which interacts weakly with $A$.
This points to the fundamental importance of the quantity $|\Omega_{\nu_X}^X|/N_X!$.
In fact, the Boltzmann entropy of the system in a macrostate $\nu$ with $N$ particles is defined as
\beq
S_{\nu}:=\ln\frac{|\Omega_{\nu}|}{N!},
\label{eq:Boltzmann}
\eeq
where we set the Boltzmann constant $k_B$ to be unity throughout the paper.
From Eq.~(\ref{eq:composite}), we have
\beq
S_{(\nu_A,\nu_B)}=S_{\nu_A}+S_{\nu_B},
\eeq
which shows the additivity of the Boltzmann entropy.

The Boltzmann entropy in thermal equilibrium is given by
\beq
S_{\nu_{\mathrm{eq}}}=\ln\frac{|\Omega_{\nu_{\mathrm{eq}}}|}{N!}\approx\ln\frac{|\Omega_{E,N,\Lambda}|}{N!}=: S_{\mathrm{mc}}(E,N,V),
\eeq
where $S_{\mathrm{mc}}(E,N,V)$ is the microcanonical entropy.
The Boltzmann entropy in thermal equilibrium is thus approximately calculated by using the microcanonical ensemble according to the standard equilibrium statistical mechanics.

We make a remark on the factor $N!$ in Eq.~(\ref{eq:Boltzmann}), which is sometimes explained as a consequence of the indistinguishability of identical particles in quantum mechanics~\cite{Callen_text,Kubo_text1}.
However, this argument does not explain why the factor $N!$ is necessary in, e.g., classical colloidal systems~\cite{Frenkel2014}, which is related to the so called \textit{Gibbs paradox} in classical statistical mechanics~\cite{van_Kampen1984,Jaynes1992}.
Also see a recent work~\cite{Murashita2017} in which the relationship between the Gibbs paradox and the fluctuation theorem is discussed.
As we have already seen, the factor $N!$ naturally arises even when particles are microscopially distinguishable.
Remember that we did not take care about which particles are located in each subsystem \textit{when we define a macrostate}.
This means that we have implicitly assumed that particles are \textit{macroscopically} indistinguishable, which leads to the factor $N!$ in Eq.~(\ref{eq:Boltzmann}).

The Boltzmann entropy $S_{\nu}$ may be interpreted as the logarithm of a quantity proportional to the probability of the macrostate $\nu$.
Then, the system will be likely to evolve to a macrostate with a larger Boltzmann entropy, and finally, reach thermal equilibrium because equilibrium macrostate $\nu_{\mathrm{eq}}$ has the largest Boltzmann entropy.
In other words, thermalization occurs through the evolution from a less probable macrostate with a small value of the Boltzmann entropy to a more probable macrostate with a larger value of the Boltzmann entropy.
This provides an intuitive explanation of why we expect the approach to thermal equilibrium in isolated systems.
This is a good picture as a starting point to understand thermalization, but we should answer why a less probable macrostate should evolve to a more probable macrostate.
In order to understand this problem, we should study the dynamics of thermalization, although it might be expected that the dynamical details should not be so important in thermalization in view of the universality of thermodynamics.
We present the theory of thermalization in Sec.~\ref{sec:thermalization}, and, in particular, we postpone  the discussion on thermalization in classical systems until Sec.~\ref{sec:classical}.

\section{Thermal equilibrium in quantum systems}
\label{sec:quantum_equilibrium}

\subsection{Preliminaries}
\label{sec:preliminary}

Here we explain the setting and notations in this paper.
We summarize our notations in Table~\ref{table:notations}.
For simplicity, we consider a lattice system (quantum spins, bosons, or fermions) in this section.

Let us denote by $\Lambda\subset\mathbb{R}^d$ a set of positions of all the sites belonging to the system, $\Lambda=\{\bm{r}_1,\bm{r}_2,\dots,\bm{r}_V\}$, where $\bm{r}_i\in\mathbb{R}^d$ is the position of the $i$th site.
The number of sites is defined by $V=|\Lambda|$, where for a set $X$, we denote $|X|$ by the number of elements in $X$.
In the one-dimensional case, we will also use $L$ instead of $V$.
A microstate is specified by a state vector $|\psi\>$ in a Hilbert space $\mathcal{H}$.
The Hamiltonian of the system is denoted by $\hat{H}$, and the corresponding eigenstates and eigenvalues are denoted by $|\phi_n\>$ and $E_n$, respectively; $\hat{H}|\phi_n\>=E_n|\phi_n\>$.
We assume that the system always has an almost definite value of the macroscopic energy $E$, which implies that the microstate of the system is always in the following Hilbert subspace
\beq
\mathcal{H}_{E,\Lambda}:=\mathrm{Span}\{|\phi_n\>\in\mathcal{H}: E_n\in[E-\Delta E,E]\}
\eeq
with some energy width $\Delta E$ that is arbitrary as long as it is macroscopically small but microscopically large\footnote
{More precisely, $\Delta E=o(V)$ should satisfy $\beta\Delta E\gg 1$, where $\beta$ is the inverse temperature corresponding to the energy $E$.
}.
We call $\mathcal{H}_{E,\Lambda}$ the \textit{energy shell} of the quantum system.
We define $D_{E,\Lambda}:=\dim\,\mathcal{H}_{E,\Lambda}$.
If we want to restrict our discussion into the subspace with a fixed value of another macroscopic conserved quantity besides energy, the energy shell should be considered in this subspace.
For example, when the total number operator of bosons or fermions denoted by $\hat{N}$ is a conserved quantity and we consider a fixed particle number $N$, then the energy shell $\mathcal{H}_{E,N,\Lambda}$ should be defined as the Hilbert subspace spanned by simultaneous eigenstates $\{|\phi_n\>\}$ of $\hat{H}$ and $\hat{N}$ satisfying $E_n\in[E-\Delta E,E]$ and $\hat{N}|\phi_n\>=N|\phi_n\>$.

We shall introduce several classes of observables.
In a lattice system,  a Hilbert space $\mathcal{H}_i$ is associated with each site $i\in\{1,2,\dots, V\}$.
We assume that the Hilbert space $\mathcal{H}_i$ is finite dimensional.
For a Bose system, the dimension of the local Hilbert space is infinite, but we make $\dim\,\mathcal{H}_i$ finite by truncating the Hilbert space so that the maximum number of bosons at each site is restricted to $n_{\mathrm{max}}<+\infty$.
The algebra of bounded linear operators on $\mathcal{H}_i$ is denoted by $\mathcal{B}_i$.
For a subset $X\subset\{1,2,\dots,V\}$, we define $\mathcal{H}_X:=\bigotimes_{i\in X}\mathcal{H}_i$ and $\mathcal{B}_X:=\bigotimes_{i\in X}\mathcal{B}_i$, which is the algebra of bounded operators on $\mathcal{H}_X$.
The dimension of the Hilbert space $\mathcal{H}_X$ is defined as $D_X:=\dim\,\mathcal{H}_X$.
It is noted that $D_X=e^{\mathcal{O}(|X|)}$.
The entire Hilbert space is given by $\mathcal{H}=\bigotimes_{i=1}^V\mathcal{H}_i$ (for Bose or Fermi systems, $\mathcal{H}$ represents the Fock space) and the algebra of the bounded linear operators on $\mathcal{H}$ is denoted by $\mathcal{B}=\bigotimes_{i=1}^V\mathcal{B}_i$.

For $\hat{O}\in\mathcal{B}$, we define the operator norm of $\hat{O}$ by
\beq
\|\hat{O}\|:=\sup_{|\psi\>\in\mathcal{H}:\<\psi|\psi\>=1}\sqrt{\<\psi|\hat{O}^{\dagger}\hat{O}|\psi\>}.
\eeq
We also define the trace norm of $\hat{O}$ by
\beq
\|\hat{O}\|_1:=\mathrm{Tr}\sqrt{\hat{O}^{\dagger}\hat{O}}.
\eeq

The support of an operator $\hat{O}\in\mathcal{B}$, which is denoted by $\mathrm{Supp}(\hat{O})$, is defined as the minimal set $X\subset\{1,2,\dots,V\}$ for which $\hat{O}=\hat{O}_X\otimes \hat{1}_{X^c}$ can be satisfied with $\hat{O}_X\in\mathcal{B}_X$ and $\hat{1}_{X^c}$ being the identity operator on $\mathcal{H}_{X^c}$, where the complement of $X$ is denoted by $X^c:=\{1,2,\dots,V\}\setminus X$.
We define $\mathcal{S}_X$ as a set of all the operators $\hat{O}\in\mathcal{B}$ with $\mathrm{Supp}(\hat{O})\subseteq X$.
The number of linearly independent operators in $\mathcal{S}_X$ is given by $D_X^2$.

Now we define \textit{local operators} on a length scale $\ell$.
Let us denote by $X_i^{(\ell)}$ the set of sites such that the distance from the site $i$ is not greater than $\ell$, i.e.,
\beq
X_i^{(\ell)}:=\{j\in\{1,2,\dots,V\}: |\bm{r}_i-\bm{r}_j|\leq\ell\}.
\eeq
The set of operators supported by $X_i^{(\ell)}$ is denoted by $\mathcal{S}_i^{(\ell)}$, i.e., $\mathcal{S}_i^{(\ell)}:=\mathcal{S}_{X_i^{(\ell)}}$.
We define
\beq
v_{\ell}:=\max_{i\in\{1,2,\dots,V\}}|X_i^{(\ell)}|,
\eeq
where $v_{\ell}=\mathcal{O}(\ell^d)$ in a $d$-dimensional lattice.
For notational simplicity, we define
\beq
\left\{
\begin{aligned}
&D_i^{(\ell)}:=D_{X_i^{(\ell)}}, \\
&D_{\ell}:=\max_{i\in\{1,2,\dots,V\}}D_i^{(\ell)}.
\end{aligned}
\right.
\eeq
It is noted that $D_{\ell}=e^{\mathcal{O}(v_{\ell})}=e^{\mathcal{O}(\ell^d)}$.

We say that $\hat{O}\in\mathcal{B}$ is a local operator on the length scale $\ell$ if $\hat{O}\in\mathcal{S}_i^{(\ell)}$ for some $i\in\{1,2,\dots,V\}$.
The set of all the local operators on the length scale $\ell$ is denoted by $\mathcal{S}_{\mathrm{loc}}^{(\ell)}=\bigcup_{i=1}^V\mathcal{S}_i^{(\ell)}$.
For a Fermi system, however, the locality of operators is ambiguous (see the remark at the end of this section).

We say that $\hat{O}\in\mathcal{B}$ is a \textit{few-body operator}\footnote
{Such an operator is also called a $k$-local operator in literature.
}
if $|\mathrm{Supp}(\hat{O})|\leq k$ with some $k=\mathcal{O}(1)$.
The set of few-body operators with a certain $k$ is denoted by $\mathcal{S}_{\mathrm{few}}^{(k)}:=\bigcup_{X\subset\{1,2,\dots,V\}, |X|=k}\mathcal{S}_X$.
If $\hat{O}$ is a local operator on length scale $\ell$, it is also a few-body operator with $k=v_{\ell}$, which implies $\mathcal{S}_{\mathrm{few}}^{(v_{\ell})}\supset\mathcal{S}_{\mathrm{loc}}^{(\ell)}$.

An operator $\hat{A}\in\mathcal{B}$ is said to be an \textit{extensive local operator} in the (sub)system $X\subseteq\{1,2,\dots,V\}$ if $\hat{A}$ is a translation-invariant sum of a local operator, i.e., $\hat{A}=\sum_{i\in X}\hat{O}_i$ with $\hat{O}_i\in\mathcal{S}_i^{(\ell)}$ being translational copies of a local operator on the length scale $\ell=\mathcal{O}(1)$.
An \textit{intensive local operator} is defined as the density of an extensive local operator $\sum_{i\in X}\hat{O}_i/|X|$.

For convenience in later analysis, we introduce a complete orthonormal basis of $\mathcal{S}_X$.
Any $\hat{O}\in\mathcal{S}_X$ can be written as $\hat{O}=\hat{O}_X\otimes \hat{1}_{X^c}$ with $\hat{O}_X\in\mathcal{B}_X$.
For $p=1,2,\dots,D_X^2$, let $\hat{O}^{(p)}=\hat{O}_X^{(p)}\otimes\hat{1}_{X^c}$ be operators satisfying $\mathrm{Tr}_X\hat{O}_X^{(p)\dagger}\hat{O}_X^{(q)}=\delta_{pq}$, where $\mathrm{Tr}_X$ denotes the trace over the Hilbert space $\mathcal{H}_X$.
Then, $\{\hat{O}^{(p)}\}_{p=1}^{D_X^2}$ is a complete orthonormal basis of $\mathcal{S}_X$ in the sense that any $\hat{O}\in\mathcal{S}_X$ can be expanded as
\beq
\hat{O}=\sum_{p=1}^{D_X^2}c_pO^{(p)}
\eeq
with $c_p=\mathrm{Tr}_X \hat{O}_X^{(p)\dagger}\hat{O}_X$.

Moreover, we can always find a complete orthonormal basis satisfying
\beq
\|\hat{O}^{(p)}\|=\frac{1}{\sqrt{D_X}}
\label{eq:orthonormal}
\eeq
for every $p$.
An example of such $\{\hat{O}^{(p)}\}_{p=1}^{D_X^2}$ is constructed as follows.
First, let us define $n,m=0,1,2,\dots,D_X-1$ by $p=D_Xn+m+1$, in which there is a one-to-one correspondence between $p$ and $(n,m)$.
For an arbitrary basis vectors $\{|k\>\}_{k=1}^{D_X}$ with $\<k|k'\>=\delta_{k,k'}$, let us define, for each $p=1,2,\dots,D_X^2$,
\beq
\hat{O}_X^{(p)}=\frac{1}{\sqrt{D_X}}\sum_{k=1}^{D_X}e^{i\frac{2\pi}{D_X}mk}|k+n\>\<k|,
\eeq
where we define $|k+D_X\>=|k\>$ for notational simplicity.
Then, a set of operators $\{\hat{O}^{(p)}\}_{p=1}^{D_X^2}$ satisfy $\mathrm{Tr}_X\,\hat{O}^{(p)\dagger}_X\hat{O}^{(q)}_X=\delta_{pq}$ and $\|\hat{O}^{(p)}\|=\|\hat{O}_X^{(p)}\|=1/\sqrt{D_X}$.
This complete set is known as the Weyl operator basis~\cite{Bennett1993}.

By using the inequalities $\sum_{p=1}^{D_X^2}|c_p|\leq D_X\left(\sum_{p=1}^{D_X^2}|c_p|^2\right)^{1/2}$ and $\sum_{p=1}^{D_X^2}|c_p|^2=\mathrm{Tr}_X\hat{O}_X^{\dagger}\hat{O}_X\leq D_X\|\hat{O}\|^2$, we have
\beq
\sum_{p=1}^{D_X^2}|c_p|\leq D_X^{3/2}\|\hat{O}\|,
\label{eq:c_p}
\eeq
which will be used later.

\begin{table}
\centering
\caption{List of notations used in this review.}
\label{table:notations}
\renewcommand{\arraystretch}{1.5} 
\begin{tabular}{l p{8 cm}} 
\hline\hline
$\Lambda\in \mathbb{R}^d$ & set of positions of all sites \\
$V=|\Lambda|$ & total number of sites\\
$\mathcal{H}_i$ & Hilbert space at site $i$ with $w=\dim\mathcal{H}_i<\infty$\\
$\mathcal{H}_X=\bigotimes_{i\in X}\mathcal{H}_i$ &
Hilbert space on region $X\subset \{1,2,\dots,V\}$
with $D_X=\dim \mathcal{H}_X$\\
$\mathcal{H}=\bigotimes_{i=1}^V\mathcal{H}_i$ & entire Hilbert space\\
$\mathcal{B}_i,\mathcal{B}_X,\mathcal{B}$ & algebras of bounded linear operators on $\mathcal{H}_i$, $\mathcal{H}_X$, $\mathcal{H}$\\
$\mathcal{S}_X$ & set of $\hat{O}\in \mathcal{B}$ such that $\mathrm{Supp}(\hat{O})\subseteq X$\\
$X_i^{(\ell)}$ & $\{j\in \{1,2,\dots,V\} : |\bm{r}_i-\bm{r}_j|\le \ell\}$\\
$\mathcal{S}_i^{(\ell)} =\mathcal{S}_{X_i^{(\ell)}}$& set of local operators on length $\ell$ at site $i$\\
$\mathcal{S}^{(\ell)}_\mathrm{loc}=\bigcup_{i=1}^V \mathcal{S}_i^{(\ell)}$ & set of local operators on length $\ell$\\
$\mathcal{S}_\mathrm{few}^{(k)}$ & set of few-body operators with $k$, $\{\hat{O}\in \mathcal{B} : |\mathrm{Supp}(\hat{O})|\le k\}$\\
$\{\hat{O}^{(p)}_X\}_{p=1}^{D_X^2}$ & orthonormal basis set of $\mathcal{B}_X$\\
$\{\hat{O}^{(p)}=\hat{O}^{(p)}_X\otimes \hat{1}_{X^c}\}_{p=1}^{D_X^2}$ & orthonormal basis set of $\mathcal{S}_X$\\[1mm]
\hline
\end{tabular}
\renewcommand{\arraystretch}{1}
\end{table}

We conclude this section by making a remark on the locality of a Fermi system.
We should take care about locality in a Fermi system because the locality of the fermion creation and annihilation operators $\hat{f}_i^{\dagger}$ and $\hat{f}_i$ is ambiguous.
In general, the support of $\hat{f}_i^{\dagger}$ and $\hat{f}_i$ is the entire system because of the fermion anti-commutation relations.
However, for a local region $X$, by properly relabeling the sites, we can always make $\hat{f}_i^{\dagger}$ and $\hat{f}_i$ local for any $i\in X$ so that $\hat{f}_i^{\dagger}, \hat{f}_i\in\mathcal{S}_{X}$.
Usually, the occupation number representation of a state in a fermion system is defined by
\beq
|n_1,n_2,\dots,n_V\>=(\hat{f}_1^{\dagger})^{n_1}(\hat{f}_2^{\dagger})^{n_2}\dots(\hat{f}_V^{\dagger})^{n_V}|0\>,
\eeq
where $n_i=0$ or 1 is the occupation number of fermions at site $i$ and $|0\>$ is the Fock vacuum state.
We define local Hilbert spaces $\mathcal{H}_i$ so that $|n_1,n_2,\dots,n_V\>=|n_1\>\otimes|n_2\>\otimes\dots\otimes|n_V\>$ in which $|n_i\>\in\mathcal{H}_i$.
If we relabel the sites so that $X=\{1,2,\dots,|X|\}$, then $\hat{f}_i^{\dagger}$ and $\hat{f}_i$ for any $i\in X$ become local operators with $\mathrm{Supp}(\hat{f}_i)=\mathrm{Supp}(\hat{f}_i^{\dagger})\subseteq X$.
By using this prescription for each local region $X\subset\{1,2,\dots,V\}$, we can talk about the locality of fermion creation and annihilation operators, although we cannot make all $\hat{f}_i^{\dagger}$ and $\hat{f}_i$ local simultaneously.

It should also be noted that when a lattice Fermi system preserves the parity of the number of fermions given by $\hat{P}_{\mathrm{F}}:=(-1)^{\hat{N}}$ with $\hat{N}$ being the total number of fermions, such a system can be treated in a similar way as quantum spin systems in many respects~\cite{Nachtergaele_arXiv2017}.
For $X$ and $Y$ being subsets of $\{1,2,\dots,V\}$, let us consider two operators $\hat{A}_X$ and $\hat{B}_Y$ given by
\begin{align}
\hat{A}_X&=\prod_{i\in X}\hat{A}_i, \quad \hat{A}_i\in\{\hat{1},\hat{c}_i,\hat{c}_i^{\dagger},\hat{c}_i^{\dagger}\hat{c}_i\}, 
\label{eq:A_X}\\
\hat{B}_Y&=\prod_{i\in Y}\hat{B}_i, \quad \hat{B}_i\in\{\hat{1},\hat{c}_i,\hat{c}_i^{\dagger},\hat{c}_i^{\dagger}\hat{c}_i\},
\end{align}
where $\hat{c}_i$ is the annihilation operator of a fermion at site $i$.
Now we assume that either $\hat{A}_X$ or $\hat{B}_Y$ commutes with $\hat{P}_{\mathrm{F}}$.
Then we have $[\hat{A}_X,\hat{B}_Y]=0$ if $X\cap Y=\emptyset$.
Although the supports of $\hat{A}_X$ and $\hat{B}_Y$ are not local in general, we can treat them as if they were local operators.
Therefore, in a Fermi system, $\mathcal{S}_X$ should be defined as the set of operators written in the form of Eq.~(\ref{eq:A_X}).


\subsection{Macroscopic thermal equilibrium in quantum systems}
\label{sec:MATE}

We can discuss thermal equilibrium in quantum systems analogously to classical systems.
In quantum systems, however, there are two different notions of thermal equilibrium, which we shall call \textit{macroscopic thermal equilibrium} (MATE) and \textit{microscopic thermal equilibrium} (MITE) following Goldstein et al.~\cite{Goldstein2015,Goldstein2017}.
MATE is a natural extension of Boltzmann's notion of thermal equilibrium to quantum systems and formulated by von Neumann~\cite{Neumann1929}, while MITE is a purely quantum notion without classical analogue.
We explain MATE in this section, and MITE in Sec.~\ref{sec:MITE}.

Similarly to the classical case, a macrostate of a quantum system is specified by a fixed set of macrovariables
\beq
\mathcal{M}=\{\hat{M}_1,\hat{M}_2,\dots,\hat{M}_K\},
\eeq
where each $\hat{M}_i$ is a self-adjoint operator representing a macroscopic quantity.
To give examples of proper macrovariables, we partition a system in the region $\Lambda$ into macroscopic subsystems in the region $\Lambda_k$ ($k$ is an index of each subsystem), where $\Lambda=\bigcup_k\Lambda_k$ and $\Lambda_k\cap\Lambda_{k'}=\emptyset$ for any $k\neq k'$.
The $k$th subsystem is macroscopic in the sense that its volume $V_k$ is comparable with the total volume $V$. (The thermodynamic limit should be taken with $V_k/V$ held fixed for every $k$.)
A natural choice of $\mathcal{M}$ is a finite set of extensive or intensive local operators in the whole region $\Lambda$ or in each macroscopic subsystem $\Lambda_k$.
Examples of macrovariables include the total number of particles, the total energy, the total momentum, and the total magnetization for the entire system or for each subsystem $k$.

A macrostate of the system is specified by a set of values of the macrovariables, but there is a difficulty in the quantum case.
Since those macrovariables are not mutually commutable in general, $[\hat{M}_i,\hat{M}_j]\neq 0$, the ``values'' of the macrovariables cannot be determined simultaneously.
Von Neumann~\cite{Neumann1929} suggested that we should approximate macrovariables $\hat{M}_i$ by mutually commutable self-adjoint operators $\tilde{M}_i$ and redefine the set of macrovariables as $\tilde{\mathcal{M}}=\{\tilde{M}_1,\tilde{M}_2,\dots,\tilde{M}_K\}$, although mathematically it is highly nontrivial whether we can actually find $\tilde{\mathcal{M}}$ with a desired property~\cite{Davidson1985,Lin1997,Ogata2013}.
Then, we can partition the energy shell\footnote
{Strictly speaking, the energy shell should be redefined by using an approximate Hamiltonian $\tilde{H}$ that commutes with all $\tilde{M}_i$, $i=1,2,\dots, K$.
}
$\mathcal{H}_{E,\Lambda}$ into the subspace of macrostates $\nu$:
\beq
\mathcal{H}_{E,\Lambda}=\bigoplus_{\nu}\mathcal{H}_{\nu},
\eeq
where a macrostate $\nu$ is specified by a set of integers $(\nu_1,\nu_2,\dots,\nu_K)$ and
\begin{align}
\mathcal{H}_{\nu}:=\mathrm{Span}\left\{|\psi\>\in\mathcal{H}_{E,\Lambda}:\tilde{M}_i|\psi\>=M_i'|\psi\>, M_i'\in((\nu_i-1)\Delta M_i, \nu_i\Delta M_i]\right.
\nonumber \\
\text{ for all } i=1,2,\dots K\Big\}
\end{align}
with some small (but microscopically large) width $\Delta M_i$.
The dimension of $\mathcal{H}_{\nu}$ is denoted by $D_{\nu}$.
For a pure state in $\mathcal{H}_{\nu}$, the measurement value of a macrovariable $\tilde{M}_i$ is almost identical to $M_i^{(\nu)}=\nu_i\Delta M_i$.

Similarly to the classical case, among the macrostates $\nu$, there is a particular macrostate $\nu_{\mathrm{eq}}$ such that
\beq
\frac{D_{\nu_{\mathrm{eq}}}}{D_{E,N,\Lambda}}\approx 1,
\label{eq:therm_typicality_q1}
\eeq
or more precisely, there exists some $\gamma>0$ such that
\beq
1-\frac{D_{\nu_{\mathrm{eq}}}}{D_{E,\Lambda}}\leq e^{-\gamma V}.
\label{eq:therm_typicality_q2}
\eeq
This particular state $\nu_{\mathrm{eq}}$ corresponds to thermal equilibrium, and Eqs.~(\ref{eq:therm_typicality_q1}) and (\ref{eq:therm_typicality_q2}) correspond to the thermodynamic typicality in quantum systems.
Equation~(\ref{eq:therm_typicality_q2}) is also referred to as the ``thermodynamic bound'' in Ref.~\cite{Tasaki2016}.

In a quantum system, however, a microstate represented by a pure quantum state $|\psi\>\in\mathcal{H}_{E,\Lambda}$ is not necessarily in a single macrostate $\nu$.
We denote the projection onto the Hilbert subspace $\mathcal{H}_{\nu}$ by $\hat{P}_{\nu}$.
If for some $\nu$ $\<\psi|\hat{P}_{\nu}|\psi\>\approx 1$ and $\<\psi|\hat{P}_{\nu'}|\psi\>\approx 0$ for all $\nu'\neq\nu$, we can say that a quantum state $|\psi\>$ is almost in a macrostate $\nu$.
When $|\psi\>$ has large overlaps with several different macrostates, the quantum fluctuation of some macrovariable $\hat{M}_i$ is larger than $\Delta M_i$ in such a quantum state\footnote
{A pure quantum state is called a cat state if some macroscopic quantity exhibits a macroscopically large fluctuation~\cite{Shimizu2005}.
}.
If the initial state $|\psi(0)\>$ is in a single macrostate $\nu_0$, quantum fluctuations of macrovariables are expected to remain very small during the time evolution, and the quantum state $|\psi(t)\>$ at time $t>0$ will also be in a single macrostate $\nu_t$.
This is a physically natural but highly nontrivial dynamical property.

Although $|\psi\>$ is not always in a single macrostate, we can always determine whether $|\psi\>$ represents thermal equilibrium.
Let us denote the projection onto the equilibrium subspace $\mathcal{H}_{\mathrm{eq}}:=\mathcal{H}_{\nu_{\mathrm{eq}}}$ by $\hat{P}_{\mathrm{eq}}$.
Then, a system is said to be in MATE if its quantum state $|\psi\>$ satisfies
\beq
\<\psi|\hat{P}_{\mathrm{eq}}|\psi\>\geq 1-\epsilon
\label{eq:MATE1}
\eeq
with some fixed small $\epsilon>0$.
We also say that a state $|\psi\>$ represents MATE if Eq.~(\ref{eq:MATE1}) is satisfied.
The thermodynamic typicality expressed by Eq.~(\ref{eq:therm_typicality_q2}) suggests that we should choose $\epsilon=e^{-\alpha V}$ with some fixed $\alpha\in(0,\gamma)$~\cite{Tasaki2016}.
Thus,
\beq
\<\psi|\hat{P}_{\mathrm{eq}}|\psi\>\geq 1-e^{-\alpha V}
\label{eq:MATE2}
\eeq
is the condition of MATE.
The condition $\alpha<\gamma$ ensures that an overwhelming majority of $|\psi\>\in\mathcal{H}_{E,\Lambda}$ represents MATE, which has rigorously been proven by Tasaki~\cite{Tasaki2016}\footnote
{It should be noted that Tasaki~\cite{Tasaki2016} employed a slightly different definition of MATE, although the difference is not essential, see Ref.~\cite{Tasaki2016} as well as Refs.~\cite{Goldstein2015,Goldstein2017}.
In his formulation, the condition under which each microstate $|\psi\>\in\mathcal{H}_{E,\Lambda}$ represents MATE is directly given without introducing Hilbert subspaces corresponding to macrostates $\nu$.
The advantage of the formulation by Tasaki is that one can avoid technical complications due to the approximation replacing $\mathcal{M}$ by $\tilde{\mathcal{M}}$.
Instead, individual nonequilibrium macrostates cannot be discussed in his formulation.}.

Similarly to the classical case, we define the Boltzmann entropy of a macrostate $\nu$ as
\beq
S_{\nu}:=\ln D_{\nu}.
\eeq
The Boltzmann entropy in thermal equilibrium is given by
\beq
S_{\mathrm{eq}}=\ln D_{\mathrm{eq}}\approx \ln D_{E,\Lambda}=:S_{\mathrm{mc}}(E,V),
\eeq
where we have used Eq.~(\ref{eq:therm_typicality_q1}) and $D_{\mathrm{eq}}:=D_{\nu_{\mathrm{eq}}}$.
The Boltzmann entropy in thermal equilibrium is essentially identical to the microcanonical entropy denoted by $S_{\mathrm{mc}}(E,V)$.

\subsection{Validity of thermodynamic typicality}
\label{sec:thermodynamic_typicality}

The thermodynamic typicality is quite natural and expected to hold in general many-body systems.
However, it is a nontrivial issue to prove the thermodynamic typicality starting from the microscopic Hamiltonian.
In fact, the thermodynamic typicality has been shown in some cases through a large deviation property in the microcanonical (or canonical) ensemble as discussed below.

For simplicity, we consider a single macrovariable $\mathcal{M}=\{\hat{m}\}$, where $\hat{m}$ is an intensive local observable, $\hat{m}=(1/V)\sum_{i=1}^V\hat{O}_i$ with $\hat{O}_i\in\mathcal{S}_i^{(\ell)}$ being translational copies of a local operator $\hat{O}$.
In translation-invariant quantum spin systems with short-range interactions, there are several rigorous results.
Let us consider the canonical ensemble $\rho_{\mathrm{can}}=e^{-\beta H}/\mathrm{Tr}\,e^{-\beta H}$ with inverse temperature $\beta$ and define $\<\hat{O}\>_{\mathrm{can}}:=\mathrm{Tr}\,\hat{O}\rho_{\mathrm{can}}$.

For $d=1$, it has been proven that $\hat{m}$ shows the following large deviation property: for any positive $\epsilon$ and any positive $\beta$, there exist $V_0$ and $\eta>0$ such that
\beq
\left\<\mathsf{P}[|\hat{m}-\<\hat{m}\>_{\mathrm{can}}|\geq\epsilon]\right\>_{\mathrm{can}}\leq e^{-\eta V}
\label{eq:LDP}
\eeq
holds for any $V\geq V_0$~\cite{Ogata2010}\footnote
{Ogata~\cite{Ogata2010} actually proved a stronger result than Eq.~(\ref{eq:LDP}), but for our purpose, Eq.~(\ref{eq:LDP}) is enough.
}.
Here, for a self-adjoint operator $\hat{A}=\sum_{n}a_n|a_n\>\<a_n|$, $\mathsf{P}[\hat{A}\geq a]$ denotes the projection operator onto the subspace spanned by the eigenstates of $\hat{A}$ with the eigenvalues greater than or equal to $a$, i.e., $\mathsf{P}[\hat{A}\geq a]=\sum_{n: a_n\geq a}|a_n\>\<a_n|$.
Equation~(\ref{eq:LDP}) implies that the large-deviation probability from the equilibrium value becomes exponentially small with increasing the system size.
Because of the ensemble equivalence~\cite{Ruelle_text}, it is also shown that we can replace the canonical ensemble in Eq.~(\ref{eq:LDP}) by the microcanonical ensemble with energy $E$ corresponding to $\beta$, i.e., $E=\<H\>_{\mathrm{can}}$~\cite{Tasaki2016}.
Equation~(\ref{eq:LDP}) is equivalent to the thermodynamic typicality in the case of a single macrovariable.

For $d\geq 2$, Eq.~(\ref{eq:LDP}) was proven for sufficiently high temperatures~\cite{Netocny2004,Lenci2005}.
It is believed that the thermodynamic typicality generally holds in physically realistic many-body Hamiltonians for any equilibrium ensemble corresponding to a single thermodynamic phase.

\subsection{Canonical typicality}
\label{sec:can_typicality}

Equations~(\ref{eq:therm_typicality2}) and (\ref{eq:therm_typicality_q2}) motivate us to define $\nu_{\mathrm{eq}}$ as MATE in classical and quantum systems, respectively, and they characterize thermal equilibrium as a typical macrostate.
In quantum systems, as we see in this section, there is another typicality statement, which leads to a different definition of thermal equilibrium referred to as MITE.

Let us partition an isolated quantum system into a small subsystem $X\subset\{1,2,\dots,V\}$ and the much larger remaining part $X^c$.
We are interested in physical quantities in $\mathcal{S}_{X}$.
The expectation value of $\hat{O}=\hat{O}_X\otimes\hat{1}_{X^c}\in\mathcal{S}_X$ in a quantum state $|\psi\>$ is written as
\beq
\<\psi|\hat{O}|\psi\>=\mathrm{Tr}_X\hat{O}_X\rho_{\psi}^X,
\label{eq:reduced}
\eeq
where $\mathrm{Tr}_X$ denotes the trace over the Hilbert space $\mathcal{H}_X$, and
\beq
\rho_{\psi}^X:=\mathrm{Tr}_{X^c}|\psi\>\<\psi|
\eeq
is the reduced density matrix of the state $|\psi\>$ in the subsystem $X$.

Although the quantum state $|\psi\>$ of the total system is pure, the state $\rho_{\psi}^X$ of the subsystem $X$ is a mixed state if and only if there is quantum entanglement between $X$ and $X^c$.
This is in clear contrast to a classical system, in which a subsystem is always in a single microstate (not a mixture of microstates) whenever the total system is in a single microstate.

Even if the two states $|\psi\>$ and $|\psi'\>$ are not close to each other, their reduced density matrices $\rho_{\psi}^X$ and $\rho_{\psi'}^X$ can be very close.
An important observation by Goldstein et al.~\cite{Goldstein2006} and Popescu et al.~\cite{Popescu2006} is that almost all the quantum states in the energy shell, $|\psi\>\in\mathcal{H}_{E,\Lambda}$, give almost the same reduced density matrix $\rho_{\mathrm{mc}}^X$, where $\rho_{\mathrm{mc}}^X$ is the reduced density matrix of the microcanonical ensemble of the total system:
\beq
\rho_{\psi}^X\approx\rho_{\mathrm{mc}}^X:=\mathrm{Tr}_{X^c}\rho_{\mathrm{mc}} \quad \text{for almost all }|\psi\>\in\mathcal{H}_{E,\Lambda}.
\label{eq:can_typicality}
\eeq
More precisely, Popescu et al.~\cite{Popescu2006} obtained the following result by using Levy's lemma.
Let us choose $|\psi\>$ randomly from $\mathcal{H}_{E,\Lambda}$ according to the uniform measure on the unit sphere.
We denote the probability of obtaining $|\psi\>$ satisfying the condition $\mathcal{C}$ by $\mu_{\psi}[\mathcal{C}]$.
Then, we obtain
\beq
\mu_{\psi}\left[\left\|\rho_{\psi}^X-\rho_{\mathrm{mc}}^X\right\|_1\geq\eta\right]\leq\eta',
\label{eq:Popescu}
\eeq
where, for an arbitrary $\epsilon>0$,
\beq
\eta=\epsilon+\sqrt{\frac{D_X^2}{D_{E,\Lambda}}}
\eeq
and
\beq
\eta'=4\exp\left(-\frac{\epsilon^2}{18\pi^3}D_{E,\Lambda}\right).
\eeq
In Eq.~(\ref{eq:Popescu}), $\|\cdot\|_1$ denotes the trace norm.

For a macroscopic system, $D_{E,\Lambda}=e^{\mathcal{O}(V)}$ and $D_X=e^{\mathcal{O}(|X|)}$.
As long as $|X|\ll V$, $\eta$ and $\eta'$ are extremely small for large $V$ under a suitable choice of $\epsilon$.
This result clearly shows that almost all $|\psi\>\in\mathcal{H}_{E,\Lambda}$ are indistinguishable from the microcanonical ensemble as long as we look at a small subsystem $X$.
This property is referred to as the \textit{canonical typicality}~\cite{Goldstein2006,Goldstein2015}.

The canonical typicality ensures that for almost all $|\psi\>\in\mathcal{H}_{E,\Lambda}$,
\beq
\<\psi|\hat{O}|\psi\>\approx\<\hat{O}\>_{\mathrm{mc}}:=\mathrm{Tr}\,\hat{O}\rho_{\mathrm{mc}}
\eeq
for every $\hat{O}\in\mathcal{S}_X$.
This implies that a single pure quantum state can reproduce the result of the microcanonical ensemble.
By using this fact, we can find efficient algorithms to simulate equilibrium states in quantum systems~\cite{Imada1986, Jaklic1994, Hams2000, Sugiura2012, Sugiura2013}.

We remark that a similar typicality statement holds for an arbitrary bounded operator $\hat{O}\in\mathcal{B}$, which may be nonlocal, i.e.,
\beq
\mu_{\psi}\left[\left|\<\psi|\hat{O}|\psi\>-\<\hat{O}\>_{\mathrm{mc}}\right|\geq\epsilon\right]\leq\frac{\|\hat{O}\|^2}{\epsilon^2D_{E,\Lambda}}.
\label{eq:expectation_typicality}
\eeq
This result was obtained by Sugita~\cite{Sugita2006,Sugita2007} and Reimann~\cite{Reimann2007}.
Equation~(\ref{eq:expectation_typicality}) tells us that the locality of an operator is not essential for typicality of thermal equilibrium at the level of expectation values.

\subsection{Microscopic thermal equilibrium in quantum systems}
\label{sec:MITE}

Equation~(\ref{eq:can_typicality}) suggests another definition of thermal equilibrium, which should be characterized by a common property shared by an overwhelming majority of microstates.
We say that a system is in MITE on the length scale $\ell$ whenever its quantum state $|\psi\>$ satisfies (see Sec.~\ref{sec:preliminary} for the definition of $X_i^{(\ell)}$)
\beq
\rho_{\psi}^{X_i^{(\ell)}}\approx\rho_{\mathrm{mc}}^{X_i^{(\ell)}} \quad\text{for every } i\in\{1,2,\dots,V\}.
\label{eq:MITE}
\eeq
Equations~(\ref{eq:MITE}) and (\ref{eq:reduced}) imply that for any local operator $\hat{O}\in\mathcal{S}_{\mathrm{loc}}^{(\ell)}$,
\beq
\<\psi|\hat{O}|\psi\>\approx\<\hat{O}\>_{\mathrm{mc}}
\label{eq:MITE2}
\eeq
if a state $|\psi\>$ represents MITE.

Typicality of MITE follows the canonical typicality.
Equation~(\ref{eq:Popescu}) implies that almost all $|\psi\>\in\mathcal{H}_{E,\Lambda}$ represent MITE for sufficiently large $V$ as long as $D_{\ell}^2\ll D_{E,\Lambda}$.
If we assume that $w:=\mathrm{dim}\,\mathcal{H}_i$ is independent of $i$ ($\mathcal{H}_i$ is the local Hilbert space at site $i$), we have $D_{\ell}=w^{v_{\ell}}$ and $D_{E,\Lambda}=e^{S_{\mathrm{mc}}(E,V)}\leq w^V$, and thus the condition $D_{\ell}^2\ll D_{E,\Lambda}$ is written as
\beq
v_{\ell}\ll \frac{S_{\mathrm{mc}}(E,V)}{2\ln w}\leq\frac{V}{2}.
\label{eq:typicality_micro}
\eeq
Usually, we consider a microscopically small subsystem corresponding to $v_\ell=\mathcal{O}(1)$, but the subsystem may be macroscopically large (e.g. we can choose $v_\ell=V/100=\mathcal{O}(V)$).

It should be emphasized that the notion of MITE is of purely quantum origin since Eq.~(\ref{eq:MITE}) can be satisfied because of quantum entanglement.
In this way, there is a link between the foundation of statistical mechanics in an isolated quantum system and quantum entanglement~\cite{Popescu2006}.

So far, we have assumed that the subsystem $X=X_i^{(\ell)}$ is spatially local.
Here, we point out that the spatial locality of the subsystem $X$ is not assumed in the derivation of Eq.~(\ref{eq:Popescu}).
Even for a subsystem on a spatially nonlocal but small region $X\subset\{1,2,\dots,V\}$ with $|X|\ll V$, one can show $\rho_{\psi}^X\approx\rho_{\mathrm{mc}}^X$ for almost all $|\psi\>\in\mathcal{H}_{E,\Lambda}$.

From this observation, one can introduce a variant of MITE without the spatial locality of the subsystem $X$.
We say that a system is in \textit{few-body thermal equilibrium} with $k$ ($k$ is some positive integer) whenever its quantum state $|\psi\>$ satisfies
\beq
\rho_{\psi}^X\approx\rho_{\mathrm{mc}}^X \quad\text{for every $X\subset\{1,2,\dots V\}$ with $|X|=k$}.
\label{eq:local_eq}
\eeq
This condition is equivalent to the following one:
\beq
\<\psi|\hat{O}|\psi\>\approx\<\hat{O}\>_{\mathrm{mc}} \quad\text{for any few-body operator }\hat{O}\in\mathcal{S}_{\mathrm{few}}^{(k)}.
\eeq

The canonical typicality in the form of Eq.~(\ref{eq:Popescu}) indicates that the spatial locality of observables is not important in characterizing thermal equilibrium, but it is still an open problem to understand the role played by spatial locality in discussing nonequilibrium dynamics.
In spite of the irrelevance of the spatial locality in the statement of the canonical typicality, we will see that spatial locality plays an important role when we discuss equilibration in integrable systems in Sec.~\ref{sec:integrable}.

\subsection{Entanglement entropy and thermodynamic entropy}
\label{sec:entangle}

Let us consider a quantum system in a state $|\psi\>$ representing MITE, and take a subsystem $X\subset\{1,2,\dots,V\}$ in which $\rho_{\psi}^X\approx\rho_{\mathrm{mc}}^X$ is satisfied.
The entanglement entropy $S_{\mathrm{ent}}$ between the subsystem $X$ and the remaining part is given by
\beq
S_{\mathrm{ent}}=S_{\mathrm{vN}}(\rho_{\psi}^X):=-\mathrm{Tr}_X\rho_{\psi}^X\ln\rho_{\psi}^X,
\eeq
where $S_{\mathrm{vN}}$ denotes the von Neumann entropy~\cite{Nielsen_text}.
By using $\rho_{\psi}^X\approx\rho_{\mathrm{mc}}^X$, we have $S_{\mathrm{ent}}\approx S_{\mathrm{vN}}(\rho_{\mathrm{mc}}^X)$.
When $X$ is sufficiently large but still much smaller than the total system, the interaction between $X$ and the remaining part is negligible, and $\rho_{\mathrm{mc}}^X\approx\rho_{\mathrm{can},X}:=e^{-\beta H_X}/\mathrm{Tr}_Xe^{-\beta H_X}$, where $\beta=\d S_{\mathrm{mc}}(E,N,V)/\d E$ is the inverse temperature and $H_X$ is the Hamiltonian of $X$.
We then obtain $S_{\mathrm{ent}}\approx S_{\mathrm{vN}}(\rho_{\mathrm{can},X})$.
It is well known that the von Neumann entropy of the canonical ensemble is almost identical to the corresponding microcanonical entropy.
Thus, the entanglement entropy between a small subsystem $X$ and the remaining part essentially gives the thermodynamic entropy of the subsystem $X$ when the total system is in MITE.

\subsection{Comparison between macroscopic thermal equilibrium and microscopic thermal equilibrium}
\label{sec:comparison}

We have introduced the notions of MATE and MITE.
In this section, we discuss the relation between them.
In particular, we argue that MATE does not follow from MITE, and vice versa in general, and that MITE is meaningful in a small system, while MATE is not.

When a system is in MATE, the expectation values of macrovariables should be approximately equal to the equilibrium value given by the microcanonical ensemble, and, moreover, quantum fluctuations of macrovariables should be very small (for a macrovariable $\hat{M}_i$, its fluctuation must be smaller than $\Delta M_i$).
It means that a \textit{single measurement} of a macrovariable almost surely yields the corresponding equilibrium value.
On the other hand, when a system is in MITE, the expectation value of any macrovariable expressed as an extensive or intensive local operator should be very close to the corresponding equilibrium value, but its quantum fluctuations may not be small.
If we perform the measurement of a macrovariable many times, the average of the measurement outcomes should coincide with the equilibrium value, but a single measurement does not necessarily yield the equilibrium value.
In this respect, MATE is stronger than MITE.

In MITE on the length scale $\ell$, every local operator $\hat{O}\in\mathcal{S}_{\mathrm{loc}}^{(\ell)}$ has the expectation value very close to the equilibrium value.
It means that the system is indistinguishable from the microcanonical ensemble even microscopically.
It is not necessarily the case in a system in MATE.
In this respect, MITE is stronger than MATE.

The notion of MATE is meaningful only for a large system.
It is because the thermodynamic typicality is valid only when the equilibrium fluctuation of a macrovariable is much smaller than the mean value.
This condition yields $V^{1/2}\gg 1$.
On the other hand, the notion of MITE is meaningful even for small systems with $V\simeq 10$.
It is because the canonical typicality expressed by Eq.~(\ref{eq:can_typicality}) makes sense whenever $D_{E,\Lambda}\gg D_X^2$.
If we choose $\ell=\mathcal{O}(1)$, then $D_X=\mathcal{O}(1)$, while $D_{E,\Lambda}=e^{\mathcal{O}(V)}$.
Therefore, MITE is meaningful when $e^{\mathcal{O}(V)}\gg 1$, a condition much weaker than $V^{1/2}\gg 1$
In physical terms, one may say that quantum thermalization occurs in the Hilbert space rather than phase space.
Indeed, Saito, Takesue, and Miyashita~\cite{Saito1996} examined the system-size dependence of quantum dynamics in a spin system, and found that $V=8$ is enough to see a good agreement between the long-time average of the expectation value of a local quantity and the prediction by equilibrium statistical mechanics.
Thus, the notion of MITE is often valid for small systems, and there is no counterpart of MITE in classical thermodynamics.

\section{Quantum thermalization}
\label{sec:thermalization}

In this section, we overview general theory of thermalization.
The importance of the eigenstate thermalization hypothesis (ETH) will be emphasized.

\subsection{Macroscopic and microscopic thermalization}
\label{sec:macro_micro_thermalization}

Let us discuss the time evolution of an isolated quantum system with Hamiltonian $\hat{H}$ starting from a pure initial state $|\psi(0)\>\in\mathcal{H}_{E,\Lambda}$ at time $t=0$.
The quantum state at time $t$ is given by 
\beq
|\psi(t)\>=e^{-i\hat{H}t}|\psi(0)\>.
\eeq
Since the time evolution is unitary, the quantum state $|\psi(t)\>$ does not reach any stationary quantum state unless the system starts from one of the energy eigenstates of the Hamiltonian.

Typicality of thermal equilibrium, i.e., the thermodynamic typicality~(\ref{eq:therm_typicality_q2}) in the case of MATE and the canonical typicality~(\ref{eq:can_typicality}) in the case of MITE, tells us that almost all quantum states in the energy shell are in thermal equilibrium.
Thus, it is expected that the system reaches thermal equilibrium after some time evolution and stays there, although $|\psi(t)\>$ remains evolving.
An approach to thermal equilibrium is called \textit{thermalization}.
Intuitively, thermalization is understood as an evolution from an atypical nonequilibrium state to a typical equilibrium state.

According to different notions of thermal equilibrium, we should distinguish between \textit{macroscopic thermalization} and \textit{microscopic thermalization}.
The former corresponds to the approach to MATE, and the latter corresponds to the approach to MITE.

A state $|\psi\>\in\mathcal{H}_{E,\Lambda}$ is in MATE whenever $\<\psi|\hat{P}_{\mathrm{eq}}|\psi\>\approx 1$.
For a more precise statement, see Eq.~(\ref{eq:MATE2}).
Thus, we say that macroscopic thermalization occurs, or the system macroscopically thermalizes if
\beq
\<\psi(t)|\hat{P}_{\mathrm{eq}}|\psi(t)\>\approx 1 \quad \text{for most of the time } t.
\label{eq:MAT1}
\eeq
This is equivalent to the following condition:
\beq
\overline{\<\psi(t)|(1-\hat{P}_{\mathrm{eq}})|\psi(t)\>}\ll 1,
\label{eq:MAT2}
\eeq
where the overbar denotes the infinite-time average: $\overline{f(t)}:=(1/T)\int_0^Tdt\,f(t)$.
We note that the condition ``for most of the time $t$'' in Eq.~(\ref{eq:MATE1}) cannot be replaced by ``for all $t>\tau_{\mathrm{rel}}$ with a relaxation time $\tau_{\mathrm{rel}}$'' because recurrence phenomena should happen if we wait for a sufficiently long time.
However, the time scale of the recurrence phenomena is usually extremely long and the recurrence is not relevant in a macroscopic system (see Sec.~\ref{sec:recurrence}).

Next, we derive the condition of microscopic thermalization.
A state $|\psi\>\in\mathcal{H}_{E,\Lambda}$ is in MITE whenever $\rho_{\psi}^{X_i^{(\ell)}}\approx\rho_{\mathrm{mc}}^{X_i^{(\ell)}}$ for every $i=1,2,\dots,V$.
We say that microscopic thermalization occurs, or the system microscopically thermalizes if, for every $i=1,2,\dots,V$,
\beq
\rho_{\psi(t)}^{X_i^{(\ell)}}\approx\rho_{\mathrm{mc}}^{X_i^{(\ell)}} \quad \text{for most of the time }t,
\label{eq:MIT1}
\eeq
which is equivalent to the following condition:\footnote
{It should be noted that Eq.~(\ref{eq:MIT2}) differs from
\[
\sup_{\hat{O}\in\mathcal{S}_{\mathrm{loc}}^{(\ell)}: \|\hat{O}\|=1}\overline{\left|\<\psi(t)|\hat{O}|\psi(t)\>-\<\hat{O}\>_{\mathrm{mc}}\right|}\ll 1.
\]
This weaker inequality only ensures that for every $\hat{O}\in\mathcal{S}_{\mathrm{loc}}^{(\ell)}$, $\<\psi(t)|\hat{O}|\psi(t)\>\approx\<\hat{O}\>_{\mathrm{mc}}$ for most of the time $t$.
On the other hand, Eq.~(\ref{eq:MIT1}) or Eq.~(\ref{eq:MIT2}) ensures that for most of the time $t$, $\<\psi(t)|\hat{O}|\psi(t)\>\approx\<\hat{O}\>_{\mathrm{mc}}$ \textit{for all} $\hat{O}\in\mathcal{S}_{\mathrm{loc}}^{(\ell)}$.
}

\beq
\overline{\sup_{\hat{O}\in\mathcal{S}_{\mathrm{loc}}^{(\ell)}: \|\hat{O}\|=1}\left|\<\psi(t)|\hat{O}|\psi(t)\>-\<\hat{O}\>_{\mathrm{mc}}\right|}\ll 1,
\label{eq:MIT2}
\eeq
When Eq.~(\ref{eq:MIT1}) is satisfied for every subsystem $X\subset\{1,2,\dots,V\}$ with $|X|=k$ instead of $X_i^{(\ell)}$, we say that few-body thermalization with $k$ occurs.

By using the triangle inequality, we have
\begin{align}
\overline{\sup_{\hat{O}\in\mathcal{S}_{\mathrm{loc}}^{(\ell)}: \|\hat{O}\|=1}\left|\<\psi(t)|\hat{O}|\psi(t)\>-\<\hat{O}\>_{\mathrm{mc}}\right|}
\leq \overline{\sup_{\hat{O}\in\mathcal{S}_{\mathrm{loc}}^{(\ell)}: \|\hat{O}\|=1}\left|\<\psi(t)|\hat{O}|\psi(t)\>-\mathrm{Tr}\,\hat{O}\rho_{\mathrm{D}}\right|}
\nonumber \\
+\sup_{\hat{O}\in\mathcal{S}_{\mathrm{loc}}^{(\ell)}: \|\hat{O}\|=1}\left|\mathrm{Tr}\,\hat{O}\rho_{\mathrm{D}}-\<\hat{O}\>_{\mathrm{mc}}\right|,
\label{eq:MIT2-2}
\end{align}
where $\rho_{\mathrm{D}}$ is the \textit{diagonal ensemble}~\cite{Rigol2008,Kollar2008} defined as
\beq
\rho_{\mathrm{D}}:=\overline{|\psi(t)\>\<\psi(t)|},
\label{eq:DE}
\eeq
which is a diagonal matrix in the energy eigenbasis\footnote
{The diagonal average $\mathrm{Tr}\,\hat{O}\rho_{\mathrm{D}}$ is nothing but the infinite-time average of $\<\psi(t)|\hat{O}|\psi(t)\>$.
}.
From Eqs.~(\ref{eq:MIT2}) and (\ref{eq:MIT2-2}), one can conclude that if both
\beq
\overline{\sup_{\hat{O}\in\mathcal{S}_{\mathrm{loc}}^{(\ell)}:\|\hat{O}\|=1}\left|\<\psi(t)|\hat{O}|\psi(t)\>-\mathrm{Tr}\,\hat{O}\rho_{\mathrm{D}}\right|}\ll 1
\label{eq:equilibration}
\eeq
and
\beq
\sup_{\hat{O}\in\mathcal{S}_{\mathrm{loc}}^{(\ell)}: \|\hat{O}\|=1}\left|\mathrm{Tr}\,\hat{O}\rho_{\mathrm{D}}-\<\hat{O}\>_{\mathrm{mc}}\right|\ll 1
\label{eq:thermalization}
\eeq
are satisfied, microscopic thermalization occurs.
The first condition~(\ref{eq:equilibration}) means that the temporal fluctuation of $\<\psi(t)|\hat{O}|\psi(t)\>$ is very small, from which it immediately follows that $\<\psi(t)|\hat{O}|\psi(t)\>\approx\mathrm{Tr}\,\hat{O}\rho_{\mathrm{D}}$ for most of the time $t$.
As long as we consider local quantities, the system relaxes to a stationary state described by $\rho_{\mathrm{D}}$.
This is called \textit{equilibration}.
The second condition~(\ref{eq:thermalization}) means that this stationary state is locally equivalent to the microcanonical ensemble~\cite{Tasaki1998,Reimann2008}.

Due to the fact that a pure state $|\psi\>\in\mathcal{H}_{E,\Lambda}$ typically represents MATE and/or MITE, it immediately follows that macroscopic and/or microscopic thermalization occurs for \textit{almost all the initial states} $|\psi(0)\>$ in the energy shell $\in\mathcal{H}_{E,\Lambda}$.
One might be tempted to conclude that this typicality argument nicely explains thermalization in isolated quantum systems.
However, this is not the whole story.
It is known that there are several important classes of physical systems, e.g., integrable systems and many-body localized systems, that fail to thermalize.
The point is that an initial state that we prepare is a nonequilibrium state, which is highly atypical.
We cannot determine whether thermalization occurs starting from such an atypical initial state solely from the typicality argument.
In other words, evolution from an atypical nonequilibrium state to a typical equilibrium state is not so trivial, although it looks quite natural.

Several works~\cite{Goldstein2010,Goldstein2015,Goldstein2017,Tasaki2016,Goldstein-Hara-Tasaki2013,Goldstein-Hara-Tasaki2015} have investigated macroscopic thermalization in isolated quantum systems, but many recent studies have focused on microscopic thermalization (or few-body thermalization)~\cite{Tasaki1998, Rigol2008, Rigol2009_PRL, Rigol2009_PRA, Linden2009, Eckstein2009, Biroli2010, Jin2010, Gogolin2011, Banuls2011}.
One would reasonably expect that microscopic thermalization automatically ensures macroscopic thermalization for a realistic Hamiltonian and a realistic initial state.
If the system shows microscopic thermalization \textit{and} the quantum fluctuations of all macrovariables remain very small during the time evolution, one can conclude macroscopic thermalization.
In other words, if one admits that the system remains in a single macrostate $\nu(t)$ for $t>0$, microscopic thermalization implies macroscopic thermalization, i.e., $\nu(t)=\nu_{\mathrm{eq}}$ for most $t$.
As discussed in Sec.~\ref{sec:MATE}, it is a highly expected yet nontrivial dynamical property that the system is almost always in a single macrostate $\nu(t)$.
In conclusion, while it is plausible that microscopic thermalization of the system automatically implies macroscopic thermalization of the system, it remains an open problem to prove that this conjecture is generally true under a suitable condition.

\subsection{Equilibration}
\label{sec:equilibration}

As we have already pointed out, some systems such as integrable systems and many-body localized systems fail to thermalize.
The absence of thermalization in those systems is attributed to the violation of the  condition~(\ref{eq:thermalization}).
In contrast, it is now recognized that the condition of equilibration~(\ref{eq:equilibration}) is generally satisfied under moderate conditions on the energy spectrum and the initial state~\cite{Peres1984,Tasaki1998,Reimann2008,Linden2009,Short2011,Short2012,Reimann2012}.

Now we assume the \textit{nonresonance condition}, which states that $E_n-E_m=E_k-E_l\neq 0$ if and only if $n=k$ and $m=l$.
The nonresonance condition requires that there is no degeneracy of energy gaps.
Reimann~\cite{Reimann2008} has proved the following inequality under the nonresonance condition: for any operator $\hat{O}\in\mathcal{B}$,
\beq
\overline{\left|\<\psi(t)|\hat{O}|\psi(t)\>-\mathrm{Tr}\,\hat{O}\rho_{\mathrm{D}}\right|^2}\leq\frac{\|\hat{O}\|^2}{D_{\mathrm{eff}}},
\label{eq:Reimann}
\eeq
where the effective dimension $D_{\mathrm{eff}}$ for the initial state $|\psi(0)\>=\sum_nc_n|\phi_n\>$ is defined as
\beq
D_{\mathrm{eff}}:=\left(\sum_n|c_n|^4\right)^{-1}.
\label{eq:effective_dim}
\eeq
Roughly speaking, the effective dimension represents the effective number of energy eigenstates that contribute to the initial state.
Typically, $D_{\mathrm{eff}}=e^{\mathcal{O}(V)}$, which is huge~\cite{Linden2009}.
In Ref.~\cite{Farelly2017}, it is shown that for any initial state that exhibits exponentially decaying correlations, $D_{\mathrm{eff}}\geq cs^3\sqrt{N}/(\ln N)^{2d}$ for $N$ spin systems, where $c>0$ is a constant independent of $N$ and $s=\sigma_E/\sqrt{N}$ with $\sigma_E^2$ being the variance of the energy in the initial state.
It is noted that usually $s$ is a quantity independent of the system size, and thus this result shows $D_{\mathrm{eff}}\gg 1$ for $N\gg 1$.

Since $D_{\mathrm{eff}}$ is a huge number, inequality~(\ref{eq:Reimann}) ensures equilibration of a local observable $\hat{O}$.
It should be emphasized that even quite a small system such as $N=10-20$ spins may have a sufficiently large effective dimension, and thus it can exhibit equilibration.
It is also emphasized that inequality~(\ref{eq:Reimann}) is true for any operator $\hat{O}$, which may be neither local nor few-body, and for any Hamiltonian with the nonresonance condition.

The derivation of Reimann's result~(\ref{eq:Reimann}) is almost straightforward.
Since 
\beq
\<\psi(t)|\hat{O}|\psi(t)\>-\mathrm{Tr}\,\hat{O}\rho_{\mathrm{D}}=\sum_{n\neq m}c_n^*c_me^{i(E_n-E_m)t}O_{nm},
\eeq
where $O_{nm}=\<\phi_n|\hat{O}|\phi_m\>$, we obtain
\beq
\overline{\left|\<\psi(t)|\hat{O}|\psi(t)\>-\mathrm{Tr}\,\hat{O}\rho_{\mathrm{D}}\right|^2}=\sum_{n\neq m}\sum_{k\neq l}c_n^*c_mc_kc_l^*\overline{e^{i[(E_n-E_m)-(E_k-E_l)]t}}O_{nm}O_{kl}^*.
\eeq
Here, only the terms with $E_n-E_m=E_k-E_l\neq 0$ remain after the infinite-time average.
By using the non-resonance condition, we only have the terms with $n=k$ and $m=l$, and thus obtain
\beq
\overline{\left|\<\psi(t)|\hat{O}|\psi(t)\>-\mathrm{Tr}\,\hat{O}\rho_{\mathrm{D}}\right|^2}=\sum_{n\neq m}|c_n|^2|c_m|^2|O_{nm}|^2.
\label{eq:time_fluct}
\eeq
By using the inequality $|c_n|^2|c_m|^2\leq(|c_n|^4+|c_m|^4)/2$, we obtain
\begin{align}
\overline{\left|\<\psi(t)|\hat{O}|\psi(t)\>-\mathrm{Tr}\,\hat{O}\rho_{\mathrm{D}}\right|^2}&\leq\frac{1}{2}\sum_n|c_n|^4\left(\hat{O}\hat{O}^{\dagger}\right)_{nn}+\frac{1}{2}\sum_m|c_m|^4\left(\hat{O}^{\dagger}\hat{O}\right)_{nn}
\nonumber \\
&\leq\sum_n|c_n|^4\|\hat{O}\|^2=\frac{\|\hat{O}\|^2}{D_{\mathrm{eff}}},
\end{align}
which is nothing but Reimann's result~(\ref{eq:Reimann}).

Inequality~(\ref{eq:Reimann}) tells us that for any \textit{fixed} operator $\hat{O}$, $\<\psi(t)|\hat{O}|\psi(t)\>\approx\mathrm{Tr}\,\hat{O}\rho_{\mathrm{D}}$ for most $t$.
One would expect that for most $t$, $\<\psi(t)|\hat{O}|\psi(t)\>\approx\mathrm{Tr}\,\hat{O}\rho_{\mathrm{D}}$ for \textit{all} $\hat{O}\in\mathcal{S}_{\mathrm{loc}}^{(\ell)}$, and thus Eq.~(\ref{eq:equilibration}) is satisfied.
Linden et al.~\cite{Linden2009} have derived the following inequality under the nonresonance condition: for any subsystem $X$,
\beq
\overline{\left\|\rho_{\psi(t)}^X-\rho_{\mathrm{D}}^X\right\|_1}\leq\sqrt{\frac{D_X^2}{D_{\mathrm{eff}}}},
\label{eq:Linden}
\eeq
which ensures that all the operators defined on a subsystem $X$ simultaneously take their equilibrium values for most $t$ unless $|X|$ is too large or $D_{\mathrm{eff}}$ is too small.
By using Eq.~(\ref{eq:Linden}) and $\mathcal{S}_{\mathrm{loc}}^{(\ell)}=\bigcup_{i=1}^V\mathcal{S}_i^{(\ell)}$, we obtain
\beq
\overline{\sup_{\hat{O}\in\mathcal{S}_{\mathrm{loc}}^{(\ell)}:\|\hat{O}\|=1}\left|\<\psi(t)|\hat{O}|\psi(t)\>-\mathrm{Tr}\,\hat{O}\rho_{\mathrm{D}}\right|}
\leq V\sqrt{\frac{D_{\ell}^2}{D_{\mathrm{eff}}}}.
\eeq
By applying it to Eq.~(\ref{eq:equilibration}), we find that
\beq
V\sqrt{\frac{D_{\ell}^2}{D_{\mathrm{eff}}}}\ll 1
\label{eq:equilibration_condition}
\eeq
is a sufficient condition of equilibration\footnote
{If both the Hamiltonian and the initial state are translation invariant, $V$ on the left-hand side of Eq.~(\ref{eq:equilibration_condition}) is unnecessary.
}.

We show that the result by Linden et al. given by Eq.~(\ref{eq:Linden}) is derived from Reimann's one given by Eq.~(\ref{eq:Reimann}).
We first rewrite $\|\rho_{\psi(t)}^X-\rho_{\mathrm{D}}^X\|_1$ as
\beq
\|\rho_{\psi(t)}^X-\rho_{\mathrm{D}}^X\|_1=\sup_{\hat{O}\in\mathcal{S}_X,\|\hat{O}\|=1}\left|\<\psi(t)|\hat{O}|\psi(t)\>-\mathrm{Tr}\,\hat{O}\rho_{\mathrm{D}}\right|.
\eeq
By using an orthonormal basis $\{\hat{O}^{(p)}\}$ satisfying Eq.~(\ref{eq:orthonormal}), we can expand $\hat{O}$ as $\hat{O}=\sum_{p=1}^{D_X^2}c_p\hat{O}^{(p)}$.
Then we have, by using the Schwartz inequality,
\begin{align}
\|\rho_{\psi(t)}^X-\rho_{\mathrm{D}}^X\|_1&=\sup_{\hat{O}\in\mathcal{S}_X,\|\hat{O}\|=1}\left|\sum_{p=1}^{D_X^2}c_p\left(\<\psi(t)|\hat{O}^{(p)}|\psi(t)\>-\mathrm{Tr}\,\hat{O}^{(p)}\rho_{\mathrm{D}}\right)\right|
\nonumber \\
&\leq\sup_{\hat{O}\in\mathcal{S}_X,\|\hat{O}\|=1}\left(\sum_{p=1}^{D_X^2}|c_p|^2\right)^{1/2}\left(\sum_{p=1}^{D_X^2}\left|\<\psi(t)|\hat{O}^{(p)}|\psi(t)\>-\mathrm{Tr}\,\hat{O}^{(p)}\rho_{\mathrm{D}}\right|^2\right)^{1/2}.
\end{align}
By using the inequality $\sum_{p=1}^{D_X^2}|c_p|^2\leq D_X$ (see Sec.~\ref{sec:preliminary}), we have
\beq
\|\rho_{\psi(t)}^X-\rho_{\mathrm{D}}^X\|_1\leq\left(D_X\sum_{p=1}^{D_X^2}\left|\<\psi(t)|\hat{O}^{(p)}|\psi(t)\>-\mathrm{Tr}\,\hat{O}^{(p)}\rho_{\mathrm{D}}\right|^2\right)^{1/2}.
\eeq
By taking the infinite-time average, we obtain
\beq
\overline{\|\rho_{\psi(t)}^X-\rho_{\mathrm{D}}^X\|_1}\leq\left(D_X\sum_{p=1}^{D_X^2}\overline{\left|\<\psi(t)|\hat{O}^{(p)}|\psi(t)\>-\mathrm{Tr}\,\hat{O}^{(p)}\rho_{\mathrm{D}}\right|^2}\right)^{1/2}.
\eeq
Here we use Reimann's result~(\ref{eq:Reimann}) and inequality~(\ref{eq:orthonormal}), which yield
\beq
\overline{\|\rho_{\psi(t)}^X-\rho_{\mathrm{D}}^X\|_1}\leq\sqrt{D_X\cdot D_X^2\cdot\frac{\|\hat{O}^{(p)}\|^2}{D_{\mathrm{eff}}}}=\sqrt{\frac{D_X^2}{D_{\mathrm{eff}}}}.
\eeq
This completes the proof of Eq.~(\ref{eq:Linden}).

The result by Reimann, i.e., Eq.~(\ref{eq:Reimann}), has been improved by later works.
Short and Farrelly~\cite{Short2012} have removed the nonresonance condition.
Let us write the Hamiltonian as $\hat{H}=\sum_{n=1}^{D_E}E_n\hat{P}_n$, where $E_n$ are distinct energy eigenvalues, $D_E$ is the number of distinct energy eigenvalues, and $\hat{P}_n$ are projectors onto the eigenspaces of $\hat{H}$ with energy $E_n$.
We define $p_n=\<\psi(0)|\hat{P}_n|\psi(0)\>$, which satisfies $p_n\geq 0$ and $\sum_{n=1}^{D_E}p_n=1$.
The effective dimension is defined as $D_{\mathrm{eff}}=\left(\sum_{n=1}^{D_E}p_n^2\right)^{-1}$.
Let us define the set $\mathcal{G}$ as $\mathcal{G}:=\{(n,m):n,m\in\{1,2,\dots,D_E\}, n\neq m\}$ and $G_{\alpha}:=E_n-E_m$ for $\alpha=(n,m)\in\mathcal{G}$.
We denote the maximum degeneracy of energy gaps by $d_G$, i.e.,
\beq
d_G:=\max_{\beta\in\mathcal{G}}|\{\alpha\in\mathcal{G}:G_{\alpha}=G_{\beta}\}|.
\eeq
Short and Farrelly~\cite{Short2012} have derived the following inequality:
\beq
\overline{\left|\<\psi(t)|\hat{O}|\psi(t)\>-\mathrm{Tr}\,\hat{O}\rho_{\mathrm{D}}\right|^2}\leq\frac{d_G}{D_{\mathrm{eff}}}\|\hat{O}\|^2.
\label{eq:Short-Farrelly}
\eeq
This result tells us that equilibration occurs unless $d_G$ is too large or $D_{\mathrm{eff}}$ is too small.
Reimann and Kastner~\cite{Reimann2012} have further generalized this result to the case in which the initial state exhibits a large population of at most one energy level, that is, $\max_np_n$ is not necessarily small.
Since $D_{\mathrm{eff}}\leq 1/(\max_np_n)^2$, the effective dimension is not necessarily large in their setting.
By requiring that the second largest $p_n$, which is denoted by $\max_n'p_n$, is sufficiently small, Reimann and Kastner have derived the following inequality~\cite{Reimann2012}:
\beq
\overline{\left|\<\psi(t)|\hat{O}|\psi(t)\>-\mathrm{Tr}\,\hat{O}\rho_{\mathrm{D}}\right|^2}\leq 6\|\hat{O}\|^2d_G\max_n{'}p_n.
\eeq

So far, the temporal fluctuation of $\<\psi(t)|\hat{O}|\psi(t)\>$ is bounded from above by using the effective dimension.
There is another way to give an upper bound of different nature.
Starting from Eq.~(\ref{eq:time_fluct}), we can find a different upper bound by using $|O_{nm}|^2\leq\max_{n\neq m}|O_{nm}|^2$:
\beq
\overline{\left|\<\psi(t)|\hat{O}|\psi(t)\>-\mathrm{Tr}\,\hat{O}\rho_{\mathrm{D}}\right|^2}\leq\max_{n\neq m}|O_{nm}|^2.
\label{eq:off_equilibration}
\eeq
Therefore, if all the off-diagonal elements of $\hat{O}$ are sufficiently small, equilibration occurs for such an operator $\hat{O}$ \textit{independently of the initial state}.

\subsection{Eigenstate thermalization hypothesis (ETH)}
\label{sec:ETH}

Typicality of thermal equilibrium implies that thermalization is understood as the evolution from an atypical nonequilibrium initial state to a typical state representing thermal equilibrium.
This typicality argument, however, does not explain the presence or absence of thermalization for a given system.
Some many-body systems fail to thermalize, although typicality of thermal equilibrium holds.
To obtain a better understanding of the situation, some criterion characterizing the presence or absence of thermalization in individual systems is desired.

It is now widely recognized that the \textit{eigenstate thermalization hypothesis} (ETH) is the most likely one among such criteria.
The ETH states that \textit{all} the energy eigenstates in an energy shell $\mathcal{H}_{E,\Lambda}$ represent thermal equilibrium~\cite{Deutsch1991,Srednicki1994,Rigol2008}, which goes back to von Neumann~\cite{Neumann1929}.
Typicality of thermal equilibrium shows that most pure states in an energy shell represent thermal equilibrium (such a pure state is sometimes said to be a ``typical state'').
The ETH insists that all the energy eigenstates belong to the set of typical states.

As we have seen in Sec.~\ref{sec:MATE} and Sec.~\ref{sec:MITE}, the notion of thermal equilibrium is not unique; MATE and MITE are different concepts.
Thus, the definition of the ETH is also not unique; we can consider the MATE-ETH and the MITE-ETH~\cite{Goldstein2015}.

The MATE-ETH states that every energy eigenstate $|\phi_n\>\in\mathcal{H}_{E,\Lambda}$ represents MATE (see Eq.~(\ref{eq:MATE2})), i.e.,
\beq
\<\phi_n|\hat{P}_{\mathrm{eq}}|\phi_n\>\approx 1.
\label{eq:MATE-ETH}
\eeq
If we assume the MATE-ETH, we can show macroscopic thermalization for any initial state within an energy shell.
By using Eq.~(\ref{eq:MATE-ETH}), we obtain $\overline{\<\psi(t)|(1-\hat{P}_{\mathrm{eq}})|\psi(t)\>}\ll 1$ for any initial state $|\psi(0)\>$.
Since $1-\hat{P}_{\mathrm{eq}}\geq 0$, this implies $\<\psi(t)|(1-\hat{P}_{\mathrm{eq}})|\psi(t)\>\ll 1$ for most of the time $t$, which shows that macroscopic thermalization occurs.

The MITE-ETH\footnote
{The MITE-ETH is sometimes called the \textit{strong} ETH in order to distinguish it from the \textit{weak} ETH, the latter of which is explained in Sec.~\ref{sec:weak_ETH}.
}
states that every energy eigenstate $|\phi_n\>\in\mathcal{H}_{E,\Lambda}$ represents MITE (see Eq.~(\ref{eq:MITE2})), i.e.,
\beq
\<\phi_n|\hat{O}|\phi_n\>\approx\<\hat{O}\>_{\mathrm{mc}}
\label{eq:MITE-ETH}
\eeq
for an arbitrary local operator $\hat{O}\in\mathcal{S}_{\mathrm{loc}}^{(\ell)}$ (or for an arbitrary $\hat{O}\in\mathcal{S}_{\mathrm{few}}^{(k)}$ if we want to consider few-body thermalization).
Equation~(\ref{eq:MITE-ETH}) is equivalent to the statement that the reduced density matrix of a small system in a single energy eigenstate looks thermal equilibrium.
If we assume the MITE-ETH, we can show microscopic thermalization for any initial state that exhibits equilibration.
In other words, the MITE-ETH guarantees Eq.~(\ref{eq:thermalization}) for any initial state $|\psi(0)\>$.

There have been many numerical evidences for the MITE-ETH in generic interacting many-body systems
although no rigorous proof has not been obtained.
Early studies were performed on the nuclear shell model~\cite{Horoi1995,Zelevinsky1996}, and recent studies motivated by ultracold atom experiments have been initiated by Rigol et al~\cite{Rigol2008}.
Concerning a number of studies that have followed to witness the MITE-ETH~\cite{Rigol2009_PRA, Biroli2010, Ikeda2011, Steinigeweg2013, Sorg2014, Khodja2015, Fratus2015, Garrison_arXiv2015, Mondaini2016, Hamazaki2016},
the MITE-ETH is now widely believed to hold true in generic nonintegrable models\footnote
{Here, we say that a system is nonintegrable when it has no local conserved quantity.
However, there are several different definitions of quantum integrability, see for example Ref.~\cite{Caux2011}.
}.
In particular, the onset of quantum chaos turned out to be the key to the ETH as verified for bosons
on a lattice~\cite{Rigol2010,Santos2010a,Santos2010} and in a double-well potential~\cite{Motohashi2011}, and spin chains~\cite{Santos2012_PRL,Santos2012_PRE}.

The validity of the ETH is investigated through the system-size dependence of the following quantity:
\beq
I_{\mathrm{ETH}}[\hat{O}]:=\max_{\phi_n\in\mathcal{H}_{E,\Lambda}}\left|\<\phi_n|\hat{O}|\phi_n\>-\<\hat{O}\>_{\mathrm{mc}}\right|,
\eeq
or
\beq
I_{\mathrm{ETH}}^\text{w}[\hat{O}]:=\left[\mathcal{N}^{-1}\sum_{\phi_n\in\mathcal{H}_{E,\Lambda}}\left(\<\phi_n|\hat{O}|\phi_n\>-\<\hat{O}\>_{\mathrm{mc}}^{E_n,\Delta E}\right)^2\right]^{1/2},
\label{eq:indicator_w}
\eeq
where $\mathcal{N}:=\sum_{\phi_n\in\mathcal{H}_{E,\Lambda}}1$ and energy eigenvalues are labeled so that $E_1\leq E_2\leq E_3\leq\dots \leq E_D$.
In Eq.~(\ref{eq:indicator_w}), $\<\hat{O}\>_{\mathrm{mc}}^{E_n,\Delta E}$ denotes the microcanonical average of $\hat{O}$ within the energy shell specified by the energy interval between $E_n-\Delta E$ and $E_n$ (the energy shell depends on $n$).
Near the edge of the spectrum, the convergence to the thermodynamic limit is slow, and hence usually we choose $\mathcal{E}$ so that it does not include the edge of the energy spectrum.
If $I_{\mathrm{ETH}}$ tends to zero in the thermodynamic limit, the MITE-ETH is numerically verified at least within an accessible system size.
We note that the vanishing of $I_{\mathrm{ETH}}^\text{w}$ does not necessarily mean the MITE-ETH because it allows the existence of rare states~\cite{Biroli2010} that do not obey the MITE-ETH.
The vanishing of $I_{\mathrm{ETH}}^\text{w}$ is related to the property referred to as the weak ETH whose implication is discussed in Sec.~\ref{sec:weak_ETH}.

In nonintegrable systems, the MITE-ETH becomes more precise exponentially with increasing $V$.
Beugeling et al.~\cite{Beugeling2014} have carefully examined $I_\text{ETH}^\text{w}$ in nonintegrable spin chains, finding that $I_\text{ETH}^\text{w}$ decays in proportion to $(\dim\,\mathcal{H})^{-1/2}$ and the exponential decay of $I_\text{ETH}^\text{w}$ has been verified by Steinigeweg et al.~\cite{Steinigeweg2014} up to 35 spins.
The ETH indicator $I_\text{ETH}$ has shown in similar systems to decay also exponentially~\cite{Kim2014}\footnote
{In Ref.~\cite{Kim2014}, $\tilde{I}_{\mathrm{ETH}} :=\max_{n:E_n/V\in\mathcal{E}}|\<\phi_n|\hat{O}|\phi_n\>-\<\phi_{n+1}|\hat{O}|\phi_{n+1}\>|$ is used as an indicator of the MITE-ETH instead of $I_{\mathrm{ETH}}$, where $\mathcal{E}$ is an arbitrary interval of the energy density.
Strictly speaking, $I_{\mathrm{ETH}}$ is more appropriate than $\tilde{I}_{\mathrm{ETH}}$ as an indicator of the ETH.
If $I_{\mathrm{ETH}}\rightarrow 0$ in the thermodynamic limit for any energy shell, we can also conclude $\tilde{I}_{\mathrm{ETH}} \rightarrow 0$.
However, the converse is not true; $\tilde{I}_{\mathrm{ETH}} \rightarrow 0$ does not imply $I_{\mathrm{ETH}} \rightarrow 0$.
},
and, in particular, in proportion to $(\dim\,\mathcal{H})^{-1/2}$~\cite{Ikeda2015}.

The ETH breaks down in integrable systems.
This fact has been revealed by Rigol et al.~\cite{Rigol2008,Rigol2009_PRL,Rigol2009_PRA} by investigating hard-core bosons in one dimension.
Biroli et al.~\cite{Biroli2010} have pointed out that the ETH holds true in the weak sense in those systems and the rare states can prevent thermalization.
The weak ETH has also been confirmed in integrable systems that are not equivalent to free particles
but solvable by the Bethe ansatz~\cite{Ikeda2013,Alba2015}.
The indicator $I_\text{ETH}^\text{w}$ decays only as a power law in the system size in integrable systems in contrast to the exponential decay seen in nonintegrable ones.
Nonthermal steady states in the absence of the MITE-ETH in these integrable systems will be discussed in detail in Sec.~\ref{sec:prethermalization}.

There are other classes of systems in which the ETH also breaks down.
One class is the many-body localized (MBL) systems~\cite{Nandkishore_review2015,Altman_review2015,Imbrie_review2017},
which are extensions of the original work of Anderson on single-particle localization
to interacting systems~\cite{Anderson1958}.
In most of these systems, the absence of the translation symmetry (e.g., a system subject to a random potential~\cite{Pal2010} or a random coupling~\cite{Khatami2012})
plays a central role although it is under debate if the MBL is possible in translation-invariant systems~\cite{Carleo2012,Grover2014,DeRoeck2015,Schiulaz2015,Yao2016,Hickey2016}.
Another new class has recently been found by Shiraishi and Mori~\cite{Shiraishi-Mori2017,Mori-Shiraishi2017}, where one can construct translation-invariant short-range interacting spin systems without any local conserved quantity which do not obey the ETH.
Remarkably, in a model discussed in Refs.~\cite{Shiraishi-Mori2017,Mori-Shiraishi2017}, $I_{\mathrm{ETH}}^{\mathrm{w}}$ decays exponentially although the MITE-ETH is violated.
Therefore, an exponential decay of $I_{\mathrm{ETH}}^{\mathrm{w}}$ does not imply the MITE-ETH.
These classes will be discussed in more detail in Sec.~\ref{sec:failure}.
The complete characterization of the systems obeying the ETH still remains an open question at present.

\subsection{Srednicki's ETH ansatz and off-diagonal ETH}
\label{sec:off_diagonal}

Let us consider a self-adjoint operator $\hat{O}\in\mathcal{S}_{\mathrm{loc}}^{(\ell)}$ (or $\hat{O}\in\mathcal{S}_{\mathrm{few}}^{(k)}$ if we want to discuss few-body thermalization).
For an energy eigenstate $|\phi_n\>$, $\<\phi_n|\hat{O}^2|\phi_n\>\leq\|\hat{O}\|^2$.
By using the completeness of the energy basis, we obtain $\sum_m|O_{nm}|^2\leq\|\hat{O}\|^2$ with $O_{nm}=\<\phi_n|\hat{O}|\phi_m\>$.
By using $|O_{nn}|^2\geq 0$, we find
\beq
\sum_{m(\neq n)}|O_{nm}|^2\leq\|\hat{O}\|^2.
\label{eq:off_ineq}
\eeq

The essential idea behind the ETH is that all the energy eigenstates within an energy shell have common features.
Indeed, the MITE-ETH is expressed as $O_{nn}\approx O_{mm}$ for any $|\phi_n\>$ and $|\phi_m\>$ in $\mathcal{H}_{E,\Lambda}$.
We shall generalize this idea to off-diagonal elements: let us assume that all the off-diagonal elements $O_{nm}$ with a given mean energy $(E_n+E_m)/2\approx\bar{E}$ and a given energy difference $E_m-E_n\approx\omega$ have almost identical magnitude, $|O_{nm}|\approx O_{\bar{E},\omega}$.
By applying this conjecture, we have
\beq
\sum_{m(\neq n):E_m-E_n\in[\omega,\omega+\Delta\omega)}|O_{nm}|^2\approx O_{\bar{E},\omega}^2e^{S_{\mathrm{mc}}(\bar{E}+\omega/2)},
\eeq
where $\bar{E}=E_n+\omega/2$ and $S_{\mathrm{mc}}(E)$ is the microcanonical entropy.
Because of Eq.~(\ref{eq:off_ineq}), this should not be greater than $\|\hat{O}\|^2F_O(\bar{E},\omega)^2$ with some function\footnote
{In quantum spin systems, we can prove $F_O(\bar{E},\omega)^2\leq e^{-\omega/\omega_0}$, where $\omega_0>0$ does not depend on the system size $V$ (or $N$) and is related to the possible maximum value of the energy per spin.
For the proof, we express an off-diagonal element as $O_{nm}=\<\phi_n|\textrm{ad}_H^kO|\phi_m\>/(E_n-E_m)^k$, where $\mathrm{ad}_H:=[H,\cdot]$ is a commutator in the Hamiltonian, and utilize some inequalities for multiple commutators in a local Hamiltonian developed in recent studies, see, for example, Ref.~\cite{Arad2016}.
}
$F_O(\bar{E},\omega)$ such that $0\leq F_O(\bar{E},\omega)\leq 1$ and $\int_{-\infty}^{\infty}d\omega\,F_O(\bar{E},\omega)^2=1$.
Thus, we obtain
\beq
O_{\bar{E},\omega}\lesssim\|\hat{O}\|e^{-\frac{1}{2}S_{\mathrm{mc}}(\bar{E}+\omega/2)}F_O(\bar{E},\omega).
\label{eq:off_typical}
\eeq
Since we have assumed that individual $|O_{nm}|$ with $(E_n+E_m)/2=\bar{E}$ and $E_n-E_m=\omega$ is very close to $O_{\bar{E},\omega}$, Eq.~(\ref{eq:off_typical}) implies
\beq
|O_{nm}|\lesssim \|\hat{O}\|e^{-\frac{1}{2}\max\{S_{\mathrm{mc}}(E_n),S_{\mathrm{mc}}(E_m)\}}F_O(\bar{E},\omega)
\label{eq:off_ETH}
\eeq
for any $n\neq m$.
Since $S_{\mathrm{mc}}=\mathcal{O}(V)$, any off-diagonal element should be exponentially small in $V$.

The above argument is justified if we postulate that a many-body Hamiltonian $\hat{H}$ resembles a random matrix~\cite{DAlessio_review2016}.
Based on the random matrix theory, Srednicki~\cite{Srednicki1999} put forth a conjecture that matrix elements of $\hat{O}$ in the energy basis take the form
\beq
O_{nm}=O(\bar{E})\delta_{n,m}+e^{-S_{\mathrm{mc}}(\bar{E})/2}f_O(\bar{E},\omega)R_{nm},
\label{eq:ETH_Srednicki}
\eeq
where $O(\bar{E})$ and $f_O(\bar{E},\omega)$ are smooth functions, and $R_{nm}$ is a numerical factor that varies considerably with $n$ and $m$.
In Ref.~\cite{Srednicki1999}, it is assumed that the real and imaginary parts of $R_{nm}=R_{mn}^*$ can be treated as independent random variables with the zero mean and the unit variance.
The function $O(\bar{E})$ determines the diagonal elements, and its smoothness corresponds to the MITE-ETH.
The function $f_O(\bar{E},\omega)$ usually decays exponentially in $\omega$.
It is noted that the difference between the factors $e^{-\frac{1}{2}\max\{S_{\mathrm{mc}}(E_n),S_{\mathrm{mc}}(E_m)\}}$ in Eq.~(\ref{eq:off_ETH}) and $e^{-S_{\mathrm{mc}}(\bar{E})/2}$ in Eq.~(\ref{eq:ETH_Srednicki}) is not important because this difference can be absorbed in the definition of $f_O(\bar{E},\omega)$.
Equation~(\ref{eq:ETH_Srednicki}) is referred to as the \textit{ETH ansatz}.
The ETH ansatz has a deep connection to quantum chaos.
We recommend Ref.~\cite{DAlessio_review2016} for the concept of quantum chaos and its relation to the ETH.

In Eq.~(\ref{eq:ETH_Srednicki}), the factor $e^{-S_{\mathrm{mc}}(\bar{E})/2}$ is equal to $1/\sqrt{D_{E,\Lambda}}$, and hence the ETH ansatz predicts that typical magnitudes of off-diagonal elements of a local operator scale with the system size as $\mathcal{O}(1/\sqrt{D_{E,\Lambda}})$, which has been numerically confirmed~\cite{Beugeling2015}.
However, it is noted that this strong scaling form of off-diagonal elements does not generally hold when the system has some symmetries and the energy shell consists of several symmetry sectors.
When a symmetry is associated with a few-body conserved quantity (e.g. the total number of particles and the magnetization), the MITE-ETH, i.e., the diagonal part of Eq.~(\ref{eq:ETH_Srednicki}), is also violated in general.
On the other hand, when each symmetry is associated with a nonlocal many-body conserved quantity (e.g., the translation invariance in a lattice system, the spin reversal symmetry, and the space inversion symmetry), the MITE-ETH usually holds in a translation invariant system (but there are exceptions~\cite{Shiraishi-Mori2017,Mori-Shiraishi2017}), while the scaling $1/\sqrt{D_{E,\Lambda}}$ of off-diagonal elements is violated even in such a case\footnote
{The factor $1/\sqrt{D_{E,\Lambda}}$ should be replaced by $1/\sqrt{D_k}$ for each symmetry sector $k$, where $D_k$ is the Hilbert-space dimension of the $k$th symmetry sector within the energy shell~\cite{Mondaini2017}.
}.
Therefore, the ETH ansatz should be investigated for each symmetry sector independently~\cite{Mondaini2017}.

Even if there are several symmetry sectors, usually the following weaker statement holds in nonintegrable systems without the separation of an energy shell into symmetry sectors:
\beq
\lim_{V\rightarrow\infty}\max_{n\neq m}|O_{nm}|=0.
\label{eq:off-diagonal_ETH}
\eeq
Since a nonequilibrium state $|\psi(0)\>$ may have non-negligible overlaps with several symmetry sectors, Eq.~(\ref{eq:off-diagonal_ETH}) has a more direct connection to thermalization compared with the ETH ansatz.
Therefore, in this review, we shall call the property of Eq.~(\ref{eq:off-diagonal_ETH}) the \textit{off-diagonal ETH}.

An immediate consequence of the off-diagonal ETH is that equilibration occurs for an \textit{arbitrary} initial state.
The off-diagonal ETH implies $\max_{n\neq m}|O_{nm}|^2\ll 1$ (usually exponentially small in $V$), and then Eq.~(\ref{eq:off_equilibration}) tells us that the temporal fluctuation of $\<\psi(t)|\hat{O}|\psi(t)\>$ is small for an arbitrary initial state in a sufficiently large system.
Therefore, $D_{\mathrm{eff}}\gg 1$ is not necessary for equilibration if the off-diagonal ETH holds.

The off-diagonal ETH implies that if the initial state has a small effective dimension, the system has equilibrated from the beginning.
This is the reason why $D_{\mathrm{eff}}\gg 1$ is not necessary for equilibration in a system that obeys the off-diagonal ETH.

We can conclude that \textit{the system microscopically thermalizes starting from an arbitrary initial state in an energy shell if its Hamiltonian satisfies both the MITE-ETH and the off-diagonal ETH}.
Under the same assumption, we can also show that \textit{any nonequilibrium state must have a sufficiently large effective dimension}.

\subsection{Von Neumann's quantum ergodic theorem and related results}
\label{sec:Neumann}

So far, the ETH has been just assumed.
An important question is why we should believe the ETH.
Numerical calculations based on the exact diagonalization have indicated that the ETH would be satisfied in several nonintegrable systems~\cite{Kim2014}, but they are limited to small systems, e.g., in a spin-1/2 system, the number of spins is restricted to $N\lesssim 20$.
Is there any \textit{theoretical} argument to support the ETH?

As early as 1929, von Neumann~\cite{Neumann1929} made an attempt in this direction.
He considered an ensemble of all possible Hamiltonians with fixed eigenvalues $\{E_i\}_{i=1}^D$, where $D=D_{E,\Lambda}$ is the dimension of a fixed energy shell $\mathcal{H}_{E,\Lambda}$ (the argument in this section is restricted to the energy shell), and proved that the \textit{normality}~\cite{Goldstein2010} is satisfied for most Hamiltonians in the ensemble.
The normality considered by von Neumann means that for all $\nu$,
\beq
\max_n\left(\<\phi_n|\hat{P}_{\nu}|\phi_n\>-\frac{D_{\nu}}{D}\right)^2
+\max_{n\neq m}|\<\phi_n|\hat{P}_{\nu}|\phi_m\>|^2\ll \frac{1}{N_M^2}\frac{D_{\nu}}{D},
\label{eq:normality}
\eeq
where $N_M$ is the number of macrostates $\nu$.
Von Neumann assumed $\ln D\ll D_{\nu}\ll D$ for all $\nu$, and hence the right-hand side of (\ref{eq:normality}) is very small; however, see below for the problem of this assumption.

We can show that the system thermalizes macroscopically starting from an arbitrary initial state in the energy shell if the Hamiltonian satisfies the normality.
Therefore, von Neumann's result suggests that typical macroscopic quantum systems thermalize.
This is the so called quantum ergodic theorem.

The first term in (\ref{eq:normality}) implies that $\<\phi_n|\hat{P}_{\nu}|\phi_n\>\approx D_{\nu}/D$ for every $n$, which means that every energy eigenstate has the microcanonical distribution over the macrostates.
This requirement is similar to but much stronger than the MATE-ETH.
The second term implies that $|\<\phi_n|\hat{P}_{\nu}|\phi_m\>|$ is very small for all $n\neq m$, which is analogous to the off-diagonal ETH.

A close connection between the quantum ergodic theorem and the MATE-ETH was pointed out by Goldstein et al.~\cite{Goldstein2010} as well as Tasaki~\cite{Tasaki_arXiv2010} (Tasaki's ``thermodynamic normality'' is essentially equivalent to the MATE-ETH).
The relationship between the quantum ergodic theorem and the MITE-ETH was discussed by Rigol and Srednicki~\cite{Rigol2012}.

The precise statement of the quantum ergodic theorem proved by von Neumann~\cite{Neumann1929} is summarized as follows.
We pick up a Hamiltonian $\hat{H}$ randomly from the ensemble of Hamiltonians.
Since we fix the eigenvalues of $\hat{H}$, choosing a Hamiltonian randomly means choosing the eigenstates $|\phi_1\>, |\phi_2\>,\dots |\phi_D\>$ randomly.
This is carried out by a random unitary transformation $\hat{H}\rightarrow \hat{U}\hat{H}\hat{U}^{\dagger}$.
Thus, randomly choosing a Hamiltonian is equivalent to randomly choosing a unitary operator according to the Haar measure.
Let $\mu_U[\mathcal{C}]$ denote the probability of finding $\hat{U}$ that satisfies the condition $\mathcal{C}$.
Then, the quantum ergodic theorem states that
\beq
\mu_U\left[\max_n\left|\<\phi_n|\hat{P}_{\nu}|\phi_n\>-\frac{D_{\nu}}{D}\right|\geq\epsilon\right]\leq\frac{2D^2}{e\sqrt{2\pi D_{\nu}}}\exp\left(-\frac{\vartheta\epsilon^2D^2}{2D_{\nu}}\right),
\label{eq:Neumann1}
\eeq
where $\vartheta$ is a numerical constant slightly less than unity and $\epsilon$ must satisfy $D^{-1}\leq\epsilon\ll D_{\nu}/D$, and
\beq
\mu_U\left[\max_{n\neq m}|\<\phi_n|P_{\nu}|\phi_m\>|\geq\epsilon\right]\leq 2e^{-4\epsilon^2(D-5/2)},
\label{eq:Neumann2}
\eeq
where $\epsilon>0$ is arbitrary.
The quantum ergodic theorem tells us that the normality, i.e., inequality~(\ref{eq:normality}), holds for most Hamiltonians.

Von Neumann characterized thermal equilibrium as the microcanonical distribution of the macrostates, but we have already seen that MATE is characterized by the thermodynamic typicality, $D_{\nu_{\mathrm{eq}}}\approx D$, and Eq.~(\ref{eq:MATE1}).
von Neumann's setting assuming $\ln D\ll D_{\nu}\ll D$ for all $\nu$ is not suitable for this case as it is~\cite{Goldstein2010,Goldstein2010_Proc}.
This difficulty can be avoided by considering further partitioning of $\mathcal{H}_{\mathrm{eq}}$ as $\mathcal{H}_{\mathrm{eq}}=\bigoplus_k\mathcal{H}_{\mathrm{eq}}^{(k)}$ so that for each $k$, $\ln D\ll \mathrm{dim}\,\mathcal{H}_{\mathrm{eq}}^{(k)}\ll D$, although there is no physical meaning of each $\mathcal{H}_{\mathrm{eq}}^{(k)}$.
From a different approach, Pauli and Fierz~\cite{Pauli1937} succeeded in removing the above condition on $D_{\nu}$; the bounds obtained by Pauli and Fierz are weaker than those by von Neumann, but \textit{there is no restriction on} $D_{\nu}$.

Anyway, it is apparent that the results by von Neumann and by Pauli and Fierz are too general for the purpose of explaining thermalization in a real macroscopic system with $D_{\nu_{\mathrm{eq}}}\approx D$.
It is much more difficult to prove the nomality than the MATE-ETH, but the MATE-ETH is sufficient to explain thermalization.
Goldstein et al.~\cite{Goldstein2010} have proved in a much simpler way that most Hamiltonians satisfy the MATE-ETH under the condition that there exists an equilibrium macrostate with $D_{\nu_{\mathrm{eq}}}\approx D$.
In other words, if we choose our Hamiltonian randomly from the ensemble of Hamiltonians with fixed eigenvalues $E_1,E_2,\dots, E_D$, the MATE-ETH is satisfied with probability very close to one.
This result may suggest that the MATE-ETH is also satisfied in a given concrete system unless we have reasons to expect otherwise.

The results by von Neumann~\cite{Neumann1929}, Pauli and Fierz~\cite{Pauli1937}, and Goldstein et al.~\cite{Goldstein2010} are only concerned with the MATE-ETH.
The generalization of these results to also covering the MITE-ETH has been carried out by Reimann~\cite{Reimann2015_PRL}.
He has derived the following inequalities for an \textit{arbitrary} self-adjoint operator $\hat{O}\in\mathcal{B}$: for any $\epsilon>0$,
\beq
\mu_U\left[\max_n|\<\phi_n|\hat{O}|\phi_n\>-\<\hat{O}\>_{\mathrm{mc}}|\geq\epsilon\right]\leq 2D\exp\left(-\frac{2\epsilon^2D}{9\pi^3\Delta_O^2}\right),
\label{eq:Reimann_ergodic1}
\eeq
and
\beq
\mu_U\left[\max_{n\neq m}|\<\phi_n|\hat{O}|\phi_m\>|\geq\epsilon\right]\leq 4D(D-1)\exp\left(-\frac{\epsilon^2D}{18\pi^3\Delta_O^2}\right),
\label{eq:Reimann_ergodic2}
\eeq
where $\Delta_O$ denotes the difference between the largest and the smallest eigenvalues of $\hat{O}$.
If we put $\hat{O}=\hat{P}_{\mathrm{eq}}$, the inequality~(\ref{eq:Reimann_ergodic1}) shows that the MATE-ETH is true for most Hamiltonians.
Putting $\hat{O}=\hat{P}_{\nu}$, we recover the quantum ergodic theorem without any restriction on $D_{\nu}$.

In order to treat the MITE-ETH, we consider all $\hat{O}\in\mathcal{S}_{\mathrm{loc}}^{(\ell)}$.
The objects of interest are the following quantities:
\beq
A_1:=\mu_U\left[\sup_{\hat{O}\in\mathcal{S}_{\mathrm{loc}}^{(\ell)}: \|\hat{O}\|=1}\max_n|\<\phi_n|\hat{O}|\phi_n\>-\<\hat{O}\>_{\mathrm{mc}}|\geq\epsilon\right],
\eeq
and
\beq
A_2:=\mu_U\left[\sup_{\hat{O}\in\mathcal{S}_{\mathrm{loc}}^{(\ell)}: \|\hat{O}\|=1}\max_{n\neq m}|\<\phi_n|\hat{O}|\phi_m\>|\geq\epsilon\right].
\eeq
If $A_1$ and $A_2$ are very small, it implies that for most Hamiltonians, $\<\phi_n|\hat{O}|\phi_n\>\approx\<\hat{O}\>_{\mathrm{mc}}$ and $|\<\phi_n|\hat{O}|\phi_m\>|\ll 1$, respectively, for any local operator $\hat{O}$ with $\|\hat{O}\|=1$.

Now we shall show that $A_1$ and $A_2$ are indeed very small.
By using $\mathcal{S}_{\mathrm{loc}}^{(\ell)}=\bigcup_{i=1}^V\mathcal{S}_i^{(\ell)}$, we have
\beq
A_1\leq\sum_{i=1}^V\mu_U\left[\sup_{\hat{O}\in\mathcal{S}_i^{(\ell)}: \|\hat{O}\|=1}\max_n|\<\phi_n|\hat{O}|\phi_n\>-\<\hat{O}\>_{\mathrm{mc}}|\geq\epsilon\right].
\eeq
As explained in Sec.~\ref{sec:preliminary}, $\hat{O}\in\mathcal{S}_i^{(\ell)}$ can be expanded as $\hat{O}=\sum_{p=1}^{D_i^{(\ell)2}}c_p\hat{O}^{(p)}$ by using a complete orthonormal basis $\{\hat{O}^{(p)}\}$ satisfying Eq.~(\ref{eq:orthonormal}).
Therefore, we have
\begin{align}
A_1&\leq\sum_{i=1}^V\mu_U\left[\sup_{\hat{O}\in\mathcal{S}_i^{(\ell)}:\|\hat{O}\|=1}\max_n\left|\sum_{p=1}^{D_i^{(\ell)2}}c_p\left(\<\phi_n|\hat{O}^{(p)}|\phi_n\>-\<\hat{O}^{(p)}\>_{\mathrm{mc}}\right)\right|\geq\epsilon\right]
\nonumber \\
&\leq\sum_{i=1}^V\mu_U\left[\sup_{\hat{O}\in\mathcal{S}_i^{(\ell)}:\|\hat{O}\|=1}\left(\sum_{p=1}^{D_i^{(\ell)2}}|c_p|\cdot\max_{n,p}\left|\<\phi_n|\hat{O}^{(p)}|\phi_n\>-\<\hat{O}^{(p)}\>_{\mathrm{mc}}\right|\right)\geq\epsilon\right].
\end{align}
By using Eq.~(\ref{eq:c_p}), we obtain
\begin{align}
A_1&\leq\sum_{i=1}^V\mu_U\left[\max_{n,p}|\<\phi_n|\hat{O}^{(p)}|\phi_n\>-\<\hat{O}^{(p)}\>_{\mathrm{mc}}|\geq\epsilon/D_i^{(\ell)3/2}\right]
\nonumber \\
&\leq\sum_{i=1}^V\sum_{p=1}^{D_i^{(\ell)2}}\mu_U\left[\max_n|\<\phi_n|\hat{O}^{(p)}|\phi_n\>-\<\hat{O}^{(p)}\>_{\mathrm{mc}}|\geq\epsilon/D_i^{(\ell)3/2}\right].
\end{align}
By applying Reimann's result given by Eq.~(\ref{eq:Reimann_ergodic1}) combined with
\beq
\Delta_{O^{(p)}}\leq 2\|\hat{O}^{(p)}\|=\frac{2}{\sqrt{D_i^{(\ell)}}}
\eeq
and $D_i^{(\ell)}\leq D_{\ell}$, we obtain the following upper bound of $A_1$:
\beq
A_1\leq 2VDD_{\ell}^2\exp\left(-\frac{\epsilon^2}{18\pi^3}\frac{D}{D_{\ell}^{2}}\right),
\label{eq:upper_A1}
\eeq
which is extremely small because $D=e^{\mathcal{O}(V)}$ and $D_{\ell}=e^{\mathcal{O}(\ell^d)}$ ($\epsilon$ should satisfy $\sqrt{D_{\ell}^{2}/D}\ll \epsilon\ll 1$).
Similarly, we can obtain the following upper bound on $A_2$ from Eq.~(\ref{eq:Reimann_ergodic2}):
\beq
A_2\leq 4VD(D-1)D_{\ell}^{2}\exp\left(-\frac{\epsilon^2}{72\pi^3}\frac{D}{D_{\ell}^{2}}\right).
\label{eq:upper_A2}
\eeq
Inequalities~(\ref{eq:upper_A1}) and (\ref{eq:upper_A2}) prove that the MITE-ETH and the off-diagonal ETH hold for most Hamiltonians.

Finally, we make a remark on the status of the quantum ergodic theorem and its generalization.
In this section, an ensemble of Hamiltonians is introduced, and it is shown that the ETH (including the off-diagonal ETH) is a typical property shared by a vast majority of Hamiltonians in the ensemble.
This fact may suggest that the ETH is also satisfied in a concrete given system.
However, we should pay attention to the fact that almost all Hamiltonians in the ensemble produced by random unitary operators are highly \textit{unphysical} ones which contain many-body long-range interactions.
It would therefore not be surprising if the properties of ``typical'' Hamiltonians do not hold for a \textit{given physically realistic} Hamiltonian which contains only few-body interactions~\cite{Hamazaki_arXiv2017}.
In other words, ``typical'' Hamiltonians produced by random unitary operators may not represent ``typical'' Hamiltonians in the real world.
Thus, although the argument based on an ensemble of Hamiltonians is interesting and intuitively appealing, and sometimes works well in practice (see Sec.~\ref{sec:timescale}), we should not overestimate its value.

\subsection{Sufficient condition of macroscopic thermalization}
\label{sec:macro_thermalization}

The ETH explains the presence or absence of thermalization to a considerable extent, but it would be hopeless to prove the MITE-ETH and the MATE-ETH under a general choice of macrovariables\footnote
{For a given system, it is sometimes possible to prove the MATE-ETH under a special choice of macrovariables.
}.
A different approach to thermalization, if any, would be helpful.

Assuming that the system satisfies the thermodynamic typicality, Tasaki~\cite{Tasaki2016} gave a sufficient condition of macroscopic thermalization.
Let us consider an isolated quantum system satisfying the thermodynamic typicality expressed by Eq.~(\ref{eq:therm_typicality_q2}), $1-D_{\nu_{\mathrm{eq}}}/D_{E,\Lambda}\leq e^{-\gamma V}$.
Tasaki proved that if the effective dimension of the initial state satisfies
\beq
D_{\mathrm{eff}}\geq e^{-\eta V}D
\label{eq:Tasaki_thermalization}
\eeq
with a constant $\eta$ satisfying $0\leq\eta<\gamma$, the system macroscopically thermalizes starting from this initial state.
Thus, Eq.~(\ref{eq:Tasaki_thermalization}) provides a sufficient condition of macroscopic thermalization.

An intuitive meaning of this sufficient condition is understood as follows.
The thermodynamic typicality~(\ref{eq:therm_typicality_q2}) implies $D_{\mathrm{neq}}\leq e^{-\gamma V}D_{E,\Lambda}$, where
\beq
D_{\mathrm{neq}}:=D_{E,\Lambda}-D_{\mathrm{eq}}
\eeq
is the dimension of the nonequilibrium subspace defined by
\beq
\mathcal{H}_{\mathrm{neq}}:=\bigoplus_{\nu\neq\nu_{\mathrm{eq}}}\mathcal{H}_{\nu}.
\eeq
On the other hand, the condition~(\ref{eq:Tasaki_thermalization}) implies $D_{\mathrm{eff}}\gg e^{-\gamma V}D_{E,\Lambda}$ because $\eta<\gamma$.
By combining it with $D_{\mathrm{neq}}\leq e^{-\gamma V}D_{E,\Lambda}$, we have $D_{\mathrm{eff}}\gg D_{\mathrm{neq}}$.
Since the effective dimension roughly corresponds to the effective number of energy eigenstates contributing to the initial state, $D_{\mathrm{eff}}\gg D_{\mathrm{neq}}$ indicates that the diagonal ensemble $\rho_{\mathrm{D}}=\overline{|\psi(t)\>\<\psi(t)|}$ must have a dominantly large weight to the equilibrium macrostate, $\mathrm{Tr}\,\hat{P}_{\mathrm{eq}}\rho_{\mathrm{D}}=\overline{\<\psi(t)|\hat{P}_{\mathrm{eq}}|\psi(t)\>}\approx 1$, which immediately implies $\<\psi(t)|\hat{P}_{\mathrm{eq}}|\psi(t)\>\approx 1$ for most $t$ in a long run.
This means that a system starting from an initial state with a sufficiently large effective dimension must thermalize macroscopically.

\subsection{Weak ETH and its implication on thermalization}
\label{sec:weak_ETH}

\subsubsection{Definition of the weak ETH}

The MITE-ETH that we have introduced in Sec. 3.3 is sometimes referred to as the strong ETH in the sense that \textit{all} the energy eigenstates in an energy shell $\mathcal{H}_{E,\Lambda}$ represent MITE.
In this section, we consider a variant of the MITE-ETH, which is referred to as the \textit{weak ETH}~\cite{Biroli2010}. 
It states that \textit{most} of the energy eigenstates in $\mathcal{H}_{E,\Lambda}$ represent MITE.
The other energy eigenstates that do not represent microscopic thermal equilibirum are called the \textit{rare states}~\cite{Biroli2010}.
While it would be very difficult (indeed, it has not been successful) to prove the MITE-ETH for a given system, the weak ETH can be proven for translation-invariant short-range interacting spin systems~\cite{Biroli2010,Iyoda2017,Mori2016_weak}.

Now we shall explain the precise statement of the weak ETH.
Let us denote by $\mathcal{E}$ a set of energy eigenstates $|\phi_n\>$ in $\mathcal{H}_{E,\Lambda}$ and denote by $\mu_{\mathcal{E}}[\mathcal{C}]$ the probability that $\mathcal{C}$ is true with respect to a uniform distribution on $\mathcal{E}$, i.e., the fraction of the energy eigenstates that satisfy the condition $\mathcal{C}$.
The weak ETH on length scale $\ell$ is then equivalent to $\Delta^{\mathrm{ETH}}_{\epsilon}\ll 1$ for a sufficiently small $\epsilon>0$, where
\begin{align}
\Delta^{\mathrm{ETH}}_{\epsilon}&:= \mu_{\mathcal{E}}\left[\sup_{\hat{O}\in\mathcal{S}_{\mathrm{loc}}^{(\ell)}: \|\hat{O}\|=1}|\<\phi_n|\hat{O}|\phi_n\>-\<\hat{O}\>_{\mathrm{mc}}|\geq\epsilon\right]
\nonumber \\
&=\mu_{\mathcal{E}}\left[\max_{i\in\{1,2,\dots,V\}}\sup_{\hat{O}\in\mathcal{S}_{X_i^{(\ell)}}: \|\hat{O}\|=1}|\<\phi_n|\hat{O}|\phi_n\>-\<\hat{O}\>_{\mathrm{mc}}|\geq\epsilon\right].
\label{eq:weak_ETH}
\end{align}
If we assume the translation invariance,
\beq
\Delta_{\epsilon}^{\mathrm{ETH}}=\mu_{\mathcal{E}}\left[\sup_{\hat{O}\in\mathcal{S}_1^{(\ell)}: \|\hat{O}\|=1}|\<\phi_n|\hat{O}|\phi_n\>-\<\hat{O}\>_{\mathrm{mc}}|\geq\epsilon\right].
\eeq
The condition $\Delta^{\mathrm{ETH}}_{\epsilon}\ll 1$ means that for most energy eigenstates $|\phi_n\>\in\mathcal{E}$, $\<\phi_n|\hat{O}|\phi_n\>\approx\<\hat{O}\>_{\mathrm{mc}}$ for \textit{all} the local operators $\hat{O}\in\mathcal{S}_{\mathrm{loc}}^{(\ell)}$.
That is, most energy eigenstates represent MITE on length scale $\ell$, which is the statement of the weak ETH.

Meanwhile, if we have, \textit{for a fixed operator} $\hat{O}\in\mathcal{B}$, $\<\phi_n|\hat{O}|\phi_n\>\approx\<\hat{O}\>_{\mathrm{mc}}$ for most energy eigenstates $|\phi_n\>\in\mathcal{E}$, we shall say that the weak-ETH condition for $\hat{O}$ holds, or $\hat{O}$ satisfies the weak-ETH condition.
The weak-ETH condition for $\hat{O}$ is expressed as $\Delta^{\mathrm{ETH}}_{\epsilon}[\hat{O}]\ll 1$, where
\beq
\Delta^{\mathrm{ETH}}_{\epsilon}[\hat{O}]:=\mu_{\mathcal{E}}\left[|\<\phi_n|\hat{O}|\phi_n\>-\<\hat{O}\>_{\mathrm{mc}}|\geq\epsilon\|\hat{O}\|\right].
\label{eq:weak_ETH_condition}
\eeq

We shall discuss the relation between the weak ETH given by Eq.~(\ref{eq:weak_ETH}) and the weak-ETH condition for $\hat{O}$ given by Eq.~(\ref{eq:weak_ETH_condition}).
We can show that, under the assumption of translation invariance,
\beq
\sup_{\hat{O}\in\mathcal{S}_{\mathrm{loc}}^{(\ell)}}\Delta_{\epsilon}^{\mathrm{ETH}}[\hat{O}]\leq\Delta_{\epsilon}^{\mathrm{ETH}}\leq D_{\ell}^2\sup_{\hat{O}\in\mathcal{S}_{\mathrm{loc}}^{(\ell)}}\Delta_{\epsilon/D_{\ell}}^{\mathrm{ETH}}[\hat{O}].
\label{eq:weak_ETH_inequality}
\eeq
The inequality $\sup_{\hat{O}\in\mathcal{S}_{\mathrm{loc}}^{(\ell)}}\Delta_{\epsilon}^{\mathrm{ETH}}[\hat{O}]\leq\Delta_{\epsilon}^{\mathrm{ETH}}$ is obvious, so we shall prove the inequality $\Delta_{\epsilon}^{\mathrm{ETH}}\leq D_{\ell}^2\sup_{\hat{O}\in\mathcal{S}_{\mathrm{loc}}^{(\ell)}}\Delta_{\epsilon/D_{\ell}}^{\mathrm{ETH}}[\hat{O}]$.
We expand $\hat{O}\in\mathcal{S}_1^{(\ell)}$ as $\hat{O}=\sum_{p=1}^{D_{\ell}^2}c_p\hat{O}^{(p)}$, and then, by using Eq.~(\ref{eq:c_p}),
\begin{align}
\Delta^{\mathrm{ETH}}_{\epsilon}
&\leq\mu_{\mathcal{E}}\left[\max_{p\in\{1,2,\dots,D_{\ell}^2\}}\left|\<\phi_n|\hat{O}^{(p)}|\phi_n\>-\<\hat{O}^{(p)}\>_{\mathrm{mc}}\right|\geq\epsilon/D_{\ell}^{3/2}\right]
\nonumber \\
&\leq D_{\ell}^{2}\max_{p\in\{1,2,\dots,D_{\ell}^2\}}\mu_{\mathcal{E}}\left[\left|\<\phi_n|\hat{O}^{(p)}|\phi_n\>-\<\hat{O}^{(p)}\>_{\mathrm{mc}}\right|\geq\epsilon\|\hat{O}^{(p)}\|/D_{\ell}\right]
\nonumber \\
&=D_{\ell}^2\max_{p\in\{1,2,\dots,D_{\ell}^2\}}\Delta^{\mathrm{ETH}}_{\epsilon/D_{\ell}}[\hat{O}^{(p)}]
\nonumber \\
&\leq D_{\ell}^2\sup_{\hat{O}\in\mathcal{S}_{\mathrm{loc}}^{(\ell)}} \Delta_{\epsilon/D_{\ell}}^{\mathrm{ETH}}[\hat{O}],
\label{eq:Delta_ETH}
\end{align}
where we have used Eq.~(\ref{eq:orthonormal}).

The inequality~(\ref{eq:weak_ETH_inequality}) tells us that if any local operator on length scale $\ell$ satisfies the weak-ETH condition, it implies that the weak ETH on length scale $\ell$ holds unless $D_{\ell}=e^{\mathcal{O}(\ell^d)}$ is not too large.

By using the Chebyshev inequality~\cite{Feller_text}, we obtain
\beq
\Delta_{\epsilon}^{\mathrm{ETH}}[\hat{O}]\leq\frac{\left(I_{\mathrm{ETH}}^{\mathrm{w}}[\hat{O}]\right)^2}{\epsilon^2\|\hat{O}\|^2},
\eeq
where $I_{\mathrm{ETH}}^{\mathrm{w}}[\hat{O}]$ is defined by Eq.~(\ref{eq:indicator_w}).
Therefore, the weak ETH can be numerically tested by computing $I_{\mathrm{ETH}}^{\mathrm{w}}[\hat{O}]$.

\subsubsection{Sufficient condition of microscopic thermalization}
\label{sec:micro_sufficient}

The weak ETH does not ensure that every initial state in an energy shell undergoes microscopic thermalization.
Instead, the weak ETH ensures that an initial state without a substantial overlap with rare states undergoes microscopic thermalization.
Since the total number of the rare states in an energy shell is much smaller than the dimension of the energy shell, the overlap with the rare states cannot be large and the system microscopically thermalizes if the initial state has a sufficiently large effective dimension $D_{\mathrm{eff}}$.
We can show
\beq
\left|\overline{\<\psi(t)|\hat{O}|\psi(t)\>}-\<\hat{O}\>_{\mathrm{mc}}\right|<\epsilon\|\hat{O}\|+2\|\hat{O}\|\sqrt{\frac{D}{D_{\mathrm{eff}}}\Delta_{\epsilon}^{\mathrm{ETH}}[\hat{O}]}
\eeq
for an arbitrary $\epsilon>0$.
The above inequality and the condition of thermalization (\ref{eq:thermalization}) imply that if
\beq
D_{\mathrm{eff}}\gg D\sup_{\hat{O}\in\mathcal{S}_{\mathrm{loc}}^{(\ell)}}\Delta_{\epsilon}^{\mathrm{ETH}}[\hat{O}]
\label{eq:sufficient_micro_thermalization}
\eeq
holds, the system thermalizes microscopically to a precision of $\epsilon$, that is, for any $\hat{O}\in\mathcal{S}_{\mathrm{loc}}^{(\ell)}$,
\beq
\left|\overline{\<\psi(t)|\hat{O}|\psi(t)\>}-\<\hat{O}\>_{\mathrm{mc}}\right|\lesssim\epsilon\|\hat{O}\|.
\eeq
From inequality~(\ref{eq:weak_ETH_inequality}), the weak ETH implies $\sup_{\hat{O}\in\mathcal{S}_{\mathrm{loc}}^{(\ell)}}\Delta_{\epsilon}^{\mathrm{ETH}}[\hat{O}]\ll 1$, and hence Eq.~(\ref{eq:sufficient_micro_thermalization}) is meaningful when the weak ETH is satisfied.
In particular, when $\ell$ is independent of the system size, we can show $\sup_{\hat{O}\in\mathcal{S}_{\mathrm{loc}}^{(\ell)}}\Delta_{\epsilon}^{\mathrm{ETH}}[\hat{O}]=e^{-\mathcal{O}(V)}$ (see Eq.~(\ref{eq:Mori_LDP}) below).
Therefore, the system microscopically thermalizes whenever $D_{\mathrm{eff}}\geq De^{-\eta' V}$ with a certain $\eta'>0$ for sufficiently large $V$.
This is analogous to the sufficient condition of macroscopic thermalization given by Eq.~(\ref{eq:Tasaki_thermalization}). 

It is shown that the weak ETH is true even for translation-invariant integrable systems, which often fail to thermalize.
Thus, the weak ETH is not sufficient to determine the presence or absence of microscopic thermalization for physically realistic initial states.
In an integrable system, a physically realistic initial state, e.g., an initial state prepared by a quantum quench, does not have a sufficiently large effective dimension and has large weight to rare energy eigenstates which do not represent MITE~\cite{Biroli2010}.
In conclusion, the weak ETH is important in the sense that it gives a sufficient condition of microscopic thermalization, but it is not at all trivial whether a physically realistic initial state satisfies this sufficient condition.

\subsubsection{Rigorous results on the weak ETH}

Biroli, Kollath, and L\"auchli~\cite{Biroli2010} has argued that the weak-ETH condition is satisfied for any local operator $\hat{O}\in\mathcal{S}_{\mathrm{loc}}^{(\ell)}$ for $\ell$ independent of the system size, but their proof is not rigorous.
Iyoda, Kaneko, and Sagawa~\cite{Iyoda2017} have made this argument rigorous and extended it to local operators on length scale $\ell\propto L^{\alpha}$ with $0\leq\alpha<1/2$, where $L$ is the length of the whole system ($L^d\sim V$).
They have proved
\beq
\Delta^{\mathrm{ETH}}_{\epsilon}[\hat{O}]\leq\frac{\mathcal{O}(V^{-\frac{1-2\alpha}{4}+\delta})}{\epsilon^2}
\label{eq:Iyoda}
\eeq
for an arbitrary $\delta>0$ and for any local operator $\hat{O}$ on length scale $\ell\propto L^{\alpha}$ under the assumption that the microcanonical ensemble defined on the energy shell exhibits the exponential decay of correlations and a mild condition on the Massieu function\footnote
{The Massieu function is defined as $\phi(\beta)=\ln Z$, where $Z$ is the partition function given by $Z=\mathrm{Tr}\,e^{-\beta\hat{H}}$.
The assumption on the Massieu function is necessary to replace the microcanonical ensemble by the canonical ensemble~\cite{Tasaki_arXiv2016}, the latter of which is mathematically tractable.
}.
The right-hand side of Eq.~(\ref{eq:Iyoda}) tends to zero in the thermodynamic limit for any $\epsilon>0$.
By using Eq.~(\ref{eq:Iyoda}), the second law of thermodynamics and the fluctuation theorem are derived for pure states~\cite{Iyoda2017}.
Applying Eq.~(\ref{eq:Iyoda}) to Eq.~(\ref{eq:Delta_ETH}), we obtain
\beq
\Delta^{\mathrm{ETH}}_{\epsilon}\leq\frac{D_{\ell}^4}{\epsilon^2}\mathcal{O}(V^{\frac{1-2\alpha}{4}+\delta}).
\label{eq:Iyoda_weak}
\eeq
Since $D_{\ell}=e^{\mathcal{O}(\ell^d)}$, the right-hand side of Eq.~(\ref{eq:Iyoda_weak}) tends to zero in the thermodynamic limit as long as $\ell\lesssim(\ln V)^{1/d}$.
Thus, Iyoda, Kaneko, and Sagawa~\cite{Iyoda2017} have proved the weak ETH on length scale $\ell\lesssim (\ln V)^{1/d}$, although they explicitly gave the proof of the weak ETH only for $\ell$ that is independent of the system size in their paper.
It should be noted that although Eq.~(\ref{eq:Iyoda_weak}) ensures the weak ETH only for $\ell\lesssim(\ln V)^{1/d}$, it ensures microscopic thermalization for $\ell\lesssim L^{\alpha}$ as long as the initial state satisfies Eq.~(\ref{eq:sufficient_micro_thermalization}).

In the case that $\ell$ is independent of the system size, Mori~\cite{Mori2016_weak} has proved that
\beq
\Delta^{\mathrm{ETH}}_{\epsilon}[\hat{O}]\leq e^{-V\eta_O(\epsilon)+o(V)}
\label{eq:Mori_LDP}
\eeq
with $\eta_O(\epsilon)>0$, under the assumption that the equilibrium ensemble has the large deviation property, Eq.~(\ref{eq:LDP}), for any local intensive operator, which can indeed been proven in many cases~\cite{Ogata2010,Netocny2004,Lenci2005} (see also Ref.~\cite{Tasaki2016}).
By applying Eq.~(\ref{eq:Mori_LDP}) to Eq.~(\ref{eq:Delta_ETH}), we obtain
\beq
\Delta^{\mathrm{ETH}}_{\epsilon}\leq D_{\ell}^2e^{-V\eta_{\ell}(\epsilon/D_{\ell})+o(V)},
\label{eq:Mori_weak}
\eeq
where $\eta_{\ell}(\epsilon):=\inf_{\hat{O}\in\mathcal{S}_{\mathrm{loc}}^{(\ell)}}\eta_{O}(\epsilon)>0$.
This bound is much tighter than Eq.~(\ref{eq:Iyoda_weak}), but is restricted to $\ell$ independent of the system size.
Inequality~(\ref{eq:Mori_weak}) shows that the weak ETH on length scale $\ell=\mathcal{O}(L^0)$ holds in translation-invariant quantum spin systems with short-range interactions.


\subsubsection{Simple application}
\label{sec:application_weak}

As an application of the weak ETH, we consider an isolated system with translation invariance constituted by the system of interest $S$ and the remaining part $B$.
We assume that the volume $v$ of $S$ is much smaller than the volume $V_B$ of $B$.
The Hamiltonian is given by $\hat{H}=\hat{H}_S+\hat{H}_B+\hat{H}_I$, where $\hat{H}_S$ and $\hat{H}_B$ are the Hamiltonians of $S$ and $B$, respectively, and $\hat{H}_I$ is the interaction Hamiltonian.

The initial state is given by a mixed state 
\beq
\rho(0)=|\phi_n^S\>\<\phi_n^S|\otimes\rho_{\mathrm{mc},B},
\label{eq:weak_initial}
\eeq
where $|\phi_n^S\>$ is an energy eigenstate of $\hat{H}_S$ and $\rho_{\mathrm{mc},B}$ is the microcanonical ensemble with respect to the Hamiltonian $\hat{H}_B$ at the energy $E_B$.
In this case, $B$ acts as a bath for $S$.
Initially, the system $S$ may be nonequilibrium while the ``bath'' $B$ is in thermal equilibrium.
We ask whether the entire system thermalizes microscopically after a sufficiently long time.

Although Eq.~(\ref{eq:sufficient_micro_thermalization}) is originally derived for a pure initial state, its generalization to a mixed initial state is straightforward.
For a mixed initial state $\rho(0)$, the effective dimension is defined by $D_\mathrm{eff}:=\sum_n\<\phi_n|\rho(0)|\phi_n\>^2$, where $|\phi_n\>$'s are energy eigenstates of the entire system.

If $\hat{H}_I$ is absent, the effective dimension $D_{\mathrm{eff}}^{(0)}$ of the initial state~(\ref{eq:weak_initial}) is given by
\beq
D_{\mathrm{eff}}^{(0)}=\left(\mathrm{Tr}_B\rho_{\mathrm{mc},B}^2\right)^{-1}= e^{S_{\mathrm{mc}}^B(E_B,V_B)},
\eeq
where $S_{\mathrm{mc}}^B(E_B,V_B)$ is the microcanonical entropy of $B$.
When there is an interaction $\hat{H}_I$, the effective dimension $D_{\mathrm{eff}}$ for the initial state~(\ref{eq:weak_initial}) is not easily calculated, but one can show that $D_{\mathrm{eff}}\geq D_{\mathrm{eff}}^{(0)}$, and thus
\beq
D_{\mathrm{eff}}\geq e^{S_{\mathrm{mc}}^B(E_B,V_B)}.
\eeq

Now we can use the sufficient condition of MITE, i.e., Eq.~(\ref{eq:sufficient_micro_thermalization}).
For a mixed initial state $\rho(0)$, the effective dimension should be
The Hilbert-space dimension $D$ of the energy shell of the entire system is given by $D=e^{S_{\mathrm{mc}}(E,V)}$, where $S_{\mathrm{mc}}(E,V)$ is the microcanonical entropy of the entire system with the energy $E$ and the volume $V$.
If we consider a sufficiently large $V_B$ with $v$ being held fixed, and assume that the interaction energy is negligible compared with the total energy, we have
\beq
S_{\mathrm{mc}}(E,V)=S_{\mathrm{mc}}^B(E_B,V_B)+o(V)
\eeq
with $o(V)\geq 0$, which yields $D_{\mathrm{eff}}/D=e^{-o(V)}$.
On the other hand, inequality~(\ref{eq:Mori_weak}) implies $\sup_{\hat{O}\in\mathcal{S}_{\mathrm{loc}}^{(\ell)}}\Delta_{\mathrm{ETH}}^{\epsilon}[\hat{O}]=e^{-\mathcal{O}(V)}$ for sufficiently large $V$.
Therefore, Eq.~(\ref{eq:sufficient_micro_thermalization}) is always satisfied, and the entire system microscopically thermalizes.
It should be emphasized that this result is valid even if the entire system is integrable.

\subsection{Failure of ETH in several systems}
\label{sec:failure}

In this section, we introduce several classes of systems which do not obey the MITE-ETH.
In integrable systems, there are an extensively large number of conserved quantities $\{\hat{Q}_k\}$, each of which is written as an extensive local operator.
Those systems clearly do not obey the MITE-ETH because $\hat{Q}_k$ itself does not satisfy $\<\phi_n|\hat{Q}_k|\phi_n\>\approx\<\hat{Q}_k\>_{\mathrm{mc}}$.
It has been numerically confirmed that integrable systems do not obey the MITE-ETH~\cite{Rigol2008,Rigol2009_PRL,Biroli2010,Santos2010,Steinigeweg2013}.
It should be noted, however, that the weak ETH is satisfied in translation-invariant integrable systems~\cite{Biroli2010,Mori2016_weak}, also see Refs.~\cite{Ikeda2013,Alba2015} for numerical confirmations.
Integrable systems often fail to thermalize, which was theoretically demonstrated in Refs.~\cite{Rigol2007,Rigol2008,Cassidy2011,Calabrese2011} and experimentally observed~\cite{Kinoshita2006,Gring2012,Smith2013} (however, see Sec.~\ref{sec:application_weak} for a case in which an integrable system thermalizes).
We will discuss the relaxation dynamics of integrable and nearly integrable systems in Sec.~\ref{sec:integrable}.

Many-body localization (MBL) gives another important class of systems violating the ETH~\cite{Basko2006,Altshuler1997,Pal2010,Serbyn2013,Huse2014,Imbrie2016}.
For example, a many-spin system subject to random fields and/or random interactions exhibits the MBL.
The phenomenology of fully many-body localized systems was discussed in Refs.~\cite{Serbyn2013,Huse2014} by considering a complete set of conserved quantities each of which is given by a quasi-local operator. 
Such conserved quantities have been constructed perturbatively in Ref.~\cite{Ros2015}.
A rigorous result by Imbrie~\cite{Imbrie2016} proves that, at least in a one-dimensional spin system with sufficiently strong randomness of the magnetic field and the exchange interaction, the MBL occurs and even the weak ETH is violated.
The exact conserved quantities have also been obtained as a by-product of the rigorous proof of the MBL~\cite{Imbrie2016}.
The MBL was realized experimentally and the absence of thermalization in such a system was observed~\cite{Schreiber2015,Kondov2015,Choi2016,Smith2016}.
In this review, we do not discuss the MBL in detail, and see Refs.~\cite{Nandkishore_review2015,Altman_review2015,Imbrie_review2017} for theoretical reviews.

Numerical studies~\cite{Rigol2008, Santos2010, Biroli2010, Roux2010, Neuenhahn2012, Steinigeweg2013, Steinigeweg2014, Beugeling2014, Kim2014} have reported that the MITE-ETH is valid if the Hamiltonian satisfies the following conditions: (i) translation invariance, which excludes the MBL due to randomness, (ii) no local conserved quantity, which excludes integrable systems, and (iii) short-range interactions.
It may then be tempting to conjecture that \textit{any} Hamiltonian satisfying the above conditions (i)-(iii) obey the MITE-ETH.

Against this conjecture, Shiraishi and Mori~\cite{Shiraishi-Mori2017} have proposed a general method to construct a Hamiltonian satisfying (i)-(iii) but violating the MITE-ETH.
The Hamiltonian considered in Ref.~\cite{Shiraishi-Mori2017} is formally expressed as
\beq
\left\{
\begin{aligned}
&\hat{H}=\hat{H}_0+\hat{H}', \\
&\hat{H}_0=\sum_{i=1}^V\hat{P}_i\hat{h}_i\hat{P}_i, \quad \hat{H}'=\sum_{i=1}^V\hat{h}_i',
\end{aligned}
\right.
\label{eq:Shiraishi-Mori}
\eeq
where $i=1,2,\dots, V$ denotes each site, and $\hat{h}_i$, $\hat{h}_i'$, and $\hat{P}_i$ are local operators acting on sites near the site $i$, i.e., they are in $\mathcal{S}_i^{(\ell)}$ with some $\ell=\mathcal{O}(1)$.
We assume that $\{\hat{P}_i\}_{i=1}^V$ are projection operators, $\hat{P}_i^2=\hat{P}_i$, such that $\mathcal{H_T}\neq\emptyset$, where $\mathcal{H_T}$ is the Hilbert subspace defined as
\beq
\mathcal{H_T}:=\{|\psi\>\in\mathcal{H}: \hat{P}_i|\psi\>=0 \text{ for all } i=1,2,\dots, V\}.
\eeq
The subspace $\mathcal{H_T}$ is nothing but the ground state subspace of the frustration-free Hamiltonian $\hat{H}_{\mathrm{ff}}:=\sum_{i=1}^V\hat{P}_i$.
A state $|\psi\>$ is said to be a target state if $|\psi\>\in\mathcal{H_T}$.
The idea by Shiraishi and Mori~\cite{Shiraishi-Mori2017} is that we can embed target states into the middle of the energy spectrum of $\hat{H}$ as its exact energy eigenstates if $\hat{H}'$ satisfies $[\hat{H}',\hat{P}_i]=0$ for all $i$.
More precisely, the Hamiltonian $\hat{H}$ of Eq.~(\ref{eq:Shiraishi-Mori}) has the orthonormal energy eigenstates $|\phi_1^{\mathcal{T}}\>, |\phi_2^{\mathcal{T}}\>, \dots, |\phi_{\mathrm{dim}\,\mathcal{H_T}}^{\mathcal{T}}\>$, each of which is a target state, $|\phi_n^{\mathcal{T}}\>\in\mathcal{H_T}$.
Since $|\phi_n^{\mathcal{T}}\>$ does not depend on $\{\hat{h}_i\}_{i=1}^V$, its local property will differ from the local property of thermal equilibrium.
Indeed, Shiraishi and Mori have proved that those target energy eigenstates $\{|\phi_n^{\mathcal{T}}\>\}_{n=1}^{\mathrm{dim}\,\mathcal{H_T}}$ do not represent MITE, which means the violation of the MITE-ETH.
Under a general choice of $\hat{h}_i$, $\hat{h}_i'$, and $\hat{P}_i$, a translation-invariant Hamiltonian~(\ref{eq:Shiraishi-Mori}) satisfies the conditions (i)-(iii) mentioned above, and hence, the method of embedding explained above provides us with a new class of Hamiltonians that violate the MITE-ETH.

In Ref.~\cite{Mori-Shiraishi2017}, nonequilibrium dynamics of a model constructed in this way has been studied, and it has been shown that this model microscopically thermalizes after any finite-temperature quench even though the MITE-ETH is violated.
It indicates that this model thermalizes for any physically realistic initial state\footnote
{The violation of the MITE-ETH implies that there must exist some initial states which do not thermalize microscopically.
Therefore, the problem is whether those initial states are likely to be realized in reality.
}.
In this sense, the MITE-ETH is not a necessary condition of microscopic thermalization.

\subsection{Periodically driven systems and Floquet ETH}
\label{sec:Floquet}

A periodically driven system, i.e., a quantum system subject to periodic driving, exhibit various nontrivial macroscopic phenomena.
To name only a few, dynamical localization~\cite{Dunlap1986,Grifoni1998,Kayanuma2008}, coherent destruction of tunneling~\cite{Grossmann1991,Grifoni1998,Kayanuma2008}, dynamical freezing~\cite{Das2010,Hegde2014}, and quantum phase transitions induced by periodic driving~\cite{Eckardt2005,Eckardt2009,Zenesini2009} are remarkable nonequilibrium phenomena.
Recent experimental advances also triggered studies of Floquet topological states~\cite{ Aidelsburger2013, Atala2013, Jotzu2014, Aidelsburger2015} (see Refs.~\cite{Lindner2011, Kitagawa2011_transport, Dora2012, Delplace2013, Grushin2014, Titum2016, Takasan2017} for some theoretical studies).
The ``Floquet time crystal'' is also a recent hot topic~\cite{Else2016,Yao2017,Zhang2017}. 
In the Floquet time crystal, a discrete time translation symmetry is spontaneously broken in the steady state, which is robust against weak perturbations of the driving protocol. 
Attempts to realize interesting properties of matter by applying periodic driving are called Floquet engineering.

A periodically driven system is described by a time-dependent Hamiltonian $\hat{H}(t)$ periodic in time, $\hat{H}(t)=\hat{H}(t+T)$, where $T$ denotes the period of driving.
We assume that $\hat{H}(t)$ is written as a sum of local operators at each time $t$.
The time evolution over a period from $t=0$ to $t=T$ is described by the \textit{Floquet operator} defined as
\beq
\hat{U}_{\mathrm{F}}:=\mathcal{T}e^{-i\int_0^Tdt\,\hat{H}(t)},
\eeq
where $\mathcal{T}$ denotes the time-ordering operator.
Since $\hat{U}_{\mathrm{F}}$ is a unitary operator, there exists a self-adjoint operator $\hat{H}_{\mathrm{F}}$ satisfying $\hat{U}_{\mathrm{F}}=e^{-i\hat{H}_{\mathrm{F}}T}$, which is called the \textit{Floquet Hamiltonian}.
Although the original Hamiltonian $H(t)$ depends on time, $\hat{H}_{\mathrm{F}}$ is time independent.
Then, as long as we consider stroboscopic times $t=nT$ with $n$ being an integer, the time evolution from $t=0$ is described by $\hat{U}_{\mathrm{F}}^n=e^{-i\hat{H}_{\mathrm{F}}nT}=e^{-i\hat{H}_{\mathrm{F}}t}$.
This is formally identical to the time evolution operator under a static Hamiltonian $\hat{H}_{\mathrm{F}}$.
This observation suggests that we can deal with the relaxation of a periodically driven system in a more or less similar way to that of a static system.
Floquet eigenstates $|\phi_n^{\mathrm{F}}\>$ are defined as the eigenstates of $\hat{H}_{\mathrm{F}}$, which play the role of energy eigenstates in a static system.

In general, it is difficult to calculate $\hat{H}_{\mathrm{F}}$ explicitly, but in the high-frequency (small-$T$) regime, the following \textit{Magnus expansion}~\cite{Blanes2009} provides a systematic expansion of $\hat{H}_{\mathrm{F}}$ with respect to $T$:
\beq
\hat{H}_{\mathrm{F}}=\sum_{m=0}^{\infty}T^m\hat{\Omega}_m.
\label{eq:FM}
\eeq
An explicit expression of $\hat{\Omega}_n$ is given in Ref.~\cite{Birula1969} by
\begin{align}
\hat{\Omega}_n=\sum_{\sigma}\frac{(-1)^{n-\theta[\sigma]}\theta[\sigma]!(n-\theta[\sigma])!}{i^n(n+1)^2n!T^{n+1}}\int_0^Tdt_{n+1}\int_0^{t_{n+1}}dt_n\dots\int_0^{t_2}dt_1
\nonumber \\
\times[\hat{H}(t_{\sigma(n+1)}),[\hat{H}(t_{\sigma(n)}),\dots[\hat{H}(t_{\sigma(2)}),\hat{H}(t_{\sigma(1)})]\dots]],
\label{eq:FM_explicit}
\end{align}
where $\theta$ is a permutation of $\{1,2,\dots,n+1\}$, and $\theta[\sigma]:=\sum_{i=1}^n\theta\left(\sigma(i+1)-\sigma(i)\right)$ with $\theta(\cdot)$ being the unit step function.
The Magnus expansion of $\hat{H}_{\mathrm{F}}$ is referred to as the \textit{Floquet-Magnus expansion}.
If the Floquet-Magnus expansion is convergent, $\hat{H}_{\mathrm{F}}$ is quasi-local in the sense that it can be well approximated by a sum of local operators, i.e., a few lowest-order terms of Eq.~(\ref{eq:FM}).
In this case, $\hat{H}_{\mathrm{F}}$ can be treated as if it were a static Hamiltonian with few-body interactions.
As a result, it is expected that a periodically driven system whose Floquet Hamiltonian has a convergent Magnus expansion will relax to a steady state locally indistinguishable from the microcanonical ensemble of $\hat{H}_{\mathrm{F}}$.

However, it is believed that the Floquet-Magnus expansion is divergent in generic many-body interacting systems.
If the Floquet-Magnus expansion is divergent, it suggests that $\hat{H}_{\mathrm{F}}$ contains highly nonlocal many-body interactions\footnote
{The divergence of the Floquet-Magnus expansion does not automatically imply a nonlocal Floquet Hamiltonian. Indeed, there are exceptions.
}.
Then, in a periodically driven nonintegrable system, there will be no local conserved quantity at all ($\hat{H}_{\mathrm{F}}$ is a conserved quantity, but is a highly nonlocal and many-body operator).
As a result, the dynamics is not restricted to an ``energy shell'', and we must consider the entire Hilbert space $\mathcal{H}$.
Along a line similar to the derivation of the canonical typicality in Sec.~\ref{sec:can_typicality}, it is shown that almost all the quantum states $|\psi\>\in\mathcal{H}$ are locally indistinguishable from the infinite-temperature ensemble $\rho_{\infty}:=\hat{1}/D$, where $D=\dim\,\mathcal{H}$ and $\hat{1}$ is the identity operator, that is,
\beq
\rho_{\psi}^X\approx\rho_{\infty}^X:=\mathrm{Tr}_{X^c}\,\rho_{\infty},
\eeq
for any small subsystem $X\subset\{1,2,\dots,V\}$.
It is naturally expected that a nonintegrable periodically driven system will eventually evolve to an infinite-temperature state~\cite{DAlessio2013,DAlessio2014,Lazarides2014}.

The notion of the MITE-ETH is naturally extended to a periodically driven system, which is referred to as the \textit{Floquet ETH}.
The Floquet ETH states that every Floquet eigenstate $|\phi_n^{\mathrm{F}}\>$ represents the infinite-temperature state in the sense that
\beq
\rho_{\phi_n^{\mathrm{F}}}^{X_i^{(\ell)}}\approx\rho_{\infty}^{X_i^{(\ell)}}
\label{eq:Floquet_ETH}
\eeq
for all $i\in\{1,2,\dots,V\}$, where $\ell$ is independent of the system size.
Equation~(\ref{eq:Floquet_ETH}) is equivalent to
\beq
\<\phi_n^{\mathrm{F}}|\hat{O}|\phi_n^{\mathrm{F}}\>\approx\mathrm{Tr}\,\hat{O}\rho_{\infty}=\frac{1}{D}\mathrm{Tr}\,\hat{O}
\label{eq:Floquet_ETH2}
\eeq
for any $\hat{O}\in\mathcal{S}_{\mathrm{loc}}^{(\ell)}$.
The Floquet ETH is numerically verified in several nonintegrable periodically driven systems~\cite{Lazarides2014,DAlessio2014,Kim2014,Ponte2015}.
A periodically driven system does not obey the Floquet ETH and does not heat up to infinite temperature when it is integrable~\cite{Russomanno2012,Lazarides2014} or exhibits the MBL~\cite{Lazarides2015,Ponte2015}.

Although the steady state of a periodically driven system obeying the Floquet ETH is locally indistinguishable from the infinite-temperature state and does not exhibit any interesting phenomenon useful for Floquet engineering, it has been argued that there exists nontrivial quasi-stationary states with long lifetimes in the high-frequency (small $T$) regime.
This phenomenon is called \textit{Floquet prethermalization}, which is discussed in Sec.~\ref{sec:Floquet_pre}.

\subsection{Thermalization in classical systems}
\label{sec:classical}

Let us discuss thermalization in a classical system and compare it with thermalization in a quantum system.
We show below that a classical system that satisfies the ergodic hypothesis thermalizes macroscopically.
On the other hand, a classical system does not thermalize microscopically in the most strict sense. 
However, if the classical dynamics satisfies the property of \textit{mixing}, which is a stronger statement than the ergodic hypothesis, local quantities reach their equilibrium values by considering either the temporal average over $\tau$ much longer than the equilibrium correlation times or the average over the initial state $\Gamma_0$ according to a distribution function $p_{\mathrm{ini}}(\Gamma_0)$ that is absolutely continuous with respect to the Lebesgue measure.
See, for example, Refs.~\cite{Lebowitz1973,Kubo_text1,Balescu_text} for the definition of the mixing.

We shall explain the above statement in detail. 
The time evolution is represented by a trajectory in the phase space $\{\Gamma_t\}_{t>0}$ parametrized by time $t$.
A physical quantity $A(\Gamma)$ at time $t$ is given by $A(\Gamma_t)$.
Its infinite-time average starting from the initial state $\Gamma_0$ is given by
\beq
\bar{A}_{\Gamma_0}:=\lim_{T\rightarrow\infty}\frac{1}{T}\int_0^Tdt\,A(\Gamma_t).
\eeq
The microcanonical average of $A(\Gamma)$ is given by
\beq
\<A\>_{\mathrm{mc}}=\frac{1}{|\Omega_{E,N,\Lambda}|}\int_{\Omega_{E,N,\Lambda}}d\Gamma\,A(\Gamma).
\eeq
If $A(\Gamma_t)$ approaches its equilibrium value and stays there for most of the time $t$, $\bar{A}_{\Gamma_0}\approx\<A\>_{\mathrm{mc}}$ must hold.

The \textit{ergodic hypothesis} states that for \textit{an arbitrary Lebesgue integrable function} $A(\Gamma)$,
\beq
\bar{A}_{\Gamma_0}=\<A\>_{\mathrm{mc}}
\label{eq:ergodic}
\eeq
holds \textit{for almost every initial state $\Gamma_0$} (``almost every'' here means that the set of exceptional initial states have vanishing Lebesgue measure)\footnote
{Sometimes the following argument is made to justify taking the infinite-time average in the ergodic hypothesis: any physical measurement is carried out during a finite time interval that is almost infinite compared with the microscopic time scales.
However, if so, we would not be able to observe any nonequilibrium state.
Therefore, we cannot accept this reasoning.
See also Ref.~\cite{Gallavotti_text}.
It is typicality of thermal equilibrium that allows us the probabilistic consideration in discussing thermal equilibrium, see Sec.~\ref{sec:equilibrium}.
}.
In the following discussions, we omit the subscript $\Gamma_0$ of $\bar{A}_{\Gamma_0}$ for notational simplicity.

The ergodic hypothesis is neither necessary nor sufficient for thermalization, but if we combine it to the thermodynamic typicality given in Eq.~(\ref{eq:therm_typicality1}), we can prove macroscopic thermalization for almost every initial state.
We choose $A(\Gamma)=P_{\nu_{\mathrm{eq}}}(\Gamma)$, where 
\beq
P_{\nu_{\mathrm{eq}}}:=\left\{
\begin{aligned}
&1 \quad\text{if }\Gamma\in\Omega_{\nu_{\mathrm{eq}}}, \\
&0 \quad\text{otherwise}.
\end{aligned}
\right.
\eeq
That is, $P_{\nu_{\mathrm{eq}}}(\Gamma)$ is the ``projection'' to the equilibrium macrostate.
The ergodic hypothesis yields
\beq
\overline{P_{\nu_{\mathrm{eq}}}}=\<P_{\nu_{\mathrm{eq}}}\>_{\mathrm{mc}}
=\frac{|\Omega_{\nu_{\mathrm{eq}}}|}{|\Omega_{E,N,\Lambda}|}
\eeq
for almost every initial state.
The thermodynamic typicality implies $|\Omega_{\nu_{\mathrm{eq}}}|/|\Omega_{E,N,\Lambda}|\approx 1$ and thus $\overline{P_{\nu_{\mathrm{eq}}}}\approx 1$.
Since $P_{\nu_{\mathrm{eq}}}(\Gamma)\leq 1$, this implies $P_{\nu_{\mathrm{eq}}}(\Gamma_t)\approx 1$ for most $t$, which proves macroscopic thermalization.

In contrast to quantum systems, classical systems cannot exhibit microscopic thermalization in the most strict sense, $O(\Gamma_t)\approx\<O\>_{\mathrm{mc}}$ for most times $t$ and all local quantities $O$.
If the ergodic hypothesis is true, we have $\bar{O}=\<O\>_{\mathrm{mc}}$ and $\overline{O^2}=\<O^2\>_{\mathrm{mc}}$.
As a result,
\beq
\overline{O^2}-\bar{O}^2=\<O^2\>_{\mathrm{mc}}-\<O\>_{\mathrm{mc}}^2,
\eeq
that is, the temporal fluctuation of $O$ is identical to the fluctuation of $O$ in the microcanonical ensemble.
For a local quantity, $\<O^2\>_{\mathrm{mc}}-\<O\>_{\mathrm{mc}}^2$ is not so small, and hence, $O(\Gamma_t)$ fluctuates largely forever.

Although any local quantity $O(\Gamma)$ that is not constant over the energy shell does not reach its equilibrium value, its time average over a certain time $\tau$,
\beq
O_{\tau}(\Gamma):=\frac{1}{\tau}\int_0^{\tau}dt\,O(\Gamma_t) \quad \text{with }\Gamma_0=\Gamma,
\eeq
can reach its equilibrium value for a sufficiently large $\tau$.
If we assume the ergodic hypothesis to be true, then we have
\begin{align}
\overline{O_{\tau}^2}-\overline{O_{\tau}}^2&=\<O_{\tau}^2\>_{\mathrm{mc}}-\<O_{\tau}\>_{\mathrm{mc}}^2
\nonumber \\
&=\frac{1}{\tau^2}\int_0^{\tau}dt_1\int_0^{\tau}dt_2\,C_{OO}(t_1-t_2),
\label{eq:classical_temporal}
\end{align}
where $C_{OO}(t)$ is the temporal auto-correlation function of $O$, i.e.,
\beq
C_{OO}(t)=\frac{1}{|\Omega_{E,N,\Lambda}|}\int_{\Omega_{E,N,\Lambda}}d\Gamma_0\,O(\Gamma_{t})O(\Gamma_0)-\<O\>_{\mathrm{mc}}^2.
\eeq
We define the equilibrium correlation time\footnote
{If the system satisfies the property of mixing, we have $\lim_{t\rightarrow\infty}C_{OO}(t)=0$, and thus, the existence of some finite equilibrium correlation time $\tau_O$ is ensured~\cite{Lebowitz1973}.
}
$\tau_O$ as a time satisfying $|C_{OO}(t)|\ll \<O^2\>_{\mathrm{mc}}$ for all $t\gg \tau_O$.
Then, Eq.~(\ref{eq:classical_temporal}) tells us that the temporal fluctuation of $O_{\tau}(\Gamma)$ is very small when $\tau\gg\tau_O$.
In a physically realistic system and an observable, $\tau_O$ will be independent of the system size\footnote
{The decay rate of the correlation function is evaluated from the Ruelle-Pollicot resonance~\cite{Pollicott1985,Pollicott1986,Ruelle1986_PRL,Ruelle1986_JSP}, which is intrinsic in dynamics and independent of $p_{\mathrm{ini}}(\Gamma)$ and $O$.
}, 
and the time average of a local quantity over a physically realistic observation time $\tau\gg\tau_O$ will reach the equilibrium value.
Thus, a classical system can thermalize with respect to a suitable time average of a local quantity.

So far, we have considered a single trajectory starting from a definite initial state $\Gamma_0$.
From now on, we consider a different setup.
We pick up an initial state $\Gamma_0$ according to some distribution function $p_{\mathrm{ini}}(\Gamma_0)$.
Then, the system evolves starting from this initial state, and we observe a quantity $O(\Gamma_t)$ at time $t$.
We repeat this protocol infinitely many times, and calculate the expectation value
\beq
\<O(\Gamma_t)\>_{\mathrm{ini}}:=\int d\Gamma_0\,p_{\mathrm{ini}}(\Gamma_0)O(\Gamma_t).
\eeq
If the system satisfies the property of mixing, it is shown that $\<O(\Gamma_t)\>_{\mathrm{ini}}\approx\<O\>_{\mathrm{mc}}$ for sufficiently large $t$, and, in particular, $\lim_{t\rightarrow\infty}\<O(\Gamma_t)\>_{\mathrm{ini}}=\<O\>_{\mathrm{mc}}$ as long as $p_{\mathrm{ini}}(\Gamma)$ is absolutely continuous with respect to the Lebesgue measure.
Thus, if we consider the distribution of initial states and observe the expectation value of a local quantity at time $t$, it eventually reaches the equilibrium value.

It is remarked that the quantity $\<O(\Gamma_t)\>_{\mathrm{ini}}$ does not show recurrence in a mixing system.
This fact is related to the dynamical instability, i.e., dynamical states that start very close to each other in phase space becomes widely separated with time, so that the recurrence time depends extremely sensitively on the initial condition~\cite{Lebowitz1973}\footnote
{Mathematically, the absence of recurrence is explained by the existence of a continuous spectrum of the classical Liouville operator $L_{\mathrm{cl}}=\{H,\cdot\}_{\mathrm{PB}}$, where $\{\cdot,\cdot\}_{\mathrm{PB}}$ denotes the Poisson bracket.
It is noted that the classical Liouville operator can have a continuous spectrum even in a finite system. This is in contrast to a finite quantum system, in which eigenvalues of the Hamiltonian are always discrete.
}.

\subsection{Thermalization in semiclassical systems}
\label{sec:semiclassical}

To understand how thermalization in quantum systems is connected to thermalization in classical systems, we should study a quantum system in the semiclassical regime, which we call a semiclassical system.
In the semiclassical regime, the Planck constant is effectively small, which is typically realized in a high-energy regime.
It is also known that fully connected spin models in the totally symmetric subspace\footnote
{The totally symmetric subspace is defined as the Hilbert subspace consisting of all the quantum states which are invariant under any permutation of spins.
}
can be regarded as semiclassical systems in which $1/V$ plays the role of the effective Planck constant.
Fully connected spin systems well isolated from the environment are now available in ion-trap experiments~\cite{Porras2004,Kim2009,Britton2012,Islam2013}.

The connection between the classical dynamics and the quantum dynamics can be nicely described by using the truncated Wigner approximation~\cite{Polkovnikov2010}.
In this approximation, we consider the dynamics of the Wigner function $f^W(\Gamma)$, which is a quasi-probability distribution in the classical phase space~\cite{Wigner1932}.
The Wigner function must have some width due to the uncertainty relation, and hence it would be plausible to assume that $f^W(\Gamma)$ is a sufficiently smooth function in phase space.

A remarkable feature of the truncated Wigner approximation is that the time evolution of $f^W(\Gamma)$ is given by the purely classical dynamics, $f_t^W(\Gamma)=f_0^W(\Gamma_{-t})$, where $f_t^W(\Gamma)$ is the Wigner function at time $t$ and $\{\Gamma_t\}_t$ is the classical trajectory evolving in the classical Hamilton equations.
In this way, the problem is reduced to that of the classical dynamics starting from an initial distribution function $p_{\mathrm{ini}}(\Gamma)=f_0^W(\Gamma)$. 
As discussed in Sec.~\ref{sec:classical}, if the classical Hamiltonian has the property of mixing, $\<O(\Gamma_t)\>_{\mathrm{ini}}=\int d\Gamma\,f_0^W(\Gamma)O(\Gamma_t)=\int d\Gamma f_t^W(\Gamma)O(\Gamma)$ relaxes to the microcanonical average $\<O\>_{\mathrm{mc}}$.
In this way, the truncated Wigner approximation predicts that a semiclassical system microscopically thermalizes if the underlying classical dynamics is mixing.

However, the validity of the truncated Wigner approximation is guaranteed only on a finite time scale\footnote
{As discussed in Sec.~\ref{sec:recurrence}, any bounded quantum system exhibits recurrence, but the truncated Wigner approximation often fails to predict recurrence.
This is an indication of the breakdown of the truncated Wigner approximation at long times.
}
which diverges only in the classical limit~\cite{Polkovnikov2010}.
To examine the steady state in a semiclassical system with a small but finite effective Planck constant, we should investigate the property of individual energy eigenstates.
The study on the property of semiclassical energy eigenstates has a long history in the field of quantum chaos~\cite{Percival1973, Berry1977, Heller1984, Peres1984, Tomsovic1993, Ketzmerick2000, Hufnagel2002, Baecker2008}.
In the early work~\cite{Percival1973}, it was argued that each energy eigenstate is classified into the regular or chaotic one, corresponding to the regular and the chaotic dynamics in the classical system.
It was argued by Berry~\cite{Berry1977} that if the classical dynamics is ergodic, the corresponding energy eigenstate will be indistinguishable from the microcanonical ensemble~\cite{Berry1977}.
In modern terminology, this means that a semiclassical system with ergodic classical dynamics will satisfy the MITE-ETH.

In a fully connected spin-1/2 system, the equivalence between the ergodicity of the underlying classical dynamics and the MITE-ETH is indeed understood as a consequence of the Wentzel-Kramers-Brillouin (WKB) approximation~\cite{Sciolla2011}.
On the other hand, in a periodically driven fully connected spin-1/2 system~\cite{Russomanno2015} or a fully connected spin-1 system~\cite{Mori2017_classical}, the WKB method is not applicable, but it is numerically shown that the correspondence between the ergodicity of classical dynamics and the MITE-ETH in the quantum Hamiltonian still holds.

It should be noted that a fully connected spin system in the totally symmetric subspace is not a genuine many-body system because it reduces to a semiclassical system with a few degrees of freedom.
It is a very important and challenging open problem to establish the semiclassical theory for genuinely many-body systems.
See, for example, Ref.~\cite{Castiglione1996} for such an attempt.

\subsection{Time scale of thermalization}
\label{sec:timescale}

So far, we have investigated the conditions under which equilibration or thermalization occurs \textit{after a sufficiently long time}, but there was no estimate on the relevant time scale.
It would be very difficult to analytically estimate the time scale in each concrete system because the time scale can depend on the details of the system such as the initial state and the observable.
The previous theoretical studies tried to find some essential features on the time scale of thermalization by investigating the relaxation from an artificially constructed nonequilibrium subspace~\cite{Goldstein-Hara-Tasaki2013} or investigating a ``typical'' time scale by introducing a random Hamiltonian or a random observable~\cite{Cramer2012,Monnai2014,Goldstein-Hara-Tasaki2015,Reimann2016}.
Recently, Garc\'ia-Pintos et al.~\cite{Garcia-Pintos2017} have obtained upper bounds on equilibration time scales that depend on the initial state and the physical observable.

Here, following Reimann~\cite{Reimann2016}, we calculate $\<\psi(t)|\hat{O}|\psi(t)\>$ which evolves under a random Hamiltonian $\hat{H}$ with a fixed energy spectrum for a fixed initial state $|\psi(0)\>\in\mathcal{H}_{E,\Lambda}$ and a fixed observable $\hat{O}\in\mathcal{B}$.
A random Hamiltonian $\hat{H}$ is generated as follows.
First, we prepare a reference Hamiltonian $\hat{H}_0$ with eigenvalues $E_n$ ($n=1,2,\dots,D$) in $\mathcal{H}_{E,\Lambda}$, where we simply write $D_{E,\Lambda}=D$.
For simplicity, we assume that there is no energy degeneracy.
We then perform a random unitary transformation, which defines a random Hamiltonian $\hat{H}=\hat{U}^{\dagger}\hat{H}_0\hat{U}$, where the unitary operator $\hat{U}$ on $\mathcal{H}_{E,\Lambda}$ is chosen according to the Haar measure.
The initial state is expanded as $|\psi(0)\>=\sum_{n=1}^Dc_n|\phi_n\>$ with $\sum_{n=1}^D|c_n|^2=1$, where $|\phi_n\>$ ($n=1,2,\dots,D$) are the energy eigenstates of $\hat{H}$.
It is noted that $c_n$ depends on the random unitary $\hat{U}$.
We obtain
\beq
\<\psi(t)|\hat{O}|\psi(t)\>=\mathrm{Tr}\,\hat{O}\rho_{\mathrm{D}}+\sum_{n\neq m}c_n^*c_m\<\phi_n|\hat{O}|\phi_m\>e^{i(E_n-E_m)t},
\label{eq:timescale1}
\eeq
where $\rho_{\mathrm{D}}$ is the diagonal ensemble defined in Eq.~(\ref{eq:DE}).
The average over random unitary operators is denoted by $[\cdot]_U$.
Since $\hat{O}$ does not depend on $\hat{U}$, we have
\beq
[\mathrm{Tr}\,\hat{O}\rho_{\mathrm{D}}]_U=\mathrm{Tr}\,\hat{O}\rho_{\mathrm{ave}},
\label{eq:timescale2}
\eeq
where $\rho_{\mathrm{ave}}:=[\rho_{\mathrm{D}}]_U$.
For $n\neq m$, we have
\begin{align}
\left[c_n^*c_m\<\phi_n|\hat{O}|\phi_m\>\right]_U&=\frac{1}{D(D-1)}\sum_{k\neq l}^D\left[c_k^*c_l\<\phi_k|\hat{O}|\phi_l\>\right]_U
\nonumber \\
&=\frac{1}{D(D-1)}\left(\sum_{k,l}^D\left[c_k^*c_l\<\phi_k|\hat{O}|\phi_l\>\right]_U -\sum_k\left[|c_k|^2\<\phi_k|\hat{O}|\phi_k\>\right]_U\right)
\nonumber \\
&=\frac{1}{D(D-1)}\left(\<\psi(0)|\hat{O}|\psi(0)\>-\mathrm{Tr}\,\hat{O}\rho_{\mathrm{ave}}\right).
\label{eq:timescale3}
\end{align}
Substituting Eqs.~(\ref{eq:timescale2}) and (\ref{eq:timescale3}) into Eq.~(\ref{eq:timescale1}), we obtain
\beq
\left[\<\psi(t)|\hat{O}|\psi(t)\>\right]_U=\mathrm{Tr}\,\hat{O}\rho_{\mathrm{ave}}+F(t)\left(\<\psi(0)|\hat{O}|\psi(0)\>-\mathrm{Tr}\,\hat{O}\rho_{\mathrm{ave}}\right),
\eeq
where
\beq
F(t):=\frac{D}{D-1}\left(|\phi(t)|^2-\frac{1}{D}\right)
\eeq
and
\beq
\phi(t):=\frac{1}{D}\sum_{n=1}^De^{iE_nt}.
\eeq

Reimann~\cite{Reimann2016} has proved that $\mathrm{Tr}\,\hat{O}\rho_{\mathrm{ave}}=\<\hat{O}\>_{\mathrm{mc}}+\mathcal{O}(1/D)$ and for $\xi(t):=\<\psi(t)|\hat{O}|\psi(t)\>-[\<\psi(t)|\hat{O}|\psi(t)\>]_U$, $[\xi(t)^2]_U=\mathcal{O}(\Delta_O^2/D)$ for arbitrary $t$, where $\Delta_O$ is the difference between the largest and smallest eigenvalues of $\hat{O}$.
Since $D=e^{\mathcal{O}(V)}\gg 1$, these results imply that $\mathrm{Tr}\,\hat{O}\rho_{\mathrm{ave}}\approx\<\hat{O}\>_{\mathrm{mc}}$ and $\<\psi(t)|\hat{O}|\psi(t)\>\approx[\<\psi(t)|\hat{O}|\psi(t)\>]_U$ for a vast majority of all unitaries $\hat{U}$.
We thus conclude that
\beq
\<\psi(t)|\hat{O}|\psi(t)\>\approx\<\hat{O}\>_{\mathrm{mc}}+F(t)(\<\psi(0)|\hat{O}|\psi(0)\>-\<\hat{O}\>_{\mathrm{mc}})
\label{eq:timescale_Reimann}
\eeq
to good approximation for almost all unitaries $\hat{U}$ and time $t$.
This is the central result obtained by Reimann~\cite{Reimann2016}.

The relaxation under a typical Hamiltonian is determined by the function $\phi(t)$.
For not too long times $t$, it is well approximated as
\beq
\phi(t)\approx\int_{E-\Delta E}^Edx\,\rho(x)e^{ixt},
\eeq
where $\rho(x)$ represents the (smoothened) density of energy levels $E_n$ normalized as $\int_{E-\Delta E}^Edx\,\rho(x)=1$.
Since $\rho(x)\propto e^{S_{\mathrm{mc}}(x)}$ with the microcanonical entropy $S_{\mathrm{mc}}(E)$, we can approximate $\rho(x)$ as $\rho(x)\approx ce^{\beta x}$, where $\beta=\d S_{\mathrm{mc}}(E)/\d E$ and $c$ is a constant determined by $\int_{E-\Delta E}^Edx\,\rho(x)=1$.
We use this to obtain
\beq
F(t)\approx\frac{1-2e^{-\beta\Delta E}\cos(\Delta Et)+e^{-2\beta\Delta E}}{(1-e^{-2\beta\Delta E})[1+(t/\beta)^2]}.
\eeq
For $\Delta E\gg\beta^{-1}$, which is a reasonable assumption in a macroscopic system, we obtain the following simple expression:
\beq
F(t)\approx \frac{1}{1+(t/\beta)^2},
\label{eq:typical_F}
\eeq
which shows the Lorentzian decay characterized by the \textit{Boltzmann time} $\tau_{\mathrm{B}}:=\beta$ (if we recover the Planck constant $\hbar$, $\tau_{\mathrm{B}}=\beta\hbar$, which is also referred to as the \textit{Planckian thermalization time} in Ref.~\cite{Hashimoto2013}).

Goldstein, Hara, and Tasaki~\cite{Goldstein-Hara-Tasaki2015} also found that the typical time scale of macroscopic thermalization is bounded from above by the Boltzmann time by showing that for almost all unitaries $U$ and a fixed nonequilibrium subspace $\mathcal{H}_{\mathrm{neq}}\subset\mathcal{H}_{E,\Lambda}$ with $D_{\mathrm{neq}}:=\dim\mathcal{H}_{\mathrm{neq}}\ll D$,
\beq
\frac{1}{\tau}\int_0^{\tau}dt\,\<\psi(t)|\hat{P}_{\mathrm{neq}}|\psi(t)\>\lesssim\frac{\tau_{\mathrm{B}}}{\tau}
\eeq
for \textit{any} initial state $|\psi(0)\>\in\mathcal{H}_{E,\Lambda}$ and any $\tau$ with $0<\tau\leq\tau_{\mathrm{B}}\min\{(D/D_{\mathrm{neq}})^{1/4},D^{1/6}\}$.
Here, $\hat{P}_{\mathrm{neq}}$ is the projection onto $\mathcal{H}_{\mathrm{neq}}$.
It is noted that this result is not obtained just by putting $\hat{O}=\hat{P}_{\mathrm{neq}}$ in Eq.~(\ref{eq:timescale_Reimann}).

We note that the Boltzmann time is quite short.
At room temperature $T\sim 300 K$, $\tau_{\mathrm{B}}\sim 2.5\times 10^{-12}$ s.
Of course, many macroscopic systems thermalize much slower than the Boltzmann time.
If the initial state has inhomogeneous local densities of globally conserved quantities such as the energy, the number of particles, and so on, the global relaxation proceeds via transport processes.
In such a case, the relaxation time should increase as the system size, but $\tau_{\mathrm{B}}$ does not.
This failure is due to the fact that the random Hamiltonian $\hat{H}=\hat{U}^{\dagger}\hat{H}_0\hat{U}$ is not at all local and it typically contains long-range many-body interactions.
Moreover, an isolated system sometimes exhibits a slow relaxation called the prethermalization as discussed in Sec.~\ref{sec:prethermalization}, and the Boltzmann time does not give a correct estimate.
We should think that a Hamiltonian realized in the real world is highly atypical from the viewpoint of the Haar measure.

However, it was argued by Reimann~\cite{Reimann2016} that the formula~(\ref{eq:timescale_Reimann}) describes the very rapid relaxation towards local equilibrium.
Remarkably, by comparing $\<\psi(t)|\hat{O}|\psi(t)\>$ given by Eqs.~(\ref{eq:timescale_Reimann}) and (\ref{eq:typical_F}) with some experimental and numerical data, he found the very good agreement of the theory with the experimental results of the rapid initial prethermalization of a coherently split Bose gas~\cite{Gring2012} and the rapid relaxation of hot electrons in the pump-probe experiments~\cite{Guidoni2002,Faure2013}, and with some numerical results given in Refs.~\cite{Thon2004,Bartsch2009,Rigol2009_PRA,Hetterich2015}.
It is an important problem to clarify when the above estimation of the relaxation time by Reimann works well and when it does not.
Numerical calculations in a concrete system will help us answer this question.

\subsection{Recurrence phenomena}
\label{sec:recurrence}

As stated above, (microscopic or macroscopic) thermalization occurs if $|\psi(t)\>$ in a quantum system or $\Gamma_t$ in a classical system represents (microscopic or macroscopic) thermal equilibrium \textit{for most} $t$.
The recurrence theorem prohibits replacing ``for most $t$'' by ``all $t\geq\tau$ with a sufficiently large $\tau$''.
Roughly speaking, the recurrence theorem states that a system will return to a state that is arbitrarily close to the initial state after a sufficiently long but finite time.

Originally, the recurrence theorem was discussed for classical systems by Poincar\'e~\cite{Poincare1890}.
The precise statement of the Poincar\'e recurrence theorem is that almost every point in phase space returns arbitrarily close to its starting point after a sufficiently long time as long as the energy shell $\Omega_{E,N,\Lambda}$ in phase space has a finite volume (Lebesgue measure).
This theorem is an important consequence of the Liouville theorem, which states that the phase-space volume is constant during the time evolution.

To prove it, let us consider a set of points $A\subset\Omega_{E,N,\Lambda}$ in the phase space with a finite Lebesgue measure.
Next, let us consider $B_0\subset A$ which is a set of points that leave $A$ after time $\tau$ and never return to $A$.
The Poincar\'e recurrence theorem claims that $|B_0|=0$, which we shall prove below\footnote
{Remember that $|B|$ for $B\subset\Lambda^{dN}\times\mathbb{R}^{dN}$ denotes the phase-space volume, or the Lebesgue measure, of the region $B$.}.
Let $B_n$ denote the set of points that $B_0$ evolves to after a time $n\tau$.
It is noted that $B_n\cap B_m=\emptyset$ for any $n>m$.
If $B_n\cap B_m\neq\emptyset$, then $B_{n-m}\cap B_0\neq\emptyset$, which, however, contradicts the assumption of $B_n\cap A\neq\emptyset$ for all $n\geq 1$.
Therefore, we have $|B_1\cup B_2\cup\dots\cup B_n|=\sum_{k=1}^n|B_k|$.
The Liouville theorem implies $|B_k|=|B_0|$ for all $k$, and hence $|B_1\cup B_2\cup\dots B_n|=n|B_0|$.
If $|B_0|>0$, it exceeds the total phase space volume of the energy shell for a sufficiently large $n$.
Thus, $|B_0|$ must be zero.

It is emphasized that, as mentioned in Sec.~\ref{sec:classical}, if we consider not a single trajectory $\{\Gamma_t\}_{t\geq 0}$ but the time evolution of a smooth distribution function in phase space, a mixing classical system does not exhibit recurrence.

Next, we explain the quantum recurrence theorem~\cite{Bocchieri1957,Percival1961}.
We consider an isolated quantum system with finite volume $V$, in which the energy spectrum is discrete.
The quantum state at time $t$ is generally expressed as $|\psi(t)\>=\sum_nc_ne^{-iE_nt}|\phi_n\>$ with $\hat{H}|\phi_n\>=E_n|\phi_n\>$.
The overlap between $|\psi(0)\>$ and $|\psi(t)\>$ is measured by the fidelity 
\beq
\mathcal{F}(t):=|\<\psi(0)|\psi(t)\>|^2=\left|\sum_n|c_n|^2e^{-iE_nt}\right|^2.
\label{eq:fidelity}
\eeq
It is shown that the fidelity is an almost periodic function, which means that for any $\epsilon>0$, there exists $\tau_{\mathrm{rec}}>0$ such that $\mathcal{F}(\tau_{\mathrm{rec}})\geq 1-\epsilon$.
This is the quantum recurrence theorem~\cite{Bocchieri1957,Percival1961}.
The quantum recurrence is also understood as a corollary of the Poincar\'e recurrence theorem~\cite{Schulman1978} since $|\psi(t)\>=\sum_nc_ne^{-iE_nt}|\phi_n\>$ can be regarded as an ensemble of classical oscillators with frequencies $E_n$.
Even if the number of energy eigenstates with nonzero $c_n$ is infinite, we can approximate $|\psi(t)\>$ as a finite sum $|\psi'(t)\>=\sum_{n=1}^{\mathcal{N}}c_n|\phi_n\>$ with a sufficiently large $\mathcal{N}$.
Thus, the number of oscillators is finite, and hence we can apply the Poincar\'e recurrence theorem.
As a result, the fidelity $\mathcal{F}(t)$ must return arbitrarily close to one after a sufficiently long but finite time.

In classical systems, the recurrence time $\tau_{\mathrm{rec}}$ typically behaves as $\tau_{\mathrm{rec}}=e^{\mathcal{O}(V)}$.
If the system is ergodic, the phase-space volume of the energy shell is exponentially large, $|\Omega_{E,N,\Lambda}|=e^{\mathcal{O}(V)}$, and hence the time required for exploring the entire region of the energy surface will scale as $\tau_{\mathrm{rec}}\sim|\Omega_{E,N,\Lambda}|=e^{\mathcal{O}(V)}$, which gives a rough estimate of the recurrence time.

On the other hand, in a quantum system, the recurrence time typically scales as $\tau_{\mathrm{rec}}=e^{\mathcal{O}(D_{\mathrm{eff}})}$, where $D_{\mathrm{eff}}=\left(\sum_n|c_n|^4\right)^{-1}$ is the effective dimension of the initial state.
Since typically $D_{\mathrm{eff}}=e^{\mathcal{O}(V)}$, $\tau_{\mathrm{rec}}$ in a quantum system is extraordinarily large (doubly exponential in $V$) even compared with the typical recurrence time in a classical system with same volume $V$~\cite{Peres1982}.

The reason why $\tau_{\mathrm{rec}}$ scales as $e^{\mathcal{O}(D_{\mathrm{eff}})}$ is explained by the fact mentioned above, i.e., the quantum state $|\psi(t)\>=\sum_nc_ne^{-iE_nt}|\phi_n\>$ can be regarded as an ensemble of classical harmonic oscillators with frequencies $E_n$~\cite{Schulman1978}.
The effective number of those oscillators is given by $D_{\mathrm{eff}}$.
If the frequencies $\{E_n\}$ are rationally independent, the recurrence time of classical harmonic oscillators is exponentially long in the number of harmonic oscillators, i.e., $\tau_{\mathrm{rec}}=e^{\mathcal{O}(D_{\mathrm{eff}})}$.

\begin{figure}[tb]
\centering
\includegraphics[width=14cm]{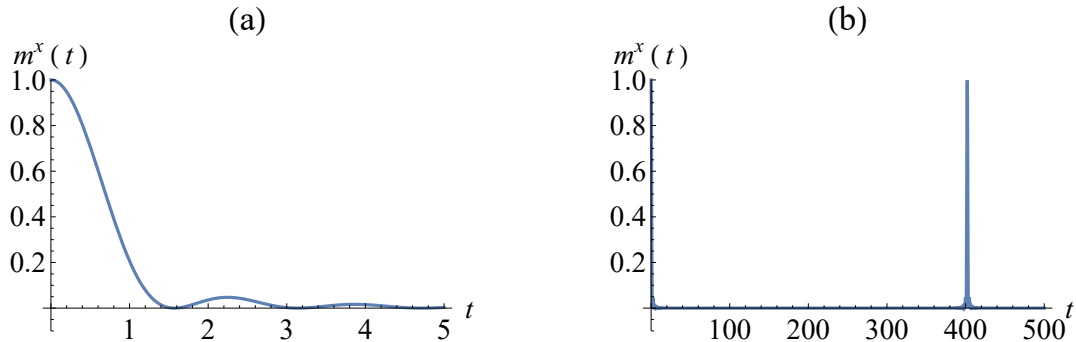}
\caption{Relaxation dynamics of $m^x(t)=\<\psi(t)|\sigma_i^x|\psi(t)\>$ in the Emch-Radin model for a fully polarized initial state along the $x$ direction.
The number of spins $V$ is set to 17.
(a) Short-time dynamics.
The system equilibrates.
(b) Long-time dynamics.
The value of $m^x(t)$ returns to the initial value, which clearly shows a recurrence phenomenon.}
\label{fig:recurrence}
\end{figure}

When the energy eigenvalues are not rationally independent, $\tau_{\mathrm{rec}}$ can be smaller than $e^{\mathcal{O}(D_{\mathrm{eff}})}$.
For example, it is shown that one-dimensional bosons exhibit the recurrence at $\tau_{\mathrm{rec}}\propto V^2$ in the weak-coupling or strong-coupling limit for an arbitrary $V$, and moreover, $\tau_{\mathrm{rec}}$ is polynomial in $V$ in the weak and the strong coupling regimes for relatively small values of $V$ (but $V$ is large enough to exhibit equilibration)~\cite{Kaminishi2015}.
In the Ising chain with exponentially decaying interactions, which was studied by Emch~\cite{Emch1966} and Radin~\cite{Radin1970}, $\tau_{\mathrm{rec}}=e^{\mathcal{O}(V)}$.

These integrable quantum systems can show recurrence phenomena on an experimentally accessible time scale.
As an example, we consider the case of $\tau_{\mathrm{rec}}=e^{\mathcal{O}(V)}$ as in the spin model studied by Emch and Radin.
This dependence is identical to the one typically expected for classical systems, but in quantum systems we can see recurrence \textit{after equilibration} in quite a small system.
For example, $V\sim 10$ spins are enough to see equilibration as mentioned in Sec.~\ref{sec:equilibration}, and $\tau_{\mathrm{rec}}$ is not too large in such a small system. Therefore, we can see both equilibration and recurrence in a small quantum system.

Figure~\ref{fig:recurrence} demonstrates (a) equilibration and (b) recurrence in the Emch-Radin model for $V=17$, whose Hamiltonian is given by $\hat{H}=\sum_{i\neq j}^V2^{-|i-j|}\sigma_i^z\sigma_j^z$.
The initial state is set as a fully-polarized state along the $x$ direction.
We clearly see both equilibration and recurrence.

Finally, it is also remarked that there are recurrence phenomena that are different from the Poincar\'e recurrence in classical systems or the fidelity recurrence ($\mathcal{F}(\tau_{\mathrm{rec}})\geq 1-\epsilon$) in quantum systems.
In classical anharmonic oscillators, the Fermi-Pasta-Ulam recurrence is known~\cite{Fermi1955}, which is interpreted as the recurrence of the energy distribution for long-wavelength harmonic modes.
Although the Poincar\'e recurrence time is exponentially long with respect to $V$, the Fermi-Pasta-Ulam recurrence time is polynomial in the system size.

\section{Prethermalization}
\label{sec:prethermalization}

In the previous section, we discussed the condition of equilibration and thermalization.
In this section, we turn our attention to the relaxation dynamics.
One might imagine that the system relaxes in a simple monotonic way, but actually a many-body quantum system often exhibits nonmonotonic nonequilibrium dynamics.
In particular, when there are well separated time scales, an intermediate stage of thermalization called \textit{prethermalization} emerges; the system first relaxes to a long-lived prethermal state before reaching the true stationary state~\cite{Berges2004}.
In this section, we discuss prethermalization in several systems.

\subsection{General aspects}

\begin{figure}[tb]
\centering
\includegraphics[width=14cm]{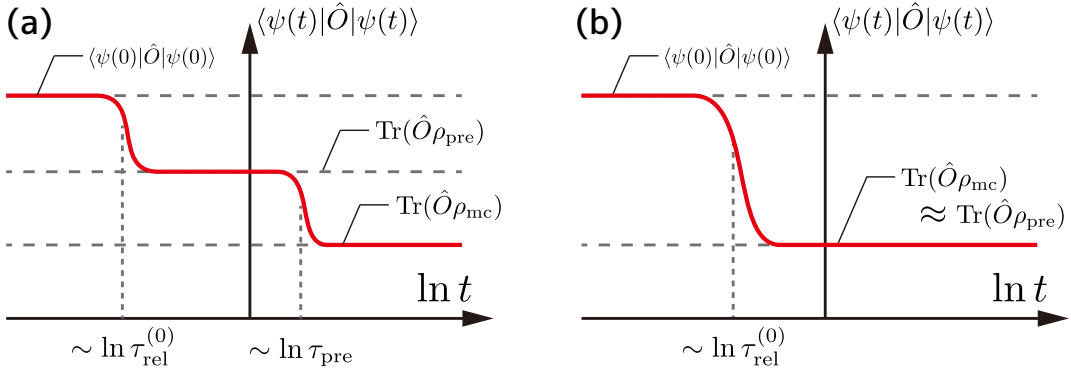}
\caption{Schematic illustration of the time evolution of the expectation value of a local operator $\hat{O}$ through the prethermalization plateau.
(a) The system first relaxes to a prethermalized state and then to a thermal state. (b) The system  directly relaxes to a thermal state if $\rho_\mathrm{mc}$ and $\rho_\mathrm{pre}$ are almost equlvalent locally.}
\label{fig:prethermalization}
\end{figure}

When there are several separated time scales, the relaxation, in general, proceeds in several steps.
When $t$ is much longer than a short time scale but much shorter than a long time scale, fast processes have already occurred but slow processes have yet to begin.
In such a situation, a prethermal state can emerge as an intermediate stationary state.

A typical situation in which a time-scale separation occurs is that the Hamiltonian is written by $\hat{H}=\hat{H}_0+\lambda \hat{V}$, where $\lambda$ is a dimensionless small parameter which controls the separation of time scales.
On a short time scale, $\hat{H}$ will be well approximated by $\hat{H}_0$.
If the relaxation time under $\hat{H}_0$, which is denoted by $\tau_{\mathrm{rel}}^{(0)}$, is not too long, the system equilibrates under $\hat{H}_0$ and the system reaches a prethermal state.
This prethermal state lasts up to a certain time $\tau_{\mathrm{pre}}$ depending on $\lambda$.
As we will discuss in Sec.~\ref{sec:Pauli}, this time scale is in many cases evaluated by the Pauli master equation and given by $\tau_{\mathrm{pre}}=\mathcal{O}(\lambda^{-2})$.
We denote by $\rho_{\mathrm{pre}}$ the density matrix which is locally equivalent to the prethermal state, i.e., $\<\psi(t)|\hat{O}|\psi(t)\>\approx\mathrm{Tr}\,\hat{O}\rho_{\mathrm{pre}}$ for any $\hat{O}\in\mathcal{S}_{\mathrm{loc}}^{(\ell)}$ with some $\ell$ and for most $t$ satisfying $\tau_{\mathrm{rel}}^{(0)}\ll t\ll \tau_{\mathrm{pre}}$ (it is noted that $\rho_{\mathrm{pre}}$ is not unique).
When $t$ is larger than $\tau_{\mathrm{pre}}$, the second relaxation occurs due to the small perturbation $\lambda \hat{V}$, and after a sufficiently long time, the system will relax to the true stationary state.

If the system eventually thermalizes, which we assume in this section, this stationary state is nothing but the equilibrium state locally equivalent to $\rho_{\mathrm{mc}}$.
If $\rho_{\mathrm{pre}}$ is locally distinct from $\rho_{\mathrm{mc}}$, i.e., $\mathrm{Tr}\,\,\hat{O}\rho_{\mathrm{pre}}\neq\mathrm{Tr}\,\hat{O}\rho_{\mathrm{mc}}$ for some $\hat{O}\in\mathcal{S}_{\mathrm{loc}}^{(\ell)}$, the system exhibits prethermalization as schematically illustrated in Fig.~\ref{fig:prethermalization} (a).
On the other hand, if $\rho_{\mathrm{pre}}$ and $\rho_{\mathrm{mc}}$ are almost equivalent locally, there is no prethermalization as shown in Fig.~\ref{fig:prethermalization} (b).

If thermalization occurs under $\hat{H}_0$, $\rho_{\mathrm{pre}}=\rho_{\mathrm{mc}}^{(0)}$, where $\rho_{\mathrm{mc}}^{(0)}$ is the microcanonical ensemble in $\hat{H}_0$.
When $\lambda$ is sufficiently small, $\rho_{\mathrm{mc}}^{(0)}$ will be very close to $\rho_{\mathrm{mc}}$.
In this case, prethermalization is not likely to occur.

Prethermalization occurs only when thermalization does \textit{not} occur under $\hat{H}_0$, i.e., $\hat{H}_0$ does not satisfy the ETH.
In this case, by assuming that there is no degeneracy in $\hat{H}_0$, the diagonal ensemble in the basis of $\hat{H}_0$,
\beq
\rho_{\mathrm{D}}^{(0)}:=\overline{e^{-i\hat{H}_0t}|\psi(0)\>\<\psi(0)|e^{i\hat{H}_0t}}=\sum_n|c_n^{(0)}|^2|\phi_n^{(0)}\>\<\phi_n^{(0)}|
\eeq
is a possible choice of $\rho_{\mathrm{pre}}$, where $|\phi_n^{(0)}\>$ is an eigenstate of $\hat{H}_0$ and the initial state is given by $|\psi(0)\>=\sum_nc_n|\phi_n^{(0)}\>$.
To construct the diagonal ensemble in the basis of $\hat{H}_0$, exponentially many parameters $\{|c_n^{(0)}|^2\}$ are necessary.

However, it is known that a prethermal state can often be expressed by a density matrix which depends on a fewer number of parameters specifying the initial state.
The absence of thermalization in $\hat{H}_0$ indicates that there are some conserved quantities $\{\hat{Q}_k\}$, each of which is expressed by the sum of local operators (or quasi-local operators, see the discussion below).
The number of such conserved quantities is usually much smaller than the dimension of the energy shell.
Then, we can construct the following \textit{generalized Gibbs ensemble} (GGE)~\cite{Rigol2007}
\beq
\rho_{\mathrm{GGE}}=\frac{e^{-\sum_k\lambda_k\hat{Q}_k}}{\mathrm{Tr}\,e^{-\sum_k\lambda_k\hat{Q}_k}},
\eeq
where the parameters $\{\lambda_k\}$ are determined so as to satisfy
\beq
\mathrm{Tr}\,\hat{Q}_k\rho_{\mathrm{GGE}}=\<\psi(0)|\hat{Q}_k|\psi(0)\>.
\eeq
The GGE can be regarded as a generalization of the usual Gibbs state $e^{-\beta \hat{H}}/\mathrm{Tr}\,e^{-\beta \hat{H}}$ for systems with multiple conserved quantities.
It is known that, at least for $\hat{H}_0$ being an integrable Hamiltonian mappable to free fermions or bosons, a prethermal state is well described by the GGE, $\rho_{\mathrm{pre}}=\rho_{\mathrm{GGE}}$, which is discussed in Sec.~\ref{sec:GGE_validity}.

A time-scale separation can also happen even if $\hat{H}$ cannot be written as $\hat{H}=\hat{H}_0+\lambda \hat{V}$ with a sufficiently small $\lambda$.
Different time scales may emerge if there exists an almost invariant nonequilibrium subspace $\mathcal{H}_0$ of the energy shell $\mathcal{H}_{E,\Lambda}$.
For a special initial state $|\psi(0)\>$ such that $|\psi(t)\>=e^{-i\hat{H}t}|\psi(0)\>$ enters the subspace $\mathcal{H}_0$ during the time evolution, the system may experience a slow dynamics, which causes the separation of time scales and prethermalization.

If there are some physical quantities that are \textit{almost conserved within the subspace $\mathcal{H}_0$}, a prethermal state will be described by the GGE associated with those nearly conserved quantities.
For example, even if the Hamiltonian is far from integrable, its low-energy excitations may be expressed by quasi-particles of the Hamiltonian $\hat{H}_0$ with weak interactions $\lambda\hat{V}$ among them.
In this case, $\mathcal{H}_0$ consists of eigenstates of the Hamiltonian $\hat{H}_0$ in the low-energy regime, in which the dynamics of the system is well approximated by that of noninteracting quasi-particles up to a certain time scale $\tau_{\mathrm{pre}}$.
Another example is discussed in Sec.~\ref{sec:KCM}.

As an interesting extreme case, if the nonequilibrium dynamics becomes \textit{self-similar} in $\mathcal{H}_0$, the time evolution in the self-similar regime is described in terms of scaling exponents and scaling functions associated with a \textit{nonthermal fixed point}~\cite{Berges2008}.
In this case, prethermalization occurs due to the critical slowing down, which is similar to critical phenomena in thermal equilibrium.

\subsection{Nearly integrable systems}
\label{sec:integrable}

In this section, we consider the Hamiltonian $\hat{H}=\hat{H}_0+\lambda \hat{V}$, where $\hat{H}_0$ is the Hamiltonian of an integrable model and $\lambda \hat{V}$ is a small integrability-breaking perturbation.
We assume that $\hat{H}$ satisfies the MITE-ETH, and thus the system eventually thermalizes microscopically.

A prethermal state is solely determined by the dynamics under the integrable Hamiltonian $\hat{H}_0$, and hence in this section we focus on the dynamics under $\hat{H}_0$.
A small perturbation $\lambda \hat{V}$ determines the time scale $\tau_{\mathrm{pre}}$ of the full thermalization.
In many cases, the decay of a prethermal state can be discussed by using the Pauli master equation, which is discussed in Sec.~\ref{sec:Pauli}.

It is noted that there is no general consensus about the definition of quantum integrability, and it is still a subject under debate~\cite{Caux2011}.
In this review, we consider integrable models which are mappable either to noninteracting models or to Bethe-ansatz solvable models~\cite{Bethe1931}.
We call the former \textit{noninteracting integrable systems} and the latter \textit{interacting integrable systems}.
In one-dimensional spin chains, the former includes the transverse-field Ising model and the quantum XY model, and the latter includes the Heisenberg model and the XXZ model.

\subsubsection{Quench dynamics of integrable systems}
\label{sec:quench_integrable}

Nonequilibrium dynamics of integrable systems has been extensively studied~\cite{Cazalilla2006, Calabrese2006, Calabrese2007, Rigol2007, Cramer2008, Gangardt2008, Eckstein2009, Iucci2009, Iucci2010, Rossini2010, Biroli2010, Fioretto2010, Calabrese2011, Rigol2011, Sato2012, Chung2012, Wright2014, Cardy2016} (for reviews, see Refs.~\cite{Calabrese_review2016,Essler-Fagotti_review2016}).
Many works consider in the setup of \textit{quantum quench}.
In a quantum quench, we prepare the ground state (or a finite-temperature equilibrium state) of a pre-quench Hamiltonian $\hat{H}_{\mathrm{ini}}$, and at a certain time, say $t=0$, suddenly quench some parameters of the Hamiltonian.
The system evolves under the post-quench Hamiltonian $\hat{H}_{\mathrm{fin}}$ for $t>0$.
We set $\hat{H}_{\mathrm{fin}}=\hat{H}_0$, namely the Hamiltonian of an integrable model.

It should be pointed out that the results in Sec.~\ref{sec:equilibration}, especially Eq.~(\ref{eq:Short-Farrelly}), does not ensure equilibration after a quench in a noninteracting integrable system because there are too many degeneracies in energy gaps.
As an example, let us consider noninteracting spinless fermions in one-dimensional chain,
\beq
\hat{H}_0=-J\sum_{x=1}^V\left(\hat{c}_x^{\dagger}\hat{c}_{x+1}+\hat{c}_{x+1}^{\dagger}\hat{c}_x\right) -\mu\sum_{x=1}^V\hat{c}_x^{\dagger}\hat{c}_x,
\label{eq:spinless}
\eeq
where $\hat{c}_x$ and $\hat{c}_x^{\dagger}$ are the annihilation and creation operators of a fermion at site $x$, respectively.
For simplicity, the number of sites $V$ is even.
This Hamiltonian is diagonalized via the Fourier transform
\beq
\hat{H}_0=\sum_k\varepsilon_k\tilde{c}_k^{\dagger}\tilde{c}_k,
\label{eq:spinless_diag}
\eeq
with the dispersion relation $\varepsilon_k=-2J\cos(k)-\mu$, where
\beq
\tilde{c}_k:=\frac{1}{\sqrt{V}}\sum_{x=1}^V\hat{c}_xe^{ikx}
\eeq
for wave numbers $k=2\pi n/V$ with integer $n=-V/2,-V/2+1,\dots,V/2-1$.
Obviously, the time evolution of an operator $\hat{O}=\tilde{c}_k\tilde{c}_{-k}$ with $k\neq 0$ is given by $\<\psi(t)|\hat{O}|\psi(t)\>=\<\psi(0)|\hat{O}|\psi(0)\>e^{-2i\varepsilon_kt}$, which does not reach a stationary value.

This observation shows that not all few-body operators $\hat{O}\in\mathcal{S}_{\mathrm{few}}^{(k)}$ equilibrate in noninteracting integrable systems.
However, if we consider local operators $\hat{O}\in\mathcal{S}_{\mathrm{loc}}^{(\ell)}$, we expect that equilibration occurs ($\tilde{c}_k\tilde{c}_{-k}\in\mathcal{S}_{\mathrm{few}}^{(k)}$ but $\notin\mathcal{S}_{\mathrm{loc}}^{(\ell)}$).

Let us consider $\hat{O}=\hat{c}_{x_1}\hat{c}_{x_2}$.
Its expectation value at time $t$ is given by
\beq
\<\psi(t)|\hat{O}|\psi(t)\>=\frac{1}{V}\sum_{k_1,k_2}e^{-i(k_1x_1+k_2x_2)}e^{-i(\varepsilon_{k_1}+\varepsilon_{k_2})t}\<\psi(0)|\tilde{c}_{k_1}\tilde{c}_{k_2}|\psi(0)\>.
\eeq
We assume that the initial state satisfies the translation invariance.
Then only the terms with $k_2=-k_1$ survive.
In the thermodynamic limit, the discrete sum over $k_1$ is replaced by an integral $(V/2\pi)\int_{-\pi}^{\pi}dk_1$, and as a result we obtain
\begin{align}
\<\psi(t)|\hat{O}|\psi(t)\>&\approx\int_{-\pi}^{\pi}\frac{dk}{2\pi}e^{-ik(x_1-x_2)}\<\psi(0)|\tilde{c}_k\tilde{c}_{-k}|\psi(0)\>e^{-2i\varepsilon_kt}
\nonumber \\
&=:\int_{-\pi}^{\pi}\frac{dk}{2\pi}g(k)e^{-2i\varepsilon_kt}.
\end{align}
We assume that $g(k)$ is a smooth function of $k$.
Then, in the limit of $t\rightarrow\infty$, this integral is evaluated by the stationary-phase approximation.
By using $\varepsilon_k=-2J\cos(k)-\mu$ and assuming $g(k)\sim k^{\alpha}$ with $\alpha\geq 0$ for small $|k|$, we obtain
\beq
\<\psi(t)|O|\psi(t)\>\sim t^{-(1+\alpha)/2}
\eeq
for large $t$.
Thus, $\<\psi(t)|O|\psi(t)\>$ shows a power-law decay towards a stationary value.

The power-law decay of local observables is a generic feature of noninteracting integrable systems~\cite{Fioretto2010, Sotiriadis2014}, while the exponential decay is generally expected in the case that a post-quench Hamiltonian is massless (e.g. at a quantum critical point)~\cite{Calabrese2006, Calabrese2007}.
It is also known that in the transverse-field Ising model, the order parameter shows an exponential decay for a quench from the ferromagnetic phase~\cite{Calabrese2011}.
In this way, equilibration generally occurs for local operators $\mathcal{S}_{\mathrm{loc}}^{(\ell)}$ even in noninteracting integrable systems.

It is noted that the relaxation dynamics in interacting integrable systems was studied by Sato et al.~\cite{Sato2012}.
In this work, the nonequilibrium dynamics of the Lieb-Liniger model, which is Bethe-ansatz solvable, is studied by using the Slavnov formula~\cite{Slavnov1989} and the Gaudin-Korepin formula~\cite{Gaudin1983,Korepin1982}.

Numerical studies have demonstrated that integrable systems equilibrate but often fail to thermalize after a quench.
The stationary state is described by the GGE, which is locally different from the microcanonical ensemble.
As discussed in Sec.~\ref{sec:weak_ETH}, even an integrable system satisfies the weak ETH, and almost every energy eigenstate in an energy shell represents MITE.
The absence of thermalization is attributed to rare eigenstates which do not represent MITE~\cite{Biroli2010}.
In other words, an initial state $|\psi(0)\>=\sum_nc_n|\phi_n\>$ has a non-negligible overlap with rare nonthermal eigenstates, i.e., $w_{\mathrm{rare}}:=\sum_{n:\text{ nonthermal }\phi_n}|c_n|^2$ is not negligible even in the thermodynamic limit.
It is noted that if an initial state is randomly chosen from the energy shell, $w_{\mathrm{rare}}$ is exponentially small with respect to $V$.
We are thus led to conclude that a quantum quench does importance sampling of exponentially rare eigenstates.

\subsubsection{GGE in noninteracting integrable systems}
\label{sec:GGE_noninteracting}

In this section, we present an example of the GGE in a translation-invariant noninteracting integrable Hamiltonian $\hat{H}_0$.
As an example, we again consider spinless free fermions given by Eq.~(\ref{eq:spinless}) or Eq.~(\ref{eq:spinless_diag}).

Obviously, the mode occupation numbers $\tilde{n}_k=\tilde{c}_k^{\dagger}\tilde{c}_k$ are conserved quantities, but not spatially local.
Instead, we can construct extensive local conserved quantities $\{\hat{Q}_n^{\pm}\}$ as
\beq
\left\{
\begin{aligned}
&\hat{Q}_n^+=2\sum_k\cos(nk)\tilde{c}_k^{\dagger}\tilde{c}_k =\sum_{x=1}^V\left(\hat{c}_x^{\dagger}\hat{c}_{x+n}+\hat{c}_{x+n}^{\dagger}\hat{c}_x\right), \\
&\hat{Q}_n^-=2\sum_k\sin(nk)\tilde{c}_k^{\dagger}\tilde{c}_k =i\sum_{x=1}^V\left(\hat{c}_x^{\dagger}\hat{c}_{x+n}-\hat{c}_{x+n}^{\dagger}\hat{c}_x\right).
\end{aligned}
\right.
\eeq
The GGE is constructed by these extensive local conserved quantities as
\beq
\rho_{\mathrm{GGE}}=\frac{e^{-\sum_n\left(\lambda_n^+\hat{Q}_n^++\lambda_n^-\hat{Q}_n^-\right)}}{\mathrm{Tr}\,e^{-\sum_n\left(\lambda_n^+\hat{Q}_n^++\lambda_n^-\hat{Q}_n^-\right)}}.
\label{eq:GGE_extensive}
\eeq
The parameters $\{\lambda_n^{\pm}\}$ are determined from the initial state $|\psi(0)\>$ so that 
\beq
\<\psi(0)|\hat{Q}_n^{\pm}|\psi(0)\>=\mathrm{Tr}\,\hat{Q}_n^{\pm}\rho_{\mathrm{GGE}}.
\eeq

Since extensive local conserved quantities $\{\hat{Q}_n^{\pm}\}$ are written by linear combinations of the mode occupation number operators $\{\tilde{n}_k\}$, the GGE is also expressed in the following form:
\beq
\rho_{\mathrm{GGE}}=\frac{e^{-\sum_k\mu_k\tilde{n}_k}}{\mathrm{Tr}\,e^{-\sum_k\mu_k\tilde{n}_k}},
\label{eq:GGE_mode}
\eeq
where the new parameters $\mu_k$ are determined by the condition 
\beq
\<\psi(0)|\tilde{n}_k|\psi(0)\>=\mathrm{Tr}\,\tilde{n}_k\rho_{\mathrm{GGE}}.
\label{eq:GGE_condition_mode}
\eeq
In this way, the GGE may be constructed from a set of mode occupation number operators $\{\tilde{n}_k\}$.

As discussed in Refs.~\cite{Fagotti2014,Essler-Fagotti_review2016}, in some special cases in which there are some symmetries in the dispersion relation $\varepsilon_k$, there are additional conserved quantities which do \textit{not} commute with the mode occupation numbers $\{\tilde{n}_k\}$.
In the above example, when $\mu=0$, we have the symmetry of $\varepsilon_k=-\varepsilon_{\pi-k}$.
In this case, the following operators commute with $\hat{H}_0$ and are therefore conserved:
\beq
\hat{Q}_n'=\sum_{x=1}^V(-1)^x\left(\hat{c}_x\hat{c}_{x+n} +\hat{c}_{x+n}^{\dagger}\hat{c}_x^{\dagger}\right)
=\sum_k\left(e^{-nik}\tilde{c}_{\pi-k}\tilde{c}_k+e^{nik}\tilde{c}_k^{\dagger}\tilde{c}_{\pi-k}^{\dagger}\right).
\label{eq:additional_conserved}
\eeq
These operators do not commute with $\tilde{n}_k$ and cannot be expressed as linear combinations of $\{\tilde{n}_k\}$\footnote
{We have a set of conserved quantities $\{\hat{Q}_n\}$ which are not mutually commutable, i.e., $[\hat{Q}_n,\hat{Q}_m]=\sum_lf_{nml}\hat{Q}_l$ with the structure constants $f_{nml}$ being not trivially equal to zero.
Such a set of conserved quantities is referred to as \textit{non-Abelian}.
}.

Since conserved quantities given by Eq.~(\ref{eq:additional_conserved}) are not translation invariant (and not extensive local operators in our terminology of Sec.~\ref{sec:preliminary}), these conserved quantities are relevant when the initial state does not have translation invariance.
An ultracold gas in a harmonic trap potential is such an example.
As a result of the presence of conserved quantities without translation invariance, the stationary state remains inhomogeneous starting from an inhomogeneous initial state.
In this case, the GGE constructed from the mode occupation numbers, Eq.~(\ref{eq:GGE_mode}) or equivalently Eq.~(\ref{eq:GGE_extensive}), cannot be used~\cite{Lancaster2010}.
Instead, we should construct the GGE by taking into account these additional conserved quantities $\{\hat{Q}_n'\}$~\cite{Fagotti2014}.

\subsubsection{Validity of GGE in noninteracting integrable systems}
\label{sec:GGE_validity}

The condition of the applicability of the GGE in noninteracting integrable systems (the pre-quench Hamiltonian may be an interacting model) was derived by Sotiriadis and Calabrese~\cite{Sotiriadis2014}.
They showed that the stationary state after equilibration in a translation-invariant noninteracting integrable system is locally equivalent to the GGE if the initial state satisfies the \textit{cluster decomposition property}.
The quantum state $|\psi\>$ is said to have the cluster decomposition property if for an arbitrary finite $\ell$,
\beq
\lim_{|\bm{r}_i-\bm{r}_j|\rightarrow\infty}\sup_{\substack{\hat{O}_i\in\mathcal{S}_i^{(\ell)},\hat{O}_j\in\mathcal{S}_j^{(\ell)}\\ \|\hat{O}_i\|=\|\hat{O}_j\|=1}}\lim_{V\rightarrow\infty} \left|\<\psi|\hat{O}_i\hat{O}_j|\psi\>-\<\psi|\hat{O}_i|\psi\>\<\psi|\hat{O}_j|\psi\>\right|=0.
\label{eq:cluster}
\eeq
Here, the thermodynamic limit is simply denoted by $\lim_{V\rightarrow\infty}$.
Roughly speaking, Eq.~(\ref{eq:cluster}) tells us that for any local operators $\hat{O}_i$ and $\hat{O}_j$ acting nontrivially on sites around $i$ and $j$, respectively, $\<\psi|\hat{O}_i\hat{O}_j|\psi\>$ is factorized as $\<\psi|\hat{O}_i|\psi\>\<\psi|\hat{O}_j|\psi\>$ when the sites $i$ and $j$ are far from each other.
This is quite a natural and physically motivated condition on an initial state prepared by a quantum quench.
It is expected that the cluster decomposition property holds for the ground state or an equilibrium state at a finite temperature of a general short-range interacting Hamiltonian.
Therefore, the result by Sotiriadis and Calabrese~\cite{Sotiriadis2014} ensures that in a quench from a generic local Hamiltonian to a noninteracting integrable Hamiltonian, the stationary state is described by the GGE as long as the system equilibrates.
How the initial state with the cluster decomposition property equilibrates to the GGE was studied in a rigorous manner by Gluza et al.~\cite{Gluza2016}, where it is argued that equilibration occurs through the ``Gaussification''.

The reason why the GGE describes the stationary state after a quench is due to the applicability of Wick's theorem in the long-time limit which is taken after the thermodynamic limit (see also Ref.~\cite{Cazalilla2012}).
To understand this aspect, let us consider spinless fermions given by Eqs.~(\ref{eq:spinless}) and (\ref{eq:spinless_diag}).

First, we consider the two-point correlation function
\begin{align}
C_t^{(2)}(x,y)&:=\<\psi(t)|\hat{c}_x^{\dagger}\hat{c}_y|\psi(t)\>
\nonumber \\
&=\frac{1}{V}\sum_{k_1,k_2}e^{i(k_1x-k_2y)}e^{i(\varepsilon_{k_1}-\varepsilon_{k_2})t}\<\psi(0)|\tilde{c}_{k_1}^{\dagger}\tilde{c}_{k_2}|\psi(0)\>.
\end{align}
We assume the translation invariance of the initial state, which implies $\<\psi(0)|\tilde{c}_{k_1}^{\dagger}\tilde{c}_{k_2}|\psi(0)\>=\<\psi(0)|\tilde{c}_{k_1}^{\dagger}\tilde{c}_{k_1}|\psi(0)\>\delta_{k_1,k_2}$, and hence
\beq
C_t^{(2)}(x,y)=\frac{1}{V}\sum_ke^{ik(x-y)}\<\psi(0)|\tilde{n}_k|\psi(0)\>,
\eeq
which is independent of $t$.
On the other hand, the two-point correlation function in the GGE is given by
\beq
C_{\mathrm{GGE}}^{(2)}(x,y)=\frac{1}{V}\sum_ke^{ik(x-y)}\mathrm{Tr}\,\tilde{n}_k\rho_{\mathrm{GGE}}.
\eeq
From Eq.~(\ref{eq:GGE_condition_mode}), these two expressions are identical:
\beq
C_t^{(2)}(x,y)=C_{\mathrm{GGE}}^{(2)}(x,y).
\label{eq:2-point_GGE}
\eeq

Next, let us consider the four-point correlation function given by
\begin{align}
&C_t^{(4)}(x_1,x_2,x_3,x_4):=\<\psi(t)|\hat{c}_{x_1}^{\dagger}\hat{c}_{x_2}^{\dagger}\hat{c}_{x_3}\hat{c}_{x_4}|\psi(t)\>
\nonumber \\
&=\frac{1}{V^2}\sum_{k_1,k_2,k_3,k_4}e^{i(k_1x_1+k_2x_2-k_3x_3-k_4x_4)}e^{i(\varepsilon_{k_1}+\varepsilon_{k_2}-\varepsilon_{k_3}-\varepsilon_{k_4})t}\<\psi(0)|\tilde{c}_{k_1}^{\dagger}\tilde{c}_{k_2}^{\dagger}\tilde{c}_{k_3}\tilde{c}_{k_4}|\psi(0)\>.
\end{align}
Because of the translation invariance of the initial state, $k_1+k_2=k_3+k_4+2\pi n$ should be satisfied for some integer $n$.
In the long-time limit, only the terms satisfying $\varepsilon_{k_1}+\varepsilon_{k_2}=\varepsilon_{k_3}+\varepsilon_{k_4}$ survive.
We assume that these two conditions imply $k_1=k_3$ and $k_2=k_4$ or $k_1=k_4$ and $k_2=k_3$ (this assumption can be replaced by a much weaker assumption, but we do not discuss it here).
Then, we obtain
\begin{align}
&\lim_{t\rightarrow\infty}\lim_{V\rightarrow\infty}C_t^{(4)}(x_1,x_2,x_3,x_4)
\nonumber \\
&=\lim_{V\rightarrow\infty}\sum_{k_1,k_2}\left(e^{ik_1(x_1-x_3)+ik_2(x_2-x_4)}+e^{ik_1(x_1-x_4)+ik_2(x_2-x_3)}\right)\<\psi(0)|\tilde{n}_{k_1}\tilde{n}_{k_2}|\psi(0)\>.
\label{eq:4-point1}
\end{align}
By substituting
\beq
\<\psi(0)|\tilde{n}_{k_1}\tilde{n}_{k_2}|\psi(0)\>=\frac{1}{V^2}\sum_{x,x',y,y'}e^{ik_1(x-x')+ik_2(y-y')}\<\psi(0)|\hat{c}_x^{\dagger}\hat{c}_{x'}\hat{c}_y^{\dagger}\hat{c}_{y'}|\psi(0)\>,
\eeq
into Eq.~(\ref{eq:4-point1}), we obtain
\begin{align}
\lim_{t\rightarrow\infty}\lim_{V\rightarrow\infty}C_t^{(4)}(x_1,x_2,x_3,x_4)=\lim_{V\rightarrow\infty}\frac{1}{V^2}\sum_{x,y}\left[\<\psi(0)|\hat{c}_x^{\dagger}\hat{c}_{x+x_1-x_3}\hat{c}_y^{\dagger}\hat{c}_{y+x_2-x_4}|\psi(0)\>
\right. \nonumber \\
\left.
+\<\psi(0)|\hat{c}_x^{\dagger}\hat{c}_{x+x_1-x_4}\hat{c}_y^{\dagger}\hat{c}_{y+x_2-x_3}|\psi(0)\>\right].
\end{align}
In the above sum, the distance between $x$ and $y$ is typically of the order of $\mathcal{O}(V)$, which is very large.
The contribution from $x$ and $y$ with $|x-y|=\mathcal{O}(1)$ is negligible in the thermodynamic limit.
Therefore, we can use the cluster decomposition property of the initial state to obtain
\begin{align}
\lim_{t\rightarrow\infty}\lim_{V\rightarrow\infty}&C_t^{(4)}(x_1,x_2,x_3,x_4)
\nonumber \\
=&\lim_{V\rightarrow\infty}\frac{1}{V^2}\sum_{x,y}\left[\<\psi(0)|\hat{c}_x^{\dagger}\hat{c}_{x+x_1-x_3}|\psi(0)\>\<\psi(0)|\hat{c}_y^{\dagger}\hat{c}_{y+x_2-x_4}|\psi(0)\>
\right. \nonumber \\
&\left.
+\<\psi(0)|\hat{c}_x^{\dagger}\hat{c}_{x+x_1-x_4}|\psi(0)\>\<\psi(0)|\hat{c}_y^{\dagger}\hat{c}_{y+x_2-x_3}|\psi(0)\>\right].
\end{align}
If we take the long-time limit after the thermodynamic limit, the four-point correlation function is factorized into the products of the two-point correlation functions, which is nothing but Wick's theorem.
Since Wick's theorem also holds in the GGE, by using Eq.~(\ref{eq:2-point_GGE}), we can conclude
\beq
\lim_{t\rightarrow\infty}\lim_{V\rightarrow\infty}C_t^{(4)}(x_1,x_2,x_3,x_4)=\lim_{V\rightarrow\infty}C_{\mathrm{GGE}}^{(4)}(x_1,x_2,x_3,x_4),
\eeq
where $C_{\mathrm{GGE}}^{(4)}$ is the four-point correlation function evaluated by the GGE.

Similarly, we can show that all higher-order correlation functions agree with the prediction by the GGE.
Owing to the cluster decomposition property, which justifies the application of Wick's theorem in the long-time limit, the GGE well describes the stationary state.

Finally, we emphasize that the GGE is valid only for local observables.
For example, if we consider an operator $\tilde{n}_{k_1}\tilde{n}_{k_2}$, which is a few-body but nonlocal operator, we generally have 
\beq
\<\psi(t)|\tilde{n}_{k_1}\tilde{n}_{k_2}|\psi(t)\>=\<\psi(0)|\tilde{n}_{k_1}\tilde{n}_{k_2}|\psi(0)\>\neq \mathrm{Tr}\,\tilde{n}_{k_1}\tilde{n}_{k_2}\rho_{\mathrm{GGE}}
\eeq
even in the thermodynamic limit.

\subsubsection{GGE in interacting integrabe systems}

So far, we have discussed noninteracting integrable systems, but the notion of the GGE is applicable to interacting integrable systems, which are not mappable to noninteracting systems but still exactly solvable by means of the Bethe ansatz~\cite{Bethe1931}.
We can construct the GGE by using local conserved quantities $\{\hat{Q}_k\}$ obtained by the Bethe ansatz.

However, in the XXZ chain, it was pointed out that the GGE constructed from a set of local conserved quantities does not adequately describe the stationary state after relaxation~\cite{Pozsgay2014,Wouters2014}.
The discrepancy between the GGE and the true stationary state was resolved by the discovery of a new family of \textit{quasi-local} conserved quantities in the XXZ chain~\cite{Mierzejewski2015,Ilievski2015_quasilocal,Ilievski2015_complete} (see Ref.~\cite{Ilievski_review2016} for a review).
Here, a quasi-local conserved quantity $\hat{Q}$ is an operator written as a (translation-invariant) sum of quasi-local operators $\hat{Q}=\sum_{i=1}^V\hat{q}_i$, where a quasi-local operator $\hat{q}_i$ at site $i$ is an operator written as a convergent sum
\beq
\hat{q}_i=\sum_{\ell=1}^{\infty}f_{\ell}\hat{O}_i^{(\ell)},
\eeq
where $\hat{O}_i^{(\ell)}\in\mathcal{S}_i^{(\ell)}$ with $\|\hat{O}_i^{(\ell)}\|=1$ and $|f_{\ell}|\leq Ce^{-\ell/\xi}$ with constants $C>0$ and $\xi>0$ independent of the system size\footnote
{In Refs.~\cite{Ilievski2015_quasilocal,Ilievski_review2016}, the quasilocality is defined in terms of the Hilbert-Schmidt norm rather than the operator norm.
This difference, however, is not essential.
}.
Since $f_{\ell}$ decays exponentially with $\ell$, we can approximate $\hat{q}_i$ by truncating the sum at $\ell=\ell_0\gg\xi$,
\beq
\hat{q}_i\approx\sum_{\ell=1}^{\ell_0}f_{\ell}\hat{O}_i^{(\ell)}\in\mathcal{S}_{\mathrm{loc}}^{(\ell_0)}.
\eeq
That is, a quasi-local operator is not local but well approximated by a local operator.
It was confirmed that the GGE constructed from all the local and quasi-local conserved quantities reproduces the stationary state in the isotropic Heisenberg chain~\cite{Ilievski2015_complete}.

It should be remarked that the GGE must be defined carefully.
In the thermodynamic limit, the number of local and quasi-local conserved quantities is infinite, and the infinite series $\sum_k\lambda_k\hat{Q}_k$ is ill-defined.
To overcome this difficulty, Pozsgay et al.~\cite{Pozsgay2017} has constructed a truncated GGE, in which we first consider a finite number of local and quasi-local conserved quantities and take a proper infinite truncation limit after the thermodynamic limit.
Ilievski et al.~\cite{Ilievski2017} has proposed a different GGE in terms of quasi-particle densities in the rapidity space which are obtained by the thermodynamic Bethe ansatz~\cite{Yang1969,Takahashi1971,Gaudin1971}.
Quasi-particle occupation number operators are nonlocal, and the GGE proposed by Ilievski et al.~\cite{Ilievski2017} corresponds to the GGE in Eq.~(\ref{eq:GGE_mode}) in terms of the mode occupation numbers in noninteracting integrable systems, where the mode occupation number operators $\tilde{n}_k$ are also nonlocal.
For more general consideration on how to construct the GGE properly, see Ref.~\cite{Doyon2017}.

\subsubsection{Generalized ETH and the method of quench action}

The concept of the MITE-ETH can be extended to integrable systems with many conserved quantities besides energy.
In this section, we assume that these conserved quantities commute with each other.
Instead of considering an energy shell, we introduce a shell of all the local and quasi-local conserved quantities $\{\hat{Q}_k\}$ defined by
\beq
\mathcal{H}_{\{q_k\}}:=\mathrm{Span}\left\{|\psi\>\in\mathcal{H}| \hat{Q}_k|\psi\>=Q_k|\psi\> \text{ with } Q_k/V\in[q_k-\delta q_k,q_k] \text{ for each }k\right\},
\eeq
where $\delta q_k$ is the width of the shell, which is sufficiently larger than $\mathcal{O}(1/V)$ but sufficiently smaller than $\mathcal{O}(1)$.
The \textit{generalized ETH}, which was proposed and numerically verified by Cassidy et al.~\cite{Cassidy2011} (also see Ref.~\cite{Vidmar_review2016}), states that all the energy eigenstates in $\mathcal{H}_{\{q_k\}}$ are locally indistinguishable from each other and locally equivalent to the \textit{generalized microcanonical ensemble}
\beq
\rho_{\mathrm{GMC}}:=\frac{\hat{1}_{\mathcal{H}_{\{q_k\}}}}{\mathrm{dim}\,\mathcal{H}_{\{q_k\}}}.
\eeq

The generalized ETH explains the validity of the GGE in an interacting integrable system after a quench as long as the initial state prepared by a quench belongs to a shell $\mathcal{H}_{\{q_k\}}$.
This condition on the initial state means that the fluctuation of any extensive (quasi-)local conserved quantity is subextensive and therefore negligible, which is ensured when the initial state has the cluster decomposition property.

Moreover, the generalized ETH suggests that even a single representative energy eigenstate in $\mathcal{H}_{\{q_k\}}$ is sufficient to describe the stationary state after a quench.
This idea was developed by Caux and Essler~\cite{Caux2013} and formulated as the \textit{quench-action method}.
The quench-action method offers an efficient way to describe the stationary state from a given simple initial state $|\psi(0)\>$.

The initial state is expanded as $|\psi(0)\>=\sum_nc_n|\phi_n\>$, where $|\phi_n\>$'s are energy eigenstates of an integrable Hamiltonian.
The expectation value of a local operator $\hat{O}$ is given by
\beq
\<\psi(t)|\hat{O}|\psi(t)\>=\sum_{n,m}c_n^*c_me^{i(E_n-E_m)t}\<\phi_n|\hat{O}|\phi_m\>.
\eeq
Assuming that the system equilibrates, we take the infinite-time average to extract the stationary state:
\beq
\overline{\<\psi(t)|\hat{O}|\psi(t)\>}=\sum_n|c_n|^2\<\phi_n|\hat{O}|\phi_n\>,
\eeq
where we assume that energy degeneracies, if exist, do not affect the stationary state.

Now we consider the thermodynamic limit, where an energy eigenstate $|\phi_n\>$ can be characterized by a function $\rho$ and we can formally make the following replacement:
\beq
|\phi_n\>\rightarrow|\rho\>, \qquad \sum_n(\cdots)\rightarrow\int\mathcal{D}\rho\,e^{S_{\rho}}(\cdots),
\eeq
where $S_{\rho}$ is the ``entropy'' of a coarse-grained state $|\rho\>$ in the sense that $e^{S_{\rho}}$ gives the number of energy eigenstates that converge to the state $|\rho\>$ in the thermodynamic limit.
The symbol $\int\mathcal{D}\rho$ denote a functional integration.
In a noninteracting integrable system $\hat{H}_0=\sum_k\varepsilon_k\tilde{c}_k^{\dagger}\tilde{c}_k$, $\rho(k)$ corresponds to the (coarse-grained) distribution of the mode occupation numbers and 
\beq
S_{\rho}=\frac{V}{2\pi}\int_{-\pi}^{\pi}dk\left[(1+\rho(k))\ln(1+\rho(k))-\rho(k)\ln\rho(k)\right].
\eeq
In an interacting integrable system which is exactly solvable by the Bethe ansatz method, $\rho(\lambda)$ is the (coarse-grained) distribution of the rapidities, and $S_{\rho}=S_{\mathrm{YY}}[\rho]$ is the so called Yang-Yang entropy~\cite{Yang1969} in the thermodynamic Bethe ansatz.

We also make the replacement of
\beq
|c_n|^2\rightarrow e^{-\Phi_{\rho}}:=e^{-S_{\rho}}\sum_{n:\phi_n\sim\rho}|c_n|^2,
\eeq
where $\phi_n\sim\rho$ here means that an eigenstate $|\phi_n\>$ converges to $|\rho\>$ in the thermodynamic limit, and hence $e^{-\Phi_{\rho}}$ is the average of $|c_n|^2$ over all $\phi_n\sim\rho$.
After these replacements, we arrive at the expression
\beq
\overline{\<\psi(t)|\hat{O}|\psi(t)\>}\approx\int\mathcal{D}\rho\,e^{S_{\rho}-\Phi_{\rho}}\<\rho|O|\rho\>.
\eeq
Here, we notice that $S_{\rho}=\mathcal{O}(V)$ and $\Phi_{\rho}$ also typically scales as $\mathcal{O}(V)$.
We can therefore evaluate the functional integral by the saddle-point method:
\beq
\overline{\<\psi(t)|\hat{O}|\psi(t)\>}\approx\<\rho_{\mathrm{sp}}|\hat{O}|\rho_{\mathrm{sp}}\>,
\label{eq:sp}
\eeq
where $\rho_{\mathrm{sp}}$ is determined by
\beq
\left.\frac{\delta}{\delta\rho}(S_{\rho}-\Phi_{\rho})\right|_{\rho=\rho_{\mathrm{sp}}}=0.
\eeq

The right-hand side of Eq.~(\ref{eq:sp}) is meaningful only in the thermodynamic limit, but it is desired to express it in terms of energy eigenstates of a finite system because numerical calculations always treat a finite system.
It is trivially done by taking some eigenstate $|\phi_{\mathrm{sp}}\>$ of a finite system that satisfies $\phi_{\mathrm{sp}}\sim\rho_{\mathrm{sp}}$.
Since it is known that
\beq
\<\phi|\hat{O}|\phi\>\approx\<\phi'|\hat{O}|\phi'\>
\eeq
for any $\phi,\phi'\sim\rho$ and any local observable $\hat{O}$, which is equivalent to the generalized ETH, we obtain
\beq
\overline{\<\psi(t)|\hat{O}|\psi(t)\>}\approx\<\phi_{\mathrm{sp}}|\hat{O}|\phi_{\mathrm{sp}}\>.
\eeq
The infinite-time average of a local quantity is calculated by a single energy eigenstate $|\phi_{\mathrm{sp}}\>$.

The quench-action method also provides us with an efficient way to calculate the relaxation dynamics.
See Ref.~\cite{Caux2013} and a recent pedagogical review~\cite{Caux_review2016}.

\subsubsection{Decay of a prethermal state: approach by the Pauli master equation}
\label{sec:Pauli}

Due to the presence of the integrability breaking perturbation $\lambda\hat{V}$, the prethermal state will eventually decay and the system will relax to thermal equilibrium.
This process is described by the Pauli master equation.

In 1928, Pauli~\cite{Pauli1928} discussed the $H$-theorem in quantum mechanics following Boltzmann's idea in an attempt to understand irreversibility starting from the reversible dynamics. 
Let us consider a quantum system with the Hamiltonian $\hat{H}=\hat{H}_0+\lambda\hat{V}$.
We assume that the diagonal part of $\hat{V}$ is absorbed in $\hat{H}_0$ and $\<\phi_n^{(0)}|\hat{V}|\phi_n^{(0)}\>=0$.
The quantum state in the interaction picture is given by $|\psi_I(t)\>=e^{i\hat{H}_0t}|\psi(t)\>$, which evolves according to
\beq
i\frac{d}{dt}|\psi_I(t)\>=\lambda\hat{V}_I(t)|\psi_I(t)\>,
\label{eq:int_picture}
\eeq
where
\beq
\hat{V}_I(t)=e^{i\hat{H}_0t}\hat{V}e^{-i\hat{H}_0t}.
\eeq
Let us consider the density matrix $\rho(t)=|\psi(t)\>\<\psi(t)|$, whose matrix elements in the basis of the eigenstates of $\hat{H}_0$ are given by
\beq
\rho_{nm}(t):=\<\phi_n^{(0)}|\rho(t)|\phi_m^{(0)}\>=e^{i(E_n^{(0)}-E_m^{(0)})t}\<\phi_n^{(0)}|\rho_I(t)|\phi_m^{(0)}\>,
\eeq
where $\rho_I(t)=|\psi_I(t)\>\<\psi_I(t)|$ is the density matrix in the interaction picture.
Here, $e^{i(E_n^{(0)}-E_m^{(0)})t}$ expresses the dynamics under $\hat{H}_0$, and the effect of the perturbation $\lambda\hat{V}$ is included in $\rho_I(t)$.
It is noted that the dynamics of $\rho_I(t)$ is very slow for small $\lambda$ compared with the dynamics under $\hat{H}_0$.
Let us consider the time $\Delta t$ which is much longer than the equilibration time $\tau_{\mathrm{rel}}^{(0)}$ under $\hat{H}_0$ but much shorter than the time scale of the motion of $\rho_I(t)$, which implies $\rho_I(t)\approx\rho_I(t')$ for $|t-t'|\leq\Delta t$.
The separation of time scales due to small $\lambda$ shows the existence of such $\Delta t$.
Under this assumption, within the time interval $[t,t+\Delta t]$, the equation of motion of $\rho(t)$ is approximately given by
\beq
i\frac{d}{dt}\rho(t)=[\hat{H}_0,\rho(t)].
\eeq
Since $\Delta t$ is much longer than the equilibration time under $\hat{H}_0$, the system equilibrates with respect to $\hat{H}_0$ within this time interval, and $\rho(t+\Delta t)$ will be locally equivalent to the diagonal ensemble in the basis of $\hat{H}_0$:
\beq
\rho_{\mathrm{D}}^{(0)}(t):=\sum_n\rho_{nn}(t)|\phi_n^{(0)}\>\<\phi_n^{(0)}|.
\eeq
This indicates that off-diagonal elements of $\rho(t)$ in the basis of $\hat{H}_0$ can be discarded at every time:
\beq
\rho(t)\rightarrow\rho_{\mathrm{D}}(t)
\label{eq:RPA}
\eeq
for all $t$.
This procedure is called the \textit{random-phase assumption} for $\rho(t)$\footnote
{At time $t+\Delta t$, we have $\rho_{nm}(t+\Delta t)\approx e^{i(E_n^{(0)}-E_m^{(0)})\Delta t}\rho_{nm}(t)$.
If $\Delta t\gg\tau_{\mathrm{rel}}^{(0)}$, the phases $\theta_{nm}:=(E_n^{(0)}-E_m^{(0)})\Delta t$ for $n\neq m$ would behave randomly, and we would be able to perform the random-phase average, which is equivalent to discarding the off-diagonal elements of $\rho(t)$.
}.

After the random-phase assumption, we can focus on the dynamics of the diagonal elements $P_n(t):=\rho_{nn}(t)$.
We again consider the time $\Delta t$ given above.
We obtain
\beq
\frac{P_n(t+\Delta t)-P_n(t)}{\Delta t}\approx\frac{1}{\Delta t}\left[\sum_m|U_{nm}(t+\Delta t,t)|^2P_m(t)-P_n(t)\right],
\eeq
where
\beq
U_{nm}(t+\Delta t,t):=\<\phi_n^{(0)}|\mathcal{T}e^{-i\int_t^{t+\Delta t}dt'\,\hat{V}_I(t')}|\phi_m^{(0)}\>.
\eeq
If we are not interested in the fast motion of $P_n(t)$, we can regard $\Delta t$ as infinitesimal and put $dP_n(t)/dt\approx[P_n(t+\Delta t)-P_n(t)]/\Delta t$.
We then obtain
\beq
\frac{dP_n(t)}{dt}=\sum_{m(\neq n)}\left[W_{nm}P_m(t)-W_{mn}P_n(t)\right],
\label{eq:Pauli}
\eeq
where
\beq
W_{nm}:=\frac{1}{\Delta t}|U_{nm}(t+\Delta t,t)|^2
\eeq
is interpreted as the transition rate from $|\phi_m^{(0)}\>$ to $|\phi_n^{(0)}\>$.

The transition rate is obtained by using the first-order perturbation theory~\cite{Dirac_text}.
The result is given by
\beq
W_{nm}=\lambda^2\left|\<\phi_n^{(0)}|\hat{V}|\phi_m^{(0)}\>\right|^2D(E_n^{(0)}-E_m^{(0)};\Delta t),
\label{eq:transition}
\eeq
where
\beq
D(\omega;\Delta t):=\frac{4\sin^2(\omega\Delta t/2)}{\omega^2\Delta t}.
\label{eq:func_D}
\eeq
The condition of $\Delta t\gg\tau_{\mathrm{rel}}^{(0)}$ implies that $\Delta t$ satisfies $(E_n^{(0)}-E_m^{(0)})\Delta t\gg 1$ for typical $n$ and $m$ with non-vanishing $\<\phi_n^{(0)}|\hat{V}|\phi_m^{(0)}\>$.
Therefore, we may take the limit of $\Delta t\rightarrow\infty$ in Eq.~(\ref{eq:func_D}):
\beq
\lim_{\Delta t\rightarrow\infty}D(E_n^{(0)}-E_m^{(0)};\Delta t)=2\pi\delta(E_n^{(0)}-E_m^{(0)})
\eeq
and therefore
\beq
W_{nm}\approx 2\pi\lambda^2\left|\<\phi_n^{(0)}|\hat{V}|\phi_m^{(0)}\>\right|^2\delta(E_n^{(0)}-E_m^{(0)}).
\label{eq:golden}
\eeq
This is nothing but Fermi's golden rule~\cite{Fermi_text}.
This is useful when we consider a semi-continuous energy spectrum, but we should go back to Eqs.~(\ref{eq:transition}) and (\ref{eq:func_D}) whenever we consider a large but finite quantum system.
The result is not sensitive to $\Delta t$ as long as it satisfies the condition mentioned above.

Equation~(\ref{eq:Pauli}) with the transition rate given by Eq.~(\ref{eq:transition}) or Eq.~(\ref{eq:golden}) is known as the Pauli master equation.
Because of the microscopic reversibility, we have $W_{nm}=W_{mn}$.
If the Pauli master equation satisfies the condition of ergodicity, i.e., for any pair of $n$ and $m$ with $E_n^{(0)}\approx E_m^{(0)}$, there exists an integer $k$ such that $(\mathsf{W}^k)_{nm}>0$, where $\mathsf{W}$ is the transition matrix whose matrix elements are given by $W_{nm}$, the Pauli master equation describes an approach to the microcanonical distribution, where all the states $|\phi_n^{(0)}\>$ of approximately the same energy are equiprobable.
Since the transition rate is proportional to $\lambda^2$, the relaxation time is proportional to $\lambda^{-2}$.

The above derivation of Eq.~(\ref{eq:Pauli}) relies on the repeated use of the random-phase assumption between the transitions by $\lambda\hat{V}$.
Pauli~\cite{Pauli1928} argued that this procedure corresponds to the hypothesis of molecular chaos for the derivation of the Boltzmann equation in classical kinetic theory of gases.

Van Hove~\cite{van_Hove1955} has laid a rigorous foundation of the Pauli master equation.
He proved that the Pauli master equation is derived from quantum mechanics in the limit of $\lambda\rightarrow 0$ and $t\rightarrow\infty$ with $\lambda^2t$ held fixed, which is referred to as the \textit{van Hove limit}.
In his derivation, the random-phase assumption is made only for the initial state, which is much more satisfactory than the above derivation which makes the random-phase assumption repeatedly at every time.
The crucial assumption made by van Hove~\cite{van_Hove1955} is the presence of the \textit{diagonal singularity} and the absence of too many energy degeneracies in $\hat{H}_0$. 
See also Ref.~\cite{Loss1986}.

To explain the diagonal singularity, for arbitrary diagonal operators $\hat{A}$ and $\hat{B}$ in the basis of  $\{|\phi_n^{(0)}\>\}$, let us consider the quantity $\sum_m\<\phi_n^{(0)}|\hat{V}\hat{A}\hat{V}\hat{B}|\phi_m^{(0)}\>$ .
We decompose it into the diagonal and the off-diagonal parts:
\begin{align}
\sum_m\<\phi_n^{(0)}|\hat{V}\hat{A}\hat{V}\hat{B}|\phi_m^{(0)}\>
&=\<\phi_n^{(0)}|\hat{V}\hat{A}\hat{V}|\phi_n^{(0)}\>\<\phi_n^{(0)}|\hat{B}|\phi_n^{(0)}\>
\nonumber \\
&\quad +\sum_{m(\neq n)}\<\phi_n^{(0)}|\hat{V}\hat{A}\hat{V}|\phi_m^{(0)}\>\<\phi_m^{(0)}|\hat{B}|\phi_m^{(0)}\>.
\label{eq:diag_sing1}
\end{align}
The diagonal singularity means that the contribution from the diagonal part is of the same order with respect to the volume $V$ as the contribution from the off-diagonal part.
This means that the diagonal part of $\hat{V}\hat{A}\hat{V}$ is singular in the thermodynamic limit\footnote
{The condition of the diagonal singularity is formally expressed as the delta-function singularity of the matrix elements of $\hat{V}\hat{A}\hat{V}$ as
$$\<\alpha|\hat{V}\hat{A}\hat{V}|\alpha'\>=\delta(\alpha-\alpha')W_A(\alpha)+Y_A(\alpha,\alpha'),$$
where $\alpha$ denotes a set of continuous quantum numbers that characterize each eigenstate of $\hat{H}_0$ in the thermodynamic limit, and $|\alpha\>$ is normalized as $\<\alpha|\alpha'\>=\delta(\alpha-\alpha')$.
The function $Y_A(\alpha,\alpha')$ may contain a weaker singularity than $\delta(\alpha-\alpha')$.
}.
Van Hove's achievement is to clarify that the diagonal singularity is responsible for irreversible thermalization.

For an ordinary noninteracting integrable Hamiltonian $\hat{H}_0$ and a local extensive operator $\hat{V}$, the diagonal singularity holds, and it is expected that the Pauli master equation generally describes the late stage (i.e., after prethermalization) thermalization dynamics of a nearly integrable system.
However, there are several exceptional cases, in which the relaxation time is \textit{not} proportional to $\lambda^{-2}$.
Such a case may happen when the initial state is the unique ground state of $\hat{H}_0$.
In this case, the transition rate from the ground state to an excited state is zero because of the  presence of $\delta(E_n^{(0)}-E_m^{(0)})$ in Eq.~(\ref{eq:golden}), and there is no dynamics under the Pauli master equation with Fermi's golden rule.
Moeckel and Kehrein~\cite{Moeckel2008} have investigated the dynamics after the quench from the ground state of the free Fermi gas to a weakly interacting Hubbard model by using the unitary perturbation theory~\cite{Hackl2008}.
On a time scale of $\mathcal{O}(\lambda^{-2})$, there is no transition but each eigenstate is dressed due to the perturbation.
Then, the dressed ground state can evolve under the Pauli master equation.
Since the dressed ground state depends on $\lambda$, the thermalization time is not simply proportional to $\lambda^{-2}$.
It is found that thermalization occurs on a time scale proportional to $\lambda^{-4}$ in Ref.~\cite{Moeckel2008}.
For a similar result on a different model, see Ref.~\cite{Nessi2014}.

Another interesting exceptional case is discussed in Sec.~\ref{sec:exp_decay}, in which the thermalization occurs on a time scale of $e^{\mathcal{O}(1/\lambda)}$.
The Pauli master equation fails there because of the presence of too many degeneracies in $\hat{H}_0$.

\subsection{Floquet prethermalization}
\label{sec:Floquet_pre}

Let us consider a periodically driven system described by the time-periodic Hamiltonian $\hat{H}(t)=\hat{H}(t+T)$.
As discussed in Sec.~\ref{sec:Floquet}, a generic nonintegrable periodically driven system is considered to obey the Floquet ETH, and the system eventually heats up to infinite temperature.
A state locally indistinguishable from the equilibrium state at infinite temperature does not exhibit any interesting phenomenon.

However, in experiments, it is known that interesting properties of matter, which are difficult to be realized in a static system, can be achieved by applying periodic driving, and this fact has triggered extensive studies on Floquet engineering (see Sec.~\ref{sec:Floquet}).
This does not contradict the Floquet ETH if the state we see in experiments is not a true stationary state but a quasi-stationary prethermal state.

Theoretically, it has recently been shown~\cite{Kuwahara2016, Mori2016_rigorous, Abanin2017_effective, Abanin2017_rigorous} that a periodically driven system generally exhibits \textit{Floquet prethermalization}, in which the system first relaxes to a prethermal state before reaching an infinite-temperature state predicted by the Floquet ETH.
The Floquet prethermalization has also been shown by numerical calculations~\cite{Bukov2015,Canovi2016,Weidinger2017,Machado_arXiv2017,Lindner2017}.

In Refs.~\cite{Kuwahara2016,Mori2016_rigorous,Abanin2017_effective,Abanin2017_rigorous}, it has been shown that generic lattice systems exhibit the Floquet prethermalization in the high-frequency regime (small periods $T$) and that the lifetime of a prethermal state is at least longer than $e^{\mathcal{O}(1/T)}$.
This result is deduced from a detailed analysis of the high-frequency expansion of the Floquet Hamiltonian, i.e., the Floquet-Magnus expansion.
By extending those rigorous analyses, Else, Bauer, and Nayak~\cite{Else2016} have proved that the Floquet time crystal, which was suggested for disordered systems exhibiting the MBL~\cite{Else2016,Yao2017}, occurs in a clean system in its prethermal regime.

As discussed in Sec.~\ref{sec:Floquet}, the Floquet ETH implies that the Floquet-Magnus expansion $\hat{H}_{\mathrm{F}}=\sum_{m=0}^{\infty}\hat{\Omega}_mT^m$ is not a convergent series.
The divergence of the Floquet-Magnus expansion is not a merely mathematical problem but has a physical consequence.
If we truncate the expansion at the $n$th order and consider the truncated Floquet Hamiltonian
\beq
\hat{H}_{\mathrm{F}}^{(n)}:=\sum_{m=0}^n\hat{\Omega}_mT^m,
\eeq
$\hat{H}_{\mathrm{F}}^{(n)}$ does not obey the Floquet ETH and cannot capture the heating up to infinite temperature.
Instead, $\hat{H}_{\mathrm{F}}^{(n)}$ is expected to obey the usual MITE-ETH since $\hat{H}_{\mathrm{F}}^{(n)}$ is a local or few-body Hamiltonian when $\hat{H}(t)$ is local or few-body, respectively.
As a result, the quantum dynamics under $\hat{H}_{\mathrm{F}}^{(n)}$ predicts a relaxation to a state that is locally indistinguishable from the microcanonical ensemble with respect to $\hat{H}_{\mathrm{F}}^{(n)}$, not to an infinite-temperature state.

Let us consider a short-range interacting system\footnote
{In fact, the strict locality of the Hamiltonian is not necessary to show the inequality~(\ref{eq:Kuwahara}).
For example, in the pair interaction potential $U(r_{ij})$, where $r_{ij}$ is the distance between sites $i$ and $j$, power-law interactions $U(r_{ij})\propto 1/r_{ij}^{\alpha}$ with $\alpha>2d$ ($d$ is the spatial dimension) are allowed in the proof of Eq.~(\ref{eq:Kuwahara}).
See Ref.~\cite{Kuwahara2016} for more detail.
}, where the Hamiltonian $\hat{H}(t)$ is written as a sum of local operators,
\beq
\hat{H}(t)=\sum_{\gamma}\hat{h}_{\gamma}(t),
\label{eq:H_driving}
\eeq
where $\hat{h}_{\gamma}(t)\in\mathcal{S}_{\mathrm{loc}}^{(\ell)}$ and we do not assume translation invariance.
We define $k$ and $g$ as
\beq
k:=\max_{\gamma}\sup_{t\in[0,T]}\left|\mathrm{Supp}(\hat{h}_{\gamma}(t))\right|
\label{eq:k}
\eeq
and
\beq
g:=\max_{i\in\{1,2,\dots,V\}}\sup_{t\in[0,T]}\sum_{\gamma: \mathrm{Supp}(\hat{h}_{\gamma})\ni i}\left\|\hat{h}_{\gamma}(t)\right\|.
\label{eq:g}
\eeq
Roughly speaking, the Hamiltonian $\hat{H}(t)$ contains up to $k$-site mutual interactions and the absolute value of the energy per site is bounded by $g$.

We can prove that at a stroboscopic time $t$ (i.e., when $t$ is an integer multiple of $T$),
\beq
\left\|e^{i\hat{H}_{\mathrm{F}}t}\hat{O}e^{-i\hat{H}_{\mathrm{F}}t}-e^{i\hat{H}_{\mathrm{F}}^{(n_0)}t}\hat{O}e^{-i\hat{H}_{\mathrm{F}}^{(n_0)}t}\right\|
\leq t\|\hat{O}\|v_{\ell}kge^{-\mathcal{O}(1/(kgT))}
\label{eq:Kuwahara}
\eeq
for any $\hat{O}\in\mathcal{S}_{\mathrm{loc}}^{(\ell)}$ (see Sec.~\ref{sec:preliminary} for the definition of $v_{\ell}$), where $n_0=\mathcal{O}(1/(kgT))$ (see Ref.~\cite{Kuwahara2016} for a more precise statement).
The inequality~(\ref{eq:Kuwahara}) shows that the stroboscopic time evolution of $\hat{O}$ is well approximated by the approximate time evolution under the truncated Floquet Hamiltonian $\hat{H}_{\mathrm{F}}^{(n_0)}$ up to an exponentially long time with respect to the frequency $\omega=2\pi/T$.

Since an effective static Hamiltonian $\hat{H}_{\mathrm{F}}^{(n_0)}$ is local, it is expected that it obeys the usual ETH.
If the relaxation time in the time evolution under $\hat{H}_{\mathrm{F}}^{(n_0)}$ is much shorter than the heating time $\tau_{\mathrm{h}}=e^{\mathcal{O}(1/(kgT))}$, the system first relaxes to thermal equilibrium with respect to $\hat{H}_{\mathrm{F}}^{(n_0)}$.
For $t\gtrsim\tau_{\mathrm{h}}$, the system will absorb the energy from the periodic driving, and finally the system heats up to infinite temperature as predicted by the Floquet ETH.
Thus, the inequality~(\ref{eq:Kuwahara}) tells us that the Floquet prethermalization generally occurs in the high-frequency regime (small $T$) and a prethermal state persists at least for an exponentially long time.
This is also important for experimental studies aiming at Floquet engineering.

If we focus on the dynamics of the effective Hamiltonian defined by a truncation of the Floquet-Magnus expansion, i.e., $\hat{O}=\hat{H}_{\mathrm{F}}^{(n)}$ with $0\leq n\leq n_0$, we can prove that at a stroboscopic time $t$
\beq
\frac{1}{V}\left\|e^{i\hat{H}_{\mathrm{F}}t}\hat{H}_{\mathrm{F}}^{(n)}e^{-i\hat{H}_{\mathrm{F}}t}-\hat{H}_F^{(n)}\right\|
\leq tkg^2e^{-\mathcal{O}(1/(kgT))}+\mathcal{O}(T^{n+1}),
\label{eq:Kuwahara2}
\eeq
where $\mathcal{O}(T^{n+1})$ on the right-hand side does not depend on $t$.
Inequality~(\ref{eq:Kuwahara2}) shows that $\hat{H}_{\mathrm{F}}^{(n)}$ is an approximately conserved quantity\footnote
{All $\hat{H}_{\mathrm{F}}^{(n)}$ with $0\leq n\leq n_0$ are approximately conserved quantities, but they are not independent of each other because all $\hat{H}_F^{(n)}$ with $n\leq n_0$ are very close to each other for small $T$, see Ref.~\cite{Mori2016_rigorous}.
}.
Especially, if $\hat{H}_{\mathrm{F}}^{(0)}=(1/T)\int_0^Tdt\,\hat{H}(t)$ is interpreted as the energy of the system, inequality~(\ref{eq:Kuwahara2}) for $n=0$ tells us that energy absorption, or heating, is exponentially slow for small $T$, which is related to an exponentially long lifetime of a prethermal state.

Interestingly, the exponentially slow energy absorption expressed by inequality~(\ref{eq:Kuwahara2}) can be proved even for long-range interacting systems~\cite{Kuwahara2016,Mori2016_rigorous}.
That is, inequality~(\ref{eq:Kuwahara2}) is true even if $\hat{h}_{\gamma}(t)$ in Eq.~(\ref{eq:H_driving}) is not a local but a few-body operator, $\hat{h}_{\gamma}(t)\in\mathcal{S}_{\mathrm{few}}^{(k)}$, where $k$ can be chosen to be identical to $k$ in Eq.~(\ref{eq:k}).
For example, we can prove inequality~(\ref{eq:Kuwahara2}) for a system of all-to-all interacting spins
\beq
\hat{H}(t)=-\frac{J(t)}{2V}\sum_{i\neq j}^VS_i^zS_j^z-h(t)\sum_{i=1}^VS_i^x
\eeq
with $J(t)=J(t+T)$ and $h(t)=h(t+T)$, where $1/V$ in front of the interaction term is necessary to make $g$ in Eq.~(\ref{eq:g}) remain finite in the thermodynamic limit (this scaling is called the Kac prescription~\cite{Kac1963,Campa_review2009}).

\subsection{Prethermal state with an exponentially long lifetime}
\label{sec:exp_decay}

In Sec.~\ref{sec:Floquet_pre}, some rigorous results for periodically driven systems were presented.
These results are also relevant to some static, time-independent systems as pointed out by Abanin et al.~\cite{Abanin2017_rigorous}.
We consider the Hamiltonian $\hat{H}=\hat{H}_0+\lambda\hat{V}$ with small $\lambda$.
The crucial assumption is that all the eigenvalues of $\hat{H}_0$ are integer multiples of some unit energy $\varepsilon$ which is independent of the system size.
First, we rescale the time $t$ as $s=\lambda t$.
Consequently, the Schr\"odinger equation $id|\psi(t)\>/dt=\hat{H}|\psi(t)\>$ is transformed to $id|\psi'(s)\>/ds=\hat{H}'|\psi'(s)\>$, where $|\psi'(s)\>=|\psi(s/\lambda)\>$ and
\beq
\hat{H}'=\frac{1}{\lambda}\hat{H}_0+\hat{V}.
\label{eq:H_exp}
\eeq

We now consider the problem in the interaction picture, in which the quantum state $|\psi_{\textrm{I}}'(s)\>=e^{\frac{i}{\lambda}\hat{H}_0s}|\psi'(s)\>$ obeys
\beq
i\frac{d}{ds}|\psi_{\mathrm{I}}'(s)\>=\hat{V}_{\mathrm{I}}(s)|\psi_{\mathrm{I}}'(s)\>,
\eeq
where
\beq
\hat{V}_{\mathrm{I}}(s)=e^{\frac{i}{\lambda}\hat{H}_0s}\hat{V}e^{-\frac{i}{\lambda}\hat{H}_0s}.
\eeq
By the assumption that all the eigenvalues of $\hat{H}_0$ are integer multiples of $\varepsilon$, we have $e^{2\pi i\hat{H}_0/\varepsilon}=1$, and hence
\beq
\hat{V}_{\mathrm{I}}(s+T)=\hat{V}_{\mathrm{I}}(s)
\eeq
with
\beq
T=\frac{2\pi\lambda}{\varepsilon},
\eeq
which is small for small $\lambda$.

The above argument shows that this kind of static systems can be transformed to a periodically driven system with a small period $T$.
Then, the results presented in Sec.~\ref{sec:Floquet_pre} are applicable.
By using an explicit expression of $\hat{V}_{\mathrm{I}}(s)$, we can construct the corresponding Floquet-Magnus expansion and truncated Floquet Hamiltonians $\hat{H}_{\mathrm{F}}^{(n)}$.
The results presented in Sec.~\ref{sec:Floquet_pre} tell us that $\hat{H}_{\mathrm{F}}^{(n)}$ is an almost conserved quantity and a prethermal state lasts at least for an exponentially long time with respect to $\omega=2\pi/T=\varepsilon/\lambda$, i.e.,
\beq
\tau_{\mathrm{pre}}\geq e^{\mathcal{O}(\varepsilon/\lambda)}.
\label{eq:exp_long}
\eeq

As discussed in Sec.~\ref{sec:Pauli}, the decay of a prethermal state is often described by the Pauli master equation, which predicts $\tau_{\mathrm{pre}}=\mathcal{O}(\lambda^{-2})$.
Equation~(\ref{eq:exp_long}) is much longer than this time scale, and hence the decay of a prethermal state in this kind of systems is not simply described by the Pauli master equation.
It is noted that the derivation of the Pauli master equation needs the assumption that $\hat{H}_0$ does not have too many energy degeneracies.
However, this assumption is not satisfied here because we assume that all the eigenvalues of a many-body Hamiltonian $\hat{H}_0$ are integer multiples of the unit energy $\varepsilon$.

An important prototypical example is the (Bose or Fermi) Hubbard model.
The Hamiltonian of the Fermi Hubbard model is given by
\beq
\hat{H}=-J\sum_{x=1}^V\sum_{\sigma=\uparrow,\downarrow}\left(\hat{c}_{x,\sigma}^{\dagger}\hat{c}_{x+1,\sigma}+\hat{c}_{x+1,\sigma}^{\dagger}\hat{c}_{x,\sigma}\right)+U\sum_{x=1}^V\hat{n}_{x,\uparrow}\hat{n}_{x,\downarrow},
\label{eq:Hubbard}
\eeq
where $\hat{c}_{x,\sigma}$ and $\hat{c}_{x,\sigma}^{\dagger}$ are the annihilation and creation operators of a fermion of spin $\sigma$ at site $x$, and $\hat{n}_{x,\sigma}=\hat{c}_{x,\sigma}^{\dagger}\hat{c}_{x,\sigma}$.
We consider the \textit{strong}-coupling regime $U\gg J$.
By identifying
\beq
\left\{
\begin{aligned}
&\hat{H}_0=\sum_{x=1}^V\hat{n}_{x,\uparrow}\hat{n}_{x,\downarrow}, \\
&\hat{V}=-J\sum_{x=1}^V\sum_{\sigma=\uparrow,\downarrow}\left(\hat{c}_{x,\sigma}^{\dagger}\hat{c}_{x+1,\sigma}+\hat{c}_{x+1,\sigma}^{\dagger}\hat{c}_{x,\sigma}\right), \\
&\lambda=\frac{1}{U},
\end{aligned}
\right.
\eeq
the Fermi-Hubbard Hamiltonian (\ref{eq:Hubbard}) is written in the form of Eq.~(\ref{eq:H_exp}).

It is obvious that all the eigenvalues of $\hat{H}_0$ are integer, and $\varepsilon=1$.
According to the general discussion leading to Eq.~(\ref{eq:exp_long}), we can conclude that the strong-coupling Hubbard model exhibits prethermalization, and a prethermal state lasts for at least an exponentially long time in $U$, $\tau_{\mathrm{pre}}\geq e^{\mathcal{O}(U/J)}$.
The same result also holds for the Bose-Hubbard model.

Kollath, L\"auchli, and Altman~\cite{Kollath2007} have studied the quench of the Bose-Hubbard model from a small-$U$ superfluid phase to a large-$U$ Mott insulator phase.
They have numerically found that the system approaches a nonequilibrium steady state and that the system does not thermalize on a numerically accessible time scale.
This result is presumably understood as a consequence of an exponentially long relaxation time $\tau_{\mathrm{pre}}\geq e^{\mathcal{O}(U/J)}$.

\subsection{Slow relaxation due to dynamical constraints}
\label{sec:KCM}

Kinetically constrained models (KCMs) are \textit{classical} lattice models for the dynamics of structural glasses.
Let us consider the stochastic dynamics of the configuration $\bm{n}=(n_1,n_2,\dots,n_V)$, where $n_i=0$ or 1 is the state of the $i$th site on a regular lattice.
We say that site $i$ is empty (occupied) when $n_i=0$ ($n_i=1$).
In a KCM, the dynamics is local and obeys a detailed-balance condition with respect to the classical Hamiltonian $H(\bm{n})$.
To be specific, we consider a trivial Hamiltonian $H(\bm{n})=-\mu\sum_{i=1}^Vn_i$, and the dynamics is given by a change of the state of a single site $i$ with probabilities
\beq
\left\{
\begin{aligned}
&p(n_i:0\rightarrow 1)=C_i\frac{1}{e^{-\beta\mu}+1}, \\
&p(n_i:1\rightarrow 0)=C_i\frac{1}{e^{\beta\mu}+1},
\label{eq:KCM}
\end{aligned}
\right.
\eeq
where $\beta$ is the inverse temperature and $C_i=0$ or 1 depending on the configuration $\bm{n}$.
A change of the state at site $i$ is prohibited when $C_i=0$, and hence $C_i$ represents a kinetic constraint.
As a simple model, let us choose $C_i$ as
\beq
C_i=1-\prod_jn_j,
\eeq
where the product runs over all nearest neighbors $j$ of site $i$.
This model is a special case of the Fredrickson-Andersen models~\cite{Fredrickson1984}.
That is, a change of the state at site $i$ is allowed only if at least one of its nearest neighbors is empty.

The dynamics given by Eq.~(\ref{eq:KCM}) satisfies a detailed balance condition with respect to the Hamiltonian $H(\bm{n})=-\mu\sum_{i=1}^Vn_i$.
Since the Hamiltonian does not contain any interaction, equilibrium states are trivial.
However, its \textit{dynamics} becomes anomalously slow due to the kinetic constraint.
Several KCMs have been studied in the context of glasses~\cite{Jackle1991, Kob1993, Jackle1994, Toninelli2006}, and it is recognized that kinetic constraints in classical systems are responsible for glassy slowing down of the dynamics.
Indeed, KCMs display several phenomenologies of glass forming liquids including super-Arrhenius relaxation behavior, dynamical heterogeneities, aging phenomena, and ergodicity-breaking transitions~\cite{Ritort_review2003,Garrahan_review2011}.

An interesting question to be addressed is whether constraints on local motion also give rise to slow relaxation in isolated quantum systems.
Lan et al.~\cite{Lan_arXiv2017} studied a quantum version of a constrained lattice gas on a one-dimensional strip of a triangular lattice.
The Hamiltonian is given by
\beq
\hat{H}=-\frac{1}{2}\sum_{\<i,j\>}\hat{C}_{ij}\left\{\lambda(\hat{\sigma}_i^+\hat{\sigma}_j^-+\hat{\sigma}_i^-\hat{\sigma}_j^+ -(1-\lambda)\left[\hat{n}_i(1-\hat{n}_j)+\hat{n}_j(1-\hat{n}_i)\right]\right\},
\label{eq:QKCM}
\eeq
where each site has two states $|0\>$ (empty) and $|1\>$ (occupied), and $\hat{\sigma}^+=|1\>\<0|$, $\hat{\sigma}^-=|0\>\<1|$, and $\hat{n}=|1\>\<1|$.
The symbol $\<i,j\>$ means that sites $i$ and $j$ constitute a nearest-neighbor pair.
The operator
\beq
\hat{C}_{ij}=1-\prod_k\hat{n}_k
\eeq
represents a \textit{dynamical constraint}, where the product is taken over all common nearest-neighbors $k$ of sites $i$ and $j$.
It is noted that the eigenvalues of $\hat{C}_{ij}$ are either 0 or 1.
Therefore, a change of the states of sites $i$ and $j$ for a quantum state $|\psi\>$ is allowed only when $\hat{C}_{ij}|\psi\>\neq 0$.
The operators $\hat{C}_{ij}$ are thus interpreted as a counterpart of classical kinetic constraints.
Lan et al.~\cite{Lan_arXiv2017} have numerically shown that the system displays a separation of time scales and a two-step relaxation is observed for $\lambda<1/2$.

It has also been shown that a quantum version of the East model, which is one of KCMs, also displays a dramatic slowing down of the dynamics~\cite{van_Horssen2015}.
In contrast to the model given by Eq.~(\ref{eq:QKCM}), the full thermalization is not achieved on numerically accessible time scales.
These results clearly show that constraints on local motion also give a mechanism of a pronounced time-scale separation in isolated quantum systems.

A quantum version of a KCM such as Eq.~(\ref{eq:QKCM}) is not described in the form of $\hat{H}=\hat{H}_0+\lambda\hat{V}$ with a Hamiltonian $\hat{H}_0$ that involves local conserved quantities.
Therefore, the presence of a time-scale separation is not apparent in the Hamiltonian but it emerges in the course of the time evolution in a manner depending on the initial state.
In Ref.~\cite{Lan_arXiv2017}, it is argued that the slow dynamics is a consequence of dynamical heterogeneities like that in classical glassy systems.

It is noted that a quantum version of KCMs such as one given by Eq.~(\ref{eq:QKCM}) is a special case of the class of Hamiltonians constructed by the method of embedding discussed in Sec.~\ref{sec:failure}.
A general Hamiltonian is given by Eq.~(\ref{eq:Shiraishi-Mori}), i.e.,
\beq
\hat{H}=\sum_{i=1}^V\hat{P}_i\hat{h}_i\hat{P}_i+\sum_{i=1}^V\hat{h}_i',
\label{eq:Shiraishi-Mori2}
\eeq
where $\hat{P}_i$ is a local projection operator and $[\hat{P}_i,\hat{h}_j']=0$ for all $i$ and $j$.
Here, $\hat{P}_i$ can be regarded as a dynamical constraint.
In Ref.~\cite{Mori-Shiraishi2017}, prethermalization in a specific model written in the form of Eq.~(\ref{eq:Shiraishi-Mori2}) has been demonstrated.
There, the separation of time scales arises due to the fact that there exists an almost invariant nonequilibrium subspace $\mathcal{H}_0$ spanned by states $|\psi\>$ with $(1/V)\sum_{i=1}^V\<\psi|\hat{P}_i|\psi\>\ll 1$.

\subsection{Prethermalization in long-range interacting systems}
\label{sec:long}

A long-range interaction is another source of a time-scale separation~\cite{Kastner2011,Kastner2012,Bachelard2013,van_den_Worm2013,Marcuzzi2013,Marcuzzi2016}.
It is known that classical Hamiltonian systems with long-range interactions sometimes display a two-step relaxation~\cite{Antoni1995}.
Here, by a long-range interaction we mean that the interaction potential decays slower than $1/r^d$ with the distance $r$, where $d$ is the spatial dimension.
When the energy and time scales are chosen so that the energy density remains finite in the thermodynamic limit, the lifetime of a quasi-stationary state diverges as $N^{\gamma}$ with $\gamma>0$.
For example, $\gamma\simeq 1.7$ in the Hamiltonian mean-field model~\cite{Yamaguchi2004}.

A diverging time scale in the thermodynamic limit also emerges in an isolated quantum system with long-range interactions.
Kastner~\cite{Kastner2011} studied the dynamics of the exactly solvable long-range interacting Emch-Radin model on a $d$-dimensional lattice, whose Hamiltonian is given by
\beq
\hat{H}=\frac{1}{2V^{1-\alpha/d}}\sum_{i\neq j}\frac{\hat{\sigma}_i^z\hat{\sigma}_j^z}{r_{ij}^{\alpha}}-h\sum_{i=1}^V\hat{\sigma}_i^z,
\label{eq:LR_Emch-Radin}
\eeq
and found diverging equilibration times for $\alpha$ satisfying $0\leq\alpha<d$.
In a later work, Kastner~\cite{Kastner2012} also found that the relaxation time $\tau_{\mathrm{rel}}$ scales as
\beq
\tau_{\mathrm{rel}}\sim\left\{
\begin{aligned}
&V^{1/2}& &\text{for } 0\leq\alpha<d/2, \\
&\left(\frac{V}{\ln V}\right)^{1/2}& &\text{for }\alpha=d/2, \\
&V^{1-\alpha/d}& &\text{for } d/2<\alpha<d.
\end{aligned}
\right.
\eeq 
At $\alpha=d/2$, the relaxation behavior changes qualitatively~\cite{Bachelard2013}.
Van den Worm et al.~\cite{van_den_Worm2013} have found that two-point correlation functions such as $\<\psi(t)|\hat{\sigma}_i^x\hat{\sigma}_j^x|\psi(t)\>$ display a two-step relaxations for $0<\alpha<d/2$.
A two-point correlation function undergoes a violent relaxation on a time scale proportional to $V^{1-\alpha/d}$, and then the second relaxation takes place on a time scale proportional to $V^{1/2}$.
See Ref.~\cite{van_den_Worm2013} for a numerical result.
Since the model of Eq.~(\ref{eq:LR_Emch-Radin}) is too simple (all $\hat{\sigma}_i^z$ commute with this Hamiltonian), the system remains out of equilibrium even after the second relaxation.

In recent years, there has been growing interest in the quantum dynamics of long-range interacting systems triggered by experimental realizations of long-range interacting systems in trapped ions~\cite{Porras2004, Kim2009, Britton2012, Islam2013}.
Several experimental reports on dynamical feature of long-range interacting systems have been reported~\cite{Richeme2014,Neyenhuis2017}.
Theoretically, long-range interacting systems exhibit novel dynamical behavior.
Long-range interactions are associated with extremely fast propagation of perturbation due to the absence of the Lieb-Robinson bound~\cite{Eisert2013}.
However, it is also known that in some cases, the propagation of perturbation is suppressed by long-range interactions.
Intuitively, one might expect that the growth of entanglement is fast in a long-range interacting system, but in some cases, the result is opposite; the entanglement grows only logarithmically with time $t$~\cite{Schachenmayer2013}.
A mechanism of the suppression of the propagation of perturbation is understood as cooprerative shielding~\cite{Santos2016}.
It is to be hoped that theoretical and experimental works would fully elucidate novel nonequilibrium phenomena of long-range interacting systems.

\subsection{Prethermalization associated with a symmetry breaking}
\label{sec:SSB}

Let us consider a quench from $\hat{H}_{\mathrm{ini}}$ to $\hat{H}_{\mathrm{fin}}$.
Following Alba and Fagotti~\cite{Alba-Fagotti2017}, we consider a situation in which the ground states of $\hat{H}_{\mathrm{ini}}$ in the thermodynamic limit exhibit a symmetry breaking but the symmetry is not broken at any positive temperature.
We assume that the symmetry can be spontaneously broken in $n$ distinct ways.
The symmetry-broken ``ground states'' are denoted by $|\Omega_k\>$ with $k=1,2,\dots,n$.

It is noted that for any finite system, in general, a symmetry-broken state $|\Omega_k\>$ is not an exact energy eigenstate of $\hat{H}_{\mathrm{ini}}$.
The exact ground state of $\hat{H}_{\mathrm{ini}}$ for a finite system, which is denoted by $|\Phi_0\>$, does not exhibit any symmetry breaking and is given by a symmetric superposition
\beq
|\Phi_0\>\approx\frac{1}{\sqrt{n}}\sum_{k=1}^n|\Omega_k\>.
\eeq
Only in the thermodynamic limit, $\{|\Omega_k\>\}$ become true ergodic ground states\footnote
{An infinite-volume state in which any intensive quantity has vanishing fluctuation is called an ergodic state.
}.

We may define $|\Omega_k\>$ as the ground state of the Hamiltonian $\hat{H}_{\mathrm{ini}}^{(k)}:=\hat{H}_{\mathrm{ini}}-h\hat{M}_k$ with a small symmetry-breaking field $-h\hat{M}_k$.
We should take the limit of $h\rightarrow+0$ after the thermodynamic limit.
Another construction of $|\Omega_k\>$ without a symmetry-breaking field is possible at least in the case of $Z_2$ symmetry breaking~\cite{Horsch1988}.
Let us denote by $\hat{M}$ the order parameter of the $Z_2$ symmetry.
We define $|\Gamma\>:=\hat{M}|\Phi_0\>/\|\hat{M}|\Phi_0\>\|$.
Then, symmetry-broken ground states $|\Omega_1\>$ and $|\Omega_2\>$ can be constructed as
\beq
|\Omega_1\>=\frac{|\Phi_0\>+|\Gamma\>}{\sqrt{2}}, \quad |\Omega_2\>=\frac{|\Phi_0\>-|\Gamma\>}{\sqrt{2}}.
\eeq
For continuous symmetries, a similar construction of symmetry-broken ground states has been conjectured but it is still a nontrivial issue~\cite{Koma1994}.

We shall prepare a low-temperature equilibrium state of $\hat{H}_{\mathrm{ini}}$ and change the Hamiltonian from $\hat{H}_{\mathrm{ini}}$ to $\hat{H}_{\mathrm{fin}}$ at time $t=0$.
If the temperature before the quench is sufficiently low but finite, the symmetry is not broken due to the assumption, and the initial state is very close to the symmetry-unbroken ground state of $\hat{H}_{\mathrm{ini}}$: 
\beq
|\psi(0)\>\approx|\Phi_0\>\approx\frac{1}{\sqrt{n}}\sum_{k=1}^n|\Omega_k\>.
\eeq
For a large system, off-diagonal matrix elements $\<\Omega_k|\hat{O}|\Omega_l\>$ ($k\neq l$) of any local operator $\hat{O}$ are vanishingly small.
Therefore, the state $|\Phi_0\>$ is locally indistinguishable from the mixture $(1/n)\sum_{k=1}^n|\Omega_k\>\<\Omega_k|$, and hence the initial state $\rho(0)=|\psi(0)\>\<\psi(0)|$ is approximately given by
\beq
\rho(0)\approx\frac{1}{n}\sum_{k=1}^n|\Omega_k\>\<\Omega_k|.
\label{eq:SSB_initial}
\eeq
The state at time $t>0$ is therefore expressed by the density matrix
\beq
\rho(t)=\frac{1}{n}\sum_{k=1}^ne^{-i\hat{H}_{\mathrm{fin}}t}|\Omega_k\>\<\Omega_k|e^{i\hat{H}_{\mathrm{fin}}t}.
\eeq

Now let us consider the situation in which $\hat{H}_{\mathrm{fin}}$ does not share the symmetry of $\hat{H}_{\mathrm{ini}}$.
In such a case, $\<\Omega_k|\hat{H}_{\mathrm{fin}}|\Omega_k\>$ may depend on $k$.
Since $|\Omega_k\>$ and $|\Omega_l\>$ with $k\neq l$ are macroscopically different, $\<\Omega_k|\hat{H}_{\mathrm{fin}}|\Omega_k\>$ and $\<\Omega_l|\hat{H}_{\mathrm{fin}}|\Omega_l\>$ are macroscopically different in general.

If $\hat{H}_{\mathrm{fin}}$ satisfies the MITE-ETH, the state $e^{-i\hat{H}_{\mathrm{fin}}t}|\Omega_k\>$ relaxes to an equilibrium state specified by the energy $\<\Omega_k|\hat{H}_{\mathrm{fin}}|\Omega_k\>$ or the corresponding inverse temperature $\beta_k$.
As a result, the system with the density matrix $\rho(t)$ will relax to a prethermal state that is described by
\beq
\rho_{\mathrm{pre}}=\frac{1}{n}\sum_{k=1}^n\frac{e^{-\beta_k\hat{H}_{\mathrm{fin}}}}{Z_k},
\label{eq:SSB_pre}
\eeq
where $Z_k=\mathrm{Tr}\,e^{-\beta_k\hat{H}_{\mathrm{fin}}}$.
Macroscopically different values of $\<\Omega_k|\hat{H}_{\mathrm{fin}}|\Omega_k\>$ and $\<\Omega_l|\hat{H}_{\mathrm{fin}}|\Omega_l\>$ imply different values of $\beta_k$ and $\beta_l$.
Thus, the prethermal state is expressed by a mixture of Gibbs states at several temperatures.

Actually, the initial state has a finite temperature and there is no long-range order.
It means that Eq.~(\ref{eq:SSB_initial}) is an approximation for the initial state.
As a result, a prethermal state described by Eq.~(\ref{eq:SSB_pre}) lasts only up to a finite time $\tau_{\mathrm{pre}}$, which increases as the temperature decreases.

This mechanism of prethermalization was found by Alba and Fagotti~\cite{Alba-Fagotti2017}.
They have demonstrated it numerically for the quench from the classical Ising model 
\beq
\hat{H}_{\mathrm{ini}}=-J\sum_{i=1}^V\hat{\sigma}_i^z\hat{\sigma}_{i+1}^z
\eeq
to the anisotropic next-nearest-neighbor Ising Hamiltonian 
\beq
\hat{H}_{\mathrm{fin}}=-\sum_{i=1}^V (J_1\sigma_i^x\sigma_{i+1}^x +J_2\sigma_i^x\sigma_{i+2}^x +h\sigma_i^z).
\eeq
In this case, $n=2$ and $|\Omega_1\>$ ($|\Omega_2\>$) is the all-up (all-down) state, respectively.

\subsection{Dynamical phase transitions}
\label{sec:DPT}

Dynamical phase transitions are nonequilibrium phase transitions in isolated quantum systems.
So far, two different notions of dynamical phase transitions, which are referred to as DPT-I and DPT-II, have been investigated.
The DPT-I denotes a phase transition in a nonthermal steady state or in a prethermal state, while the DPT-II denotes a nonequilibrium transition in a transient state.

The DPT-I has been mainly studied in several mean-field models (fully connected models)~\cite{Sciolla2010,Sciolla2011}.
As an example, let us consider the fully connected transverse-field Ising model
\beq
\hat{H}(\Gamma)=-\frac{J}{2V}\sum_{i,j=1}^V\hat{S}_i^z\hat{S}_j^z-\Gamma\sum_{i=1}^V\hat{S}_i^x,
\eeq
where $\hat{\bm S}_i$ is the spin-1/2 operator at site $i$ and $J>0$.
At $\Gamma=\Gamma_{\mathrm c}:=J/2$, a quantum phase transition occurs.
For $\Gamma<\Gamma_{\mathrm c}$, the ground state is ferromagnetic, while for $\Gamma>\Gamma_{\mathrm c}$, the ground state is paramagnetic.

We consider a quench of the transverse field from the initial value $\Gamma=\Gamma_{\mathrm{ini}}$ to the final value $\Gamma=\Gamma_{\mathrm{fin}}$.
Because of the site-permutation symmetry, we can consider in the sector of the maximum total spin $S=V/2$ (totally symmetric subspace).
Any state in this sector is written as a superposition of the orthonormal basis states $|q\>$ with $Vq=-V/2, -V/2+1, \dots, V/2$ that satisfy $\hat{\bm S}_{\mathrm{tot}}^2|q\>=(V/2)(V/2+1)|q\>$ and $\hat{S}_{\mathrm{tot}}^z|q\>=Vq|q\>$, where $\hat{\bm S}_{\mathrm{tot}}=\sum_{i=1}^V\hat{\bm S}_i$.
The wave function corresponding to a quantum state $|\psi(t)\>$ is introduced by $\psi_t(q)=\<q|\psi(t)\>$.
It is shown that, for large $V$, we have $\<q|\hat{S}_{\mathrm{tot}}^z=Vq\<q|$, $\<q|\hat{S}_{\mathrm{tot}}^x\approx V\sqrt{1/4-q^2}\cos p \<q|$, and $\<q|\hat{S}_{\mathrm{tot}}^y\approx V\sqrt{1/4-q^2}\sin p\<q|$, where $p=-(i/V)\d/\d q$ is the momentum conjugate to $q$.
As a result, the Schr\"odinger equation $id|\psi(t)\>/dt=\hat{H}(\Gamma)|\psi(t)\>$ is reduced to
\beq
\frac{i}{V}\frac{\d}{\d t}\psi_t(q)=\left[-\frac{J}{2}q^2-\Gamma\sqrt{\frac{1}{4}-q^2}\cos p\right]\psi_t(q),
\eeq
which is the semiclassical equation with an effective Planck constant $\hbar_{\mathrm{eff}}=1/V$.
Thus, this model can be regarded as a semiclassical system discussed in Sec.~\ref{sec:semiclassical}.

The classical Hamiltonian $H_{\mathrm{cl}}(q,p;\Gamma)=-Jq^2/2-\Gamma\sqrt{1/4-q^2}\cos p$ has a separatrix for $\Gamma<\Gamma_{\mathrm c}$ at $q=p=0$ with the classical energy $\varepsilon^*(\Gamma):=H_{\mathrm{cl}}(0,0;\Gamma)=-\Gamma/2$.
In this case, the ergodicity is broken below the energy $\varepsilon^*(\Gamma)$ in the classical approximation.

Let the energy of the initial state after a quench be denoted by $E_{\mathrm{fin}}=\<\psi(0)|\hat{H}(\Gamma_{\mathrm{fin}})|\psi(0)\>$.
Now we fix $\Gamma_{\mathrm{ini}}<\Gamma_{\mathrm c}$ and control the value of $\Gamma_{\mathrm{fin}}$ freely.
A dynamical phase transition occurs at $\Gamma_{\mathrm{fin}}=\Gamma_{\mathrm d}$ satisfying $E_{\mathrm{fin}}=V\varepsilon^*(\Gamma_{\mathrm d})=-V\Gamma_{\mathrm d}/2$.
For large $V$, the initial state is localized at the minimum of $H_{\mathrm{cl}}(q,p;\Gamma_{\mathrm{ini}})$, given by $q=\pm\sqrt{1/4-(\Gamma_{\mathrm{ini}}/J)^2}$ and $p=0$.
This gives $E_{\mathrm{fin}}\approx VH_{\mathrm{cl}}(\sqrt{1/4-(\Gamma_{\mathrm{ini}}/J)^2},0;\Gamma_{\mathrm{fin}})$ (the equality becomes exact in the thermodynamic limit), and hence we obtain
\beq
\Gamma_{\mathrm d}=\frac{\Gamma_{\mathrm{ini}}+\Gamma_{\mathrm c}}{2}.
\eeq
For $\Gamma_{\mathrm{fin}}<\Gamma_{\mathrm d}$, the stationary state reached after the quench is ferromagnetic, while for $\Gamma_{\mathrm{fin}}>\Gamma_{\mathrm d}$, it is paramagnetic.
The type of transitions is the same as the equilibrium phase transition, but the dynamical transition point is shifted from the equilibrium transition point, and hence this transition is purely dynamical.

This result indicates that the DPT-I is intimately related to underlying equilibrium quantum phase transitions.
Studies on several mean-field models suggest that the existence of a quantum phase transition is important for the existence of the DPT-I~\cite{Sciolla2011}.
Since a quantum state after a quench populates excited states of the post-quench Hamiltonian, it would also be expected that the DPT-I is related to a finite-temperature equilibrium phase transition.
However, it has been reported that the DPT-I occurs even when there is no finite-temperature equilibrium phase transition in the transverse-field Ising model with power-law interactions~\cite{Halimeh2017}.
The relation between the DPT-I and underlying equilibrium phase transitions are not completely figured out.

The DPT-I has been studied by Sciolla and Biroli for several mean-field models~\cite{Sciolla2010,Sciolla2011}.
Gambassi and Calabrese~\cite{Gambassi2011} have studied quench dynamics of the scalar $\phi^4$ theory and obtained the mean-field nonequilibrium phase diagram by mapping the problem to the one of classical phase transitions in films.
Moreover, they have argued that the DPT-I is generic beyond the mean-field theory based on this mapping.
Beyond the mean-field theory, Sciolla and Biroli~\cite{Sciolla2013} have studied the $O(N)$ model ($\phi^4$ theory of $N$-component vector field $\phi$ with $O(N)$ symmetry) with the leading order corrections in $1/N$.
They have found the DPT-I in this model and the critical out-of-equilibrium dynamics with diverging time and length scales, which resembles the coarsening dynamics in classical systems~\cite{Bray1994}.
Tsuji, Eckstein, and Werner~\cite{Tsuji2013} have studied the dynamics of the Hubbard model by using the nonequilibrium dynamical mean-field theory, and found the nonthermal antiferromagnetic order in a prethermal state.
Chiocchetta et al.~\cite{Chiocchetta2015,Chiocchetta2016} have performed the renormalization-group analysis of nonequilibrium dynamics after a quench close to a dynamical critical point in the $O(N)$ model, and have shown that an interesting short-time scaling is observed within the prethermal regime.
In the same model, it has also been shown that, depending on the state of the system before the quench, the evolution of the order parameter displays a temporal crossover between universal dynamical scaling regimes associated with different renormalization-group fixed points~\cite{Chiocchetta2017}.
In this way, an isolated quantum system can display intriguing nonequilibrium dynamics associated with a dynamical phase transition in a prethermal regime.

Next, we briefly review the DPT-II, which is another notion of dynamical phase transitions.
The DPT-II is characterized by a cusp singularity of the Loschmidt return rate $r(t)$ defined as
\beq
r(t)=-\lim_{V\rightarrow\infty}\frac{1}{V}\ln\mathcal{F}(t)=-\lim_{V\rightarrow\infty}\frac{1}{V}\ln\left|\<\psi(0)|\psi(t)\>\right|^2,
\eeq
where $\mathcal{F}(t)$ is the fidelity (or also called the Loschmidt echo) appearing in Eq.~(\ref{eq:fidelity}).
The singularity in $r(t)$ results from the zeros of the Loschmidt amplitude
\beq
\mathcal{G}(t)=\<\psi(0)|\psi(t)\>=\<\psi(0)|e^{-i\hat{H}t}|\psi(0)\>
\eeq
as a function of complex time $t\in\mathbb{C}$ (the fidelity is given by $\mathcal{F}(t)=|\mathcal{G}(t)|^2$).
As the system size $V$ increases, zeros of the fidelity accumulate to form lines or areas.
If such a line or boundary of such an area crosses the real-time axis, the DPT-II occurs.
This is analogous to the connection between an equilibrium phase transition captured by a singularity of the free energy density $f(\beta)=-\lim_{V\rightarrow\infty}(1/V\beta)\ln Z(\beta)$ and Fisher zeros of a complex partition function $Z(\beta)=\mathrm{Tr}\,e^{-\beta\hat{H}}$ with $\beta\in\mathbb{C}$.
Indeed, one will see a formal similarity between $\mathcal{G}(t)$ and $Z(\beta)$.

The notion of the DPT-II  was introduced by Heyl, Polkovnikov, and Kehrein~\cite{Heyl2013}.
They have shown that the DPT-II occurs for quenches across the quantum critical point in the nearest-neighbor transverse-field Ising chain, while there is no DPT-II for quenches within the same phase.
It indicates that the DPT-II is also intimately related to the underlying equilibrium phase transitions.
For topological systems of noninteracting fermions, this connection is clearly understood~\cite{Vajna2015,Huang2016}.
However, Halimeh and Zauner-Stauber~\cite{Halimeh2017_dynamical} have shown that the DPT-II is not related to equilibrium phase transitions in the transverse-field Ising model with power-law interactions.

The DPT-I and the DPT-II are different concepts, but some connections between them have been reported.
Homrighausen et al.~\cite{Homrighausen2017} have numerically shown that a DPT-II transition point between the ``regular'' and the ``anomalous'' phases~\cite{Halimeh2017} coincides with a DPT-I transition point in fully connected transverse-field Ising model.
Weidinger et al.~\cite{Weidinger2017_dynamical} have shown that the DPT-II occurs if the quench crosses the transition point of the DPT-I in the $O(N)$ model, but singularities in the Loschmidt return rate is subextensive in $N$.

\subsection{Entanglement prethermalization after a coherent splitting}
\label{sec:EP}

\subsubsection{Setup}
\label{sec:EP_setup}

\begin{figure}[tb]
\centering
\includegraphics[width=12cm]{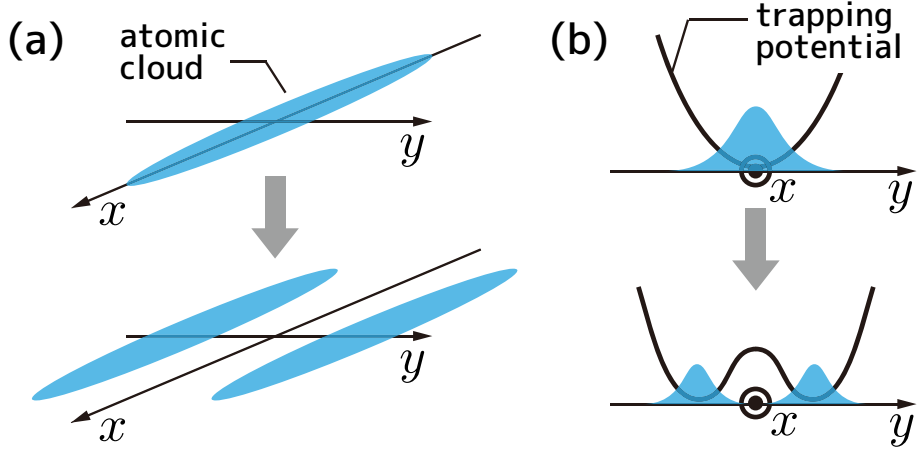}
\caption{Schematic illustration of the splitting process.
(a) A quasi-1D system is coherently split into two subsystems.
(b) Coherent splitting is realized by deforming the trapping potential into a double-well potential in the transverse direction ($y$ axis).}
\label{fig:split}
\end{figure}

In cold-atom experiments, prethermalization was observed in coherently split 1D Bose gases~\cite{Gring2012, Smith2013, Langen2013, Langen2015}.
In experiments, one first prepares an equilibrium state of a quasi one-dimensional Bose gas, and then the transverse confinement potential with the frequency $\omega_{\perp}$ is deformed to a double-well potential as in Fig.~\ref{fig:split}, which creates two identical subsystems which we call ``left'' and ``right''.
The splitting process is performed much faster than a typical time scale of the dynamics along the longitudinal direction, which we choose as the $x$ axis, but much slower than that along the transverse directions, which we choose as the $y$ and $z$ axes; the time scale in the transverse direction is given by $\omega_{\perp}^{-1}$.
Then, theoretically, the Bose field operator $\hat{\psi}(x)$ transforms upon the splitting process in the following way:
\beq
\hat{\psi}(x)\rightarrow\frac{\hat{\psi}_{\mathrm{L}}(x)+\hat{\psi}_{\mathrm{R}}(x)}{\sqrt{2}},
\label{eq:split}
\eeq
where $\hat{\psi}_{\mathrm{L}}(x)$ and $\hat{\psi}_{\mathrm{R}}(x)$ are the Bose field operators of the left and right subsystems, respectively (see Fig.~\ref{fig:split}).
We assume that quantum tunneling between the left and right subsystems is negligible and therefore these subsystems evolve independently.

We denote by $\hat{H}[\hat{\psi}]$ the Hamiltonian before the splitting, where $[\hat{\psi}]$ indicates that the Hamiltonian is written in terms of the Bose field operators $\hat{\psi}(x)$ and $\hat{\psi}^{\dagger}(x)$.
Under the mapping of Eq.~(\ref{eq:split}), this Hamiltonian is transformed as
\beq
\hat{H}[\hat{\psi}]\rightarrow\hat{H}\left[(\hat{\psi}_{\mathrm{L}}+\hat{\psi}_{\mathrm{R}})/\sqrt{2}\right].
\eeq
On the other hand, since we assume that the left and right systems are identical and independent after the splitting, the Hamiltonian after the splitting is given by $\hat{H}'[\hat{\psi}_{\mathrm{L}}]+\hat{H}'[\hat{\psi_{\mathrm{R}}}]$.
Therefore, the splitting process is theoretically modeled by a quantum quench in which the Hamiltonian is suddenly switched from
\beq
\hat{H}_{\mathrm{ini}}=\hat{H}\left[(\hat{\psi}_{\mathrm{L}}+\hat{\psi}_{\mathrm{R}})/\sqrt{2}\right]
\eeq
to
\beq
\hat{H}_{\mathrm{fin}}=\hat{H}'[\hat{\psi}_{\mathrm{L}}]+\hat{H}'[\hat{\psi}_{\mathrm{R}}].
\eeq
The initial state $|\psi(0)\>$ is an equilibrium state (or the ground state) of $\hat{H}_{\mathrm{ini}}$ and should satisfy
\beq
\frac{\hat{\psi}_{\mathrm{L}}(x)-\hat{\psi}_{\mathrm{R}}(x)}{\sqrt{2}}|\psi(0)\>=0
\label{eq:quasi_1D}
\eeq
for all $x$, which expresses the fact that there is no excitation along the transverse directions in a quasi-1D system.

In experiments, after an evolution time $t$, we turn off the trapping potentials and measure the interference contrast between the two gases~\cite{Gring2012}.
Gring et al.~\cite{Gring2012} observed a quasi-stationary state in which the interference contrast is stronger than what is expected from equilibrium statistical mechanics.
The observed interference contrast in a quasi-stationary state is well fitted by that in an equilibrium state at a certain effective temperature.
Thus, coherently split one-dimensional Bose gases exhibit prethermalization.

\subsubsection{A toy model analysis of entanglement prethermalization}
\label{sec:EP_simple}

Before analyzing  many-body Bose systems, we consider the quench problem in a simple model consisting of two identical bosons to explain general features of \textit{entanglement prethermalization} (EP)~\cite{Kaminishi2015_entanglement,Ikeda2017,Kaminishi_arXiv2017}.

Following Ref.~\cite{Ikeda2017}, we consider the initial Hamiltonian before the quench given by
\beq
\hat{H}_{\mathrm{ini}}=\frac{1}{2}(\hat{p}_1^2+\hat{p}_2^2)+\frac{1}{2}(\hat{x}_1^2+\hat{x}_2^2)+\frac{\alpha^2}{2}(\hat{x}_1-\hat{x}_2)^2,
\eeq
where $\hat{x}_i$ and $\hat{p}_i$ are the canonical coordinate and momentum of the $i$th particle ($i=1,2$), which satisfy the canonical commutation relations.
The mass $m$ of bosons and the frequency $\omega$ of the harmonic potential are set to unity.
At time $t=0$, we suddenly turn off the interaction between two particles, which corresponds to the splitting process in experiments, and thus the Hamiltonian after the quench is given by
\beq
\hat{H}_{\mathrm{fin}}=\frac{1}{2}(\hat{p}_1^2+\hat{x}_1^2)+\frac{1}{2}(\hat{p}_2^2+\hat{x}_2^2).
\eeq
By introducing the annihilation operators $\hat{a}_i=(\hat{x}_i+i\hat{p}_i)/\sqrt{2}$ $(i=1,2)$ and their linear combinations $\hat{a}_{\pm}=(\hat{a}_1\pm\hat{a}_2)/\sqrt{2}$, $\hat{H}_{\mathrm{fin}}$ is diagonalized as
\beq
\hat{H}_{\mathrm{fin}}=\hat{a}_1^{\dagger}\hat{a}_1+\hat{a}_2^{\dagger}\hat{a}_2
=\hat{a}_+^{\dagger}\hat{a}_++\hat{a}_-^{\dagger}\hat{a}_-.
\label{eq:H_two}
\eeq

We denote by $|0\>$ the Fock vacuum defined by $\hat{a}_1|0\>=\hat{a}_2|0\>=0$.
The energy eigenstates are expressed in two ways:
\beq
|m,n\>:=\frac{(\hat{a}_1^{\dagger})^m}{\sqrt{m!}}\frac{(\hat{a}_2^{\dagger})^n}{\sqrt{n!}}|0\>,
\quad
|m,n\kket:=\frac{(\hat{a}_+^{\dagger})^m}{\sqrt{m!}}\frac{(\hat{a}_-^{\dagger})^n}{\sqrt{n!}}|0\>,
\eeq
both of which have the energy eigenvalue $m+n$.
The Hamiltonian has many energy degeneracies.

The initial Hamiltonian $\hat{H}_{\mathrm{ini}}$ is diagonalized by the center-of-mass motion and the relative motion, and the corresponding annihilation operators are defined by
\beq
\left\{
\begin{aligned}
&\hat{a}_{\mathrm{CM}}=\hat{a}_+, \\
&\hat{a}_{\mathrm{rel}}=\cosh(r)\hat{a}_-+\sinh(r)\hat{a}_-^{\dagger},
\end{aligned}
\right.
\eeq
where $r$ is determined by
\beq
e^{4r}=1+\alpha^2.
\eeq
We thus obtain
\beq
\hat{H}_{\mathrm{ini}}=\hat{a}_{\mathrm{CM}}^{\dagger}\hat{a}_{\mathrm{CM}} +\sqrt{1+\alpha^2}\hat{a}_{\mathrm{rel}}^{\dagger}\hat{a}_{\mathrm{rel}}.
\label{eq:H_two_ini}
\eeq

We assume that the initial state $|\psi(0)\>$ is given by the ground state of $\hat{H}_{\mathrm{ini}}$ satisfying $\hat{a}_{\mathrm{CM}}|\psi(0)\>=\hat{a}_{\mathrm{rel}}|\psi(0)\>=0$.
This state is known as a squeezed vacuum and expressed by
\beq
|\psi(0)\>=\frac{1}{\sqrt{\cosh(r)}}e^{-\frac{\tanh(r)}{2}(a_-^{\dagger})^2}|0\>.
\eeq
The time evolution is obtained as
\beq
|\psi(t)\>=e^{-i\hat{H}_{\mathrm{fin}}t}|\psi(0)\>=\sum_{N=0}^{\infty}(-1)^N\sqrt{q_N}e^{-2iNt}|\Phi_N\>,
\label{eq:psi_Ikeda}
\eeq
where
\beq
|\Phi_N\>=|0,2N\kket=\sum_{m=0}^{2N}c_{m,2N-m}|m,2N-m\>
\eeq
with $c_{m,n}=(-1)^m(m+n)!/\sqrt{m!n!}$ and
\beq
q_N=\frac{1}{\cosh(r)}\frac{(2N)!}{(N!)^2}\left(\frac{\tanh(r)}{2}\right)^{2N}.
\eeq

Now we focus on the infinite-time average\footnote
{This simple system does not equilibrate due to equal spacings of energy eigenvalues.
In more realistic physical systems, equilibration occurs, and an equilibrated state is described by the diagonal ensemble, which is the infinite-time average of the density matrix (see Sec.~\ref{sec:equilibration}).
}.
The infinite-time average of the density matrix, i.e., the diagonal ensemble, is given by
\beq
\rho_{\mathrm{D}}=\overline{|\psi(t)\>\<\psi(t)|}=\sum_{N=0}^{\infty}q_N|\Phi_N\>\<\Phi_N|.
\label{eq:diag_Ikeda}
\eeq
After the infinite-time average, the coherence is lost between different degenerate subspaces, whereas it is not within each degenerate subspace.

We shall explain below the following property of the diagonal ensemble given by Eq.~(\ref{eq:diag_Ikeda}): each particle is described by a conventional statistical ensemble, whereas the entire system is not.
Let us consider the reduced density matrix of the particle $i=1$,
\beq
\rho_{\mathrm{D}}^{(1)}=\mathrm{Tr}_2\,\rho_{\mathrm{D}}=\sum_{m=0}^{\infty}w_m|m\>\<m|,
\eeq
where $w_m=\sum_{N=\lfloor m/2\rfloor}^{\infty}q_N|c_{m,2N-m}|^2$.
The trace over the Hilbert space of the second particle is denoted by $\mathrm{Tr}_2$.
The asymptotic behavior of $w_m$ for $m\gg 1$ is given by
\begin{align}
w_m&\simeq \tilde{w}_m=\frac{e^{-\beta m-\frac{1}{2}\ln m}}{\sqrt{2\pi\left[\cosh(r)-\frac{1}{2}\sinh(r)\right]}},
\label{eq:w_Ikeda}
\\
\beta&=\ln[2\coth(r)-1].
\end{align}
It is noted that the second term in the exponent of Eq.~(\ref{eq:w_Ikeda}) is negligibly small compared with the first one for $m\gg 1$.
Thus, Eq.~(\ref{eq:w_Ikeda}) shows that $w_m$ is asymptotically thermal for large $m$ (numerically, $w_m\simeq\tilde{w}_m$ already at $m\sim 10$). 
In this way, the reduced density matrix of each particle is approximately identical to the canonical ensemble.

Meanwhile, the diagonal ensemble of the entire system is not given by the (micro)canonical ensemble, and the difference between them affect physical observables like $\hat{O}=(\hat{x}_1-\hat{x}_2)^2$~\cite{Ikeda2017}.
Importantly, the diagonal ensemble given by Eq.~(\ref{eq:diag_Ikeda}) is \textit{not} diagonal in the basis of $|n,m\>$.
Off-diagonal elements of $\rho_{\mathrm{D}}$ in the basis of $|n,m\>$ carry the memory of the initial entanglement between the two particles.

It is noted that in the basis of $|m,n)$, the state $|\psi(t)\>$ is written by Eq.~(\ref{eq:psi_Ikeda}) with $|\Phi_N\>=|0,2N)$, and hence there is no entanglement between the degrees of freedom of the center-of-mass motion and that of the relative motion.

Important conserved quantities are therefore $\hat{a}_+\hat{a}_+$ and $\hat{a}_-\hat{a}_-$.
If we construct the GGE by using these conserved quantities, the GGE is given by
\beq
\rho_{\mathrm{GGE}}^{(\pm)}=\frac{e^{-\beta_+\hat{a}_+^{\dagger}\hat{a}_+-\beta_-\hat{a}_-^{\dagger}\hat{a}_-}}{\mathrm{Tr}\,e^{-\beta_+\hat{a}_+^{\dagger}\hat{a}_+-\beta_-\hat{a}_-^{\dagger}\hat{a}_-}},
\eeq
where $\beta_{\pm}$ are determined from $\mathrm{Tr}\,\hat{a}_{\pm}^{\dagger}\hat{a}_{\pm}\rho_{\mathrm{GGE}}^{(\pm)}=\mathrm{Tr}\,\hat{a}_{\pm}^{\dagger}\hat{a}_{\pm}\rho_{\mathrm{D}}$ to be $\beta_+=\infty$ and $\beta_-=2\ln\coth(r)$.
It is found that this GGE approximately describes the infinite-time average of physical quantities~\cite{Ikeda2017}.
It is noted that in terms of $\hat{a}_{\mathrm{L}}$ and $\hat{a}_{\mathrm{R}}$, the GGE is rewritten as
\beq
\rho_{\mathrm{GGE}}^{(\pm)}=\frac{\exp\left[-\frac{\beta_++\beta_-}{2}(\hat{a}_{\mathrm{L}}^{\dagger}\hat{a}_{\mathrm{L}} +\hat{a}_{\mathrm{R}}^{\dagger}\hat{a}_{\mathrm{R}})-\frac{\beta_+-\beta_-}{2}(\hat{a}_{\mathrm{L}}^{\dagger}\hat{a}_{\mathrm{R}} +\hat{a}_{\mathrm{R}}^{\dagger}\hat{a}_{\mathrm{L}})\right]}{\mathrm{Tr}\,\exp\left[-\frac{\beta_++\beta_-}{2}(\hat{a}_{\mathrm{L}}^{\dagger}\hat{a}_{\mathrm{L}} +\hat{a}_{\mathrm{R}}^{\dagger}\hat{a}_{\mathrm{R}})-\frac{\beta_+-\beta_-}{2}(\hat{a}_{\mathrm{L}}^{\dagger}\hat{a}_{\mathrm{R}} +\hat{a}_{\mathrm{R}}^{\dagger}\hat{a}_{\mathrm{L}})\right]},
\eeq
and correlations between the left and right subsystems remain stationary, which reflects the memory of the entanglement between them in the initial state.

If we add some perturbation, this stationary state becomes unstable.
For example, if the Hamiltonian after the splitting is given, instead of Eq.~(\ref{eq:H_two}), by
\beq
\hat{H}'_{\mathrm{fin}}=(1+\epsilon)\hat{a}_1^{\dagger}\hat{a}_1+\hat{a}_2^{\dagger}\hat{a}_2
\label{eq:H_asym}
\eeq
with small $\epsilon>0$, the system shows two-step relaxation.
After the first relaxation, the system reaches a prethermal state described by Eq.~(\ref{eq:diag_Ikeda}), and then, after the second relaxation, the system reaches a stationary state described by
\beq
\rho_{\mathrm{D}}^{\mathrm{d}}=\sum_{N=0}^{\infty}q_N\sum_{m=0}^{2N}|c_{m,2N-m}|^2|m,2N-m\>\<m,2N-m|,
\eeq
which is the diagonal part of $\hat{\rho}_{\mathrm{D}}$ in the basis of $|m,n\>$.
The second relaxation occurs on a time scale proportional to $1/\epsilon$ due to lifting the energy degeneracies by the asymmetry of two bosons~\cite{Ikeda2017}.

In this way, under a small perturbation, the state described by Eq.~(\ref{eq:diag_Ikeda}) becomes a quasi-stationary state with a finite lifetime.
Therefore, when the initial entanglement between two subsystems is protected by approximate energy degeneracies, prethermalization occurs and a quasi-stationary state appears.
This phenomenon is called the EP, first discussed in the Lieb-Liniger model in Ref.~\cite{Kaminishi2015}.

\subsubsection{EP in the Tomonaga-Luttinger model}
\label{sec:EP_TL}

The EP occurs when the subsystems are initially entangled and there exist degenerate subspaces in which the entanglement is protected.
It is expected that the experimentally observed prethermalization in coherently split one-dimensional Bose gases is explained by the EP.
In this section, we study the time evolution after a coherent splitting by using the Tomonaga-Luttinger (TL) model~\cite{Tomonaga1950,Luttinger1963}, which is one of the most powerful models for investigating the nonequilibrium dynamics in one-dimensional systems~\cite{Giamarchi_text,Cazalilla2006,Foini2015,Calzona2017}.

The TL model is regarded as a low-energy effective theory of the Lieb-Liniger model~\cite{Lieb1963}, which describes a one-dimensional Bose gas with contact interactions.
Therefore, we first formulate the problem in terms of the LL Hamiltonian, and then treat it in the TL model by performing the low-energy approximation.
The analysis in the Lieb-Liniger model is discussed in Sec.~\ref{sec:EP_LL}. 
The Hamiltonian of the Lieb-Liniger model is given by
\beq
\hat{H}_{\mathrm{LL}}[\hat{\psi}]=\int_{-L/2}^{L/2}dx\left[-\frac{1}{2m}\hat{\psi}^{\dagger}(x)\d_x^2\hat{\psi}(x)+g\hat{\psi}^{\dagger}(x)\hat{\psi}^{\dagger}(x)\hat{\psi}(x)\hat{\psi}(x)\right],
\eeq
where we impose the periodic boundary condition and $\hat{\psi}(x)$ is the annihilation operator of a boson at position $x$.
The mass of a boson is denoted by $m$ and the density $(1/L)\int_{-L/2}^{L/2}dx\,\hat{\psi}^{\dagger}(x)\hat{\psi}(x)$ is fixed at $\rho_0$.

The coupling constant $g$ is related to the $s$-wave scattering length $a_s$ of the original 3D system and the frequency $\omega_{\perp}$ of the transverse confinement potential as
\beq
g=\frac{a_s}{ma_{\perp}^2}\frac{1}{1-C(a_s/a_{\perp})},
\eeq
where $a_{\perp}=1/\sqrt{m\omega_{\perp}}$ and $C\simeq 1.0326$~\cite{Olshanii1998, Cazalilla_review2011}.
It is noted that $a_s$ can experimentally be controlled by a Feshbach resonance.
In this section, we consider a repulsive weak interaction with $0<mg/\rho_0\ll 1$.

As discussed in Sec.~\ref{sec:EP_setup}, the coherent splitting corresponds to the quench from $\hat{H}_{\mathrm{ini}}=\hat{H}_{\mathrm{LL}}[(\hat{\psi}_{\mathrm{L}}+\hat{\psi}_{\mathrm{R}})/\sqrt{2}]$ to $\hat{H}_{\mathrm{fin}} =\hat{H}_{\mathrm{LL}}'[\hat{\psi}_{\mathrm{L}}]+\hat{H}_{\mathrm{LL}}'[\hat{\psi}_{\mathrm{R}}]$, where $\hat{H}_{\mathrm{LL}}[\hat{\psi}]$ is given by the Lieb-Liniger Hamiltonian with a new coupling constant $g'$.
If the transverse confinement frequency $\omega_{\perp}$ changes in the splitting process, $g'$ will also change.
In this section, we set $g'=g$, and hence $\hat{H}_{\mathrm{LL}}'[\hat{\psi}]=\hat{H}_{\mathrm{LL}}[\hat{\psi}]$, but this special choice of $g'$ is not essential.

In the weak-coupling and low-energy regime, we can obtain the TL Hamiltonian by putting
\beq
\hat{\psi}^{\dagger}(x)=\sqrt{\rho+\hat{n}(x)}e^{-i\hat{\theta}(x)}
\eeq
and expanding the Hamiltonian with respect to $\hat{n}(x)$ and $\d_x\hat{\theta}(x)$ up to the second order, where $\rho$ is the average density of bosons.
The commutation relations are given by $[\hat{n}(x),\hat{n}(x')]=[\hat{\theta}(x),\hat{\theta}(x')]=0$ and $[\hat{n}(x),\hat{\theta}(x')]=i\delta(x-x')$.
It is noted that a derivative $\d_x\hat{n}(x)$ is regarded as a higher-order term compared with $\hat{n}(x)$ in the low-energy (i.e., long-wavelength) approximation.
In this approximation, we obtain
\beq
\hat{H}_{\mathrm{LL}}[\hat{\psi}]\approx\hat{H}_{\mathrm{TL}}[\hat{n},\hat{\theta}]:=\int_{-L/2}^{L/2}dx \, \left\{\frac{\rho}{2m}\left[\d_x\hat{\theta}(x)\right]^2+g\hat{n}(x)^2\right\}.
\label{eq:LL-TL}
\eeq
The TL Hamiltonian is denoted by $\hat{H}_{\mathrm{TL}}[\hat{n},\hat{\theta}]$.

Since we have the two types of the Bose field operators $\hat{\psi}_{\mathrm{L}}(x)$ and $\hat{\psi}_{\mathrm{R}}(x)$, we have two sets of the canonical conjugate operators $\{\hat{n}_{\mathrm{L}}(x),\hat{\theta}_{\mathrm{L}}(x)\}$ and $\{\hat{n}_{\mathrm{R}}(x),\hat{\theta}_{\mathrm{R}}(x)\}$.
The average density of bosons in each subsystem is given by $\rho_0/2$, and hence the post-quench Hamiltonian is obtained by putting $\rho=\rho_0/2$ in Eq.~(\ref{eq:LL-TL}):
\beq
\hat{H}_{\mathrm{fin}}\simeq\sum_{\mathrm{X}=\mathrm{L},\mathrm{R}}\int_{-L/2}^{L/2}dx\left\{\frac{\rho_0}{4m}\left[\d_x\hat{\theta}_{\mathrm{X}}(x)\right]^2+g\hat{n}_{\mathrm{X}}(x)^2\right\}.
\label{eq:fin_TL}
\eeq
The pre-quench Hamiltonian is more complicated, but it is shown that
\beq
\hat{H}_{\mathrm{ini}}=\hat{H}_{\mathrm{LL}}\left[\frac{\hat{\psi}_{\mathrm{L}}+\hat{\psi}_{\mathrm{R}}}{\sqrt{2}}\right]\approx\int_{-L/2}^{L/2}dx\left\{\frac{\rho_0}{8m}\left[\d_x\hat{\theta}_{\mathrm{c}}(x)\right]^2+4g\hat{n}_{\mathrm{c}}(x)^2\right\}
\label{eq:ini_TL}
\eeq
and the condition of Eq.~(\ref{eq:quasi_1D}) is written as
\beq
\left[\hat{n}_{\mathrm{s}}(x)+i\frac{\rho_0}{2}\hat{\theta}_{\mathrm{s}}(x)\right]|\psi(0)\>=0
\label{eq:quasi_1D_TL}
\eeq
for all $x$, where we introduce the new operators:
\beq
\left\{
\begin{aligned}
&\hat{n}_{\mathrm{c}}(x)=\frac{\hat{n}_{\mathrm{L}}(x)+\hat{n}_{\mathrm{R}}(x)}{2}, & &\hat{\theta}_{\mathrm{c}}(x)=\hat{\theta}_{\mathrm{L}}(x)+\hat{\theta}_{\mathrm{R}}(x), \\
&\hat{n}_{\mathrm{s}}(x)=\frac{\hat{n}_{\mathrm{L}}(x)-\hat{n}_{\mathrm{R}}(x)}{2}, & &\hat{\theta}_{\mathrm{s}}(x)=\hat{\theta}_{\mathrm{L}}(x)-\hat{\theta}_{\mathrm{R}}(x).
\end{aligned}
\right.
\eeq
These new operators obey the commutation relations
\beq
[\hat{n}_{\alpha}(x),\hat{n}_{\alpha'}(x')]=[\hat{\theta}_{\alpha}(x),\hat{\theta}_{\alpha'}(x')]=0, \quad [\hat{n}_{\alpha}(x),\hat{\theta}_{\alpha'}(x')]=i\delta_{\alpha,\alpha'}\delta(x-x').
\eeq
We call c and s ``charge'' and ``spin'' components following Kitagawa et al.~\cite{Kitagawa2011}.
By performing the Fourier transformation, the post-quench Hamiltonian is diagonalized as
\beq
\hat{H}_{\mathrm{fin}}=\sum_k\omega_k\left(\hat{b}_k^{\mathrm{L}\dagger}\hat{b}_k^{\mathrm{L}} +\hat{b}_k^{\mathrm{R}\dagger}\hat{b}_k^{\mathrm{R}}\right),
\label{eq:TL_fin_diag}
\eeq
where
\begin{align}
\omega_k&=|k|\sqrt{\frac{\rho_0g}{m}}, \\
\hat{b}_k^{\mathrm{X}}&=\left[\frac{1}{|k|}\sqrt{\frac{mg}{\rho_0}}\right]^{1/2}\tilde{n}_{\mathrm{X}}(k)+i\left[\frac{|k|}{4}\sqrt{\frac{\rho_0}{mg}}\right]^{1/2}\tilde\theta_{\mathrm{X}}(k) \quad (\mathrm{X}=\mathrm{L} \text{ or } \mathrm{R}),
\end{align}
and
\begin{align}
\tilde{n}_{\mathrm{X}}(k)&=\frac{1}{\sqrt{L}}\int_{-L/2}^{L/2}dx\,\hat{n}_{\mathrm{X}}(x)e^{-ikx}, \\
\tilde{\theta}_{\mathrm{X}}(k)&=\frac{1}{\sqrt{L}}\int_{-L/2}^{L/2}dx\,\hat{\theta}_{\mathrm{X}}(x)e^{-ikx}.
\end{align}
It should be noted that there is an ultraviolet cutoff of the wave number $k_{\mathrm{c}}=2\pi/\xi_{\mathrm{h}}$, where $\xi_{\mathrm{h}}=\pi\sqrt{2/(m\rho_0g)}$ is the healing length~\cite{Kitagawa2011} (remember that the density of bosons in each subsystem is $\rho_0/2$ here).

Similarly to the initial state considered in Sec.~\ref{sec:EP_simple}, the initial state here, which is the ground state of $\hat{H}_{\mathrm{ini}}$ given by Eq.~(\ref{eq:ini_TL}) and satisfies Eq.~(\ref{eq:quasi_1D_TL}), is given by a two-mode squeezed vacuum,
\beq
|\psi(0)\>=|\psi_{\mathrm{c}}(0)\>\otimes|\psi_{\mathrm{s}}(0)\>,
\eeq
where
\begin{align}
|\psi_{\mathrm{c}}(0)\>&=\prod_{k:0<|k|<k_{\mathrm{c}}}\frac{1}{\cosh(r_{\mathrm{c}})}e^{-\tanh(r_{\mathrm{c}})\hat{b}_k^{\mathrm{c}\dagger}\hat{b}_{-k}^{\mathrm{c}\dagger}}|0_{\mathrm{c}}\>, \\
|\psi_{\mathrm{s}}(0)\>&=\prod_{k:0<|k|<k_{\mathrm{c}}}\frac{1}{\cosh[r_{\mathrm{s}}(k)]}e^{-\tanh[r_{\mathrm{s}}(k)]\hat{b}_k^{\mathrm{s}\dagger}\hat{b}_{-k}^{\mathrm{s}\dagger}}|0_{\mathrm{s}}\>
\end{align}
with
\beq
e^{r_{\mathrm{c}}}=2^{1/4}, \quad e^{r_{\mathrm{s}}(k)}=\left(\frac{|k|^2}{4mg\rho_0}\right)^{1/4}.
\eeq
Here, $\hat{b}_k^{\mathrm{c}}$ and $\hat{b}_k^{\mathrm{s}}$ are defined by
\beq
\hat{b}_k^{\mathrm{c}}:=\frac{\hat{b}_k^{\mathrm{L}}+\hat{b}_k^{\mathrm{R}}}{\sqrt{2}}, \quad \hat{b}_k^{\mathrm{s}}:=\frac{\hat{b}_k^{\mathrm{L}}-\hat{b}_k^{\mathrm{R}}}{\sqrt{2}}.
\eeq
Since the post-quench Hamiltonian (\ref{eq:TL_fin_diag}) is also written as
\beq
\hat{H}_{\mathrm{fin}}=\sum_k\left(\hat{b}_k^{\mathrm{c}\dagger}\hat{b}_k^{\mathrm{c}} +\hat{b}_k^{\mathrm{s}\dagger}\hat{b}_k^{\mathrm{s}}\right)
=:\hat{H}_{\mathrm{c}}+\hat{H}_{\mathrm{s}},
\eeq
the quantum state at time $t$ is given by
\beq
|\psi(t)\>=e^{-i\hat{H}_{\mathrm{fin}}t}|\psi(0)\>=e^{-i\hat{H}_{\mathrm{c}}t}|\psi_{\mathrm{c}}(0)\>\otimes e^{-i\hat{H}_{\mathrm{s}}t}|\psi_{\mathrm{s}}(0)\>,
\eeq
in which the charge and spin components are always decoupled.
Therefore, they will evolve to their own GGEs independently, and the stationary state is described by the density matrix
\beq
\rho_{\mathrm{ss}}=\rho_{\mathrm{GGE}}^{\mathrm{c}}\otimes\rho_{\mathrm{GGE}}^{\mathrm{s}}=:\rho_{\mathrm{GGE}}^{\mathrm{cs}},
\eeq
where
\beq
\rho_{\mathrm{GGE}}^{\mathrm{X}}= \frac{e^{-\sum_k\lambda_k^{\mathrm{X}}\hat{b}_k^{\mathrm{X}\dagger}\hat{b}_k^{\mathrm{X}}}}{\mathrm{Tr}\,e^{-\sum_k\lambda_k\hat{b}_k^{\mathrm{X}\dagger}\hat{b}_k^{\mathrm{X}}}} \quad (\mathrm{X}=\mathrm{c},\mathrm{s}),
\label{eq:GGE_cs}
\eeq
with $\{\lambda_k\}$ determined by $\<\psi(0)|\hat{b}_k^{\mathrm{X}\dagger}\hat{b}_k^{\mathrm{X}}|\psi(0)\>=\mathrm{Tr}_{\mathrm{X}}\hat{b}_k^{\mathrm{X}\dagger}\hat{b}_k^{\mathrm{X}}\rho_{\mathrm{GGE}}^{\mathrm{X}}$.

It is noted that the left and right subsystems are not decoupled in the stationary state, which means that although the time evolutions of the left and right subsystems are independent of each other, the stationary state is not described by the GGEs of the left and right subsystems:
\beq
\rho_{\mathrm{ss}}\neq \rho_{\mathrm{GGE}}^{\mathrm{L}}\otimes\rho_{\mathrm{GGE}}^{\mathrm{R}}=:\rho_{\mathrm{GGE}}^{\mathrm{LR}},
\eeq
where $\rho_{\mathrm{GGE}}^{\mathrm{L}}$ and $\rho_{\mathrm{GGE}}^{\mathrm{R}}$ are defined similarly to Eq.~(\ref{eq:GGE_cs}).
The correlation in the stationary state between the left and right subsystems is due to the entanglement in the initial state.

\begin{figure}
\centering
\includegraphics[width=14cm]{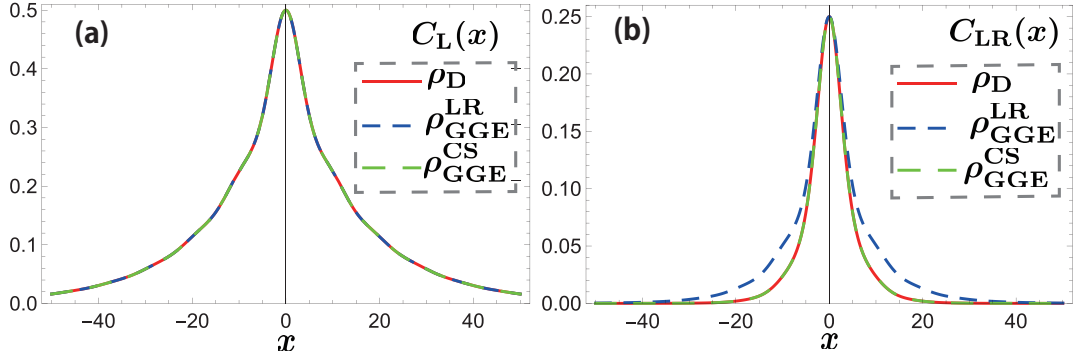}
\caption{Numerical results for (a) the auto-correlation function and (b) the cross-correlation function in the TL model with the parameters $\rho_0=1$, $2m=1$, $g=0.5$, and $L=10^4$.
Red curves are those calculated by the diagonal ensemble $\rho_{\mathrm{D}}$, i.e., the infinite-time averages.
The blue curves are the average in $\rho_{\mathrm{GGE}}^{\mathrm{LR}}$.
The green curves are the average in $\rho_{\mathrm{GGE}}^{\mathrm{cs}}$.
In (a), all the curves are collapsed into a single curve.
}
\label{fig:EP_TL}
\end{figure}

We now consider the two correlation functions in the stationary state.
One is the auto-correlation function $C_{\mathrm{L}}(x)$ defined as
\begin{align}
C_{\mathrm{L}}(x):=\<\hat{\psi}_{\mathrm{L}}^{\dagger}(x)\hat{\psi}_{\mathrm{L}}(0)\>
&\approx\frac{\rho_0}{2}\left\<e^{-i(\hat{\theta}_{\mathrm{L}}(x)-\hat{\theta}_{\mathrm{L}}(0))}\right\>
\nonumber \\
&=\frac{\rho_0}{2}e^{-\frac{1}{2}\left\<(\hat{\theta}_{\mathrm{L}}(x)-\hat{\theta}_{\mathrm{L}}(0))^2\right\>},
\label{eq:auto}
\end{align}
where the bracket denotes the average in the stationary state.
The last equality in Eq.~(\ref{eq:auto}) follows from the fact that $|\psi(t)\>$ is a Gaussian state for any $t$.
The other one is the cross-correlation function $C_{\mathrm{LR}}(x)$ defined by
\begin{align}
C_{\mathrm{LR}}(x):=\<\hat{\psi}_{\mathrm{L}}^{\dagger}(x)\hat{\psi}_{\mathrm{R}}^{\dagger}(0)\hat{\psi}_{\mathrm{L}}(0)\hat{\psi}_{\mathrm{R}}(x)\>
&\approx\frac{\rho_0^2}{4}\left\<e^{-i(\hat{\theta}_{\mathrm{L}}(x)-\hat{\theta}_{\mathrm{R}}(x))+i(\hat{\theta}_{\mathrm{L}}(0)-\hat{\theta}_{\mathrm{R}}(0))}\right\>
\nonumber \\
&=\frac{\rho_0^2}{4}\left\<e^{-i(\hat{\theta}_{\mathrm{s}}(x)-\hat{\theta}_{\mathrm{s}}(0))}\right\>
\nonumber \\
&=\frac{\rho_0^2}{4}e^{-\frac{1}{2}\left\<(\hat{\theta}_{\mathrm{s}}(x)-\hat{\theta}_{\mathrm{s}}(0))^2\right\>}
\end{align}
The auto-correlation function measures correlations within a subsystem L, while the cross-correlation function characterizes the interference contrast between the subsystems, which can be measured experimentally.

It is found that $\rho_{\mathrm{GGE}}^{\mathrm{LR}}$ well reproduces the auto-correlation function in the stationary state, while it does not reproduce the cross-correlation function.
On the other hand, $\rho_{\mathrm{GGE}}^{\mathrm{cs}}$ reproduces both correlation functions.
See Fig.~\ref{fig:EP_TL} for a numerical demonstration.
Because of the initial entanglement between the left and right subsystems, $\rho_{\mathrm{GGE}}^{\mathrm{LR}}$ cannot describe the interference between them.

\begin{figure}[tb]
\centering
\includegraphics[width=8cm]{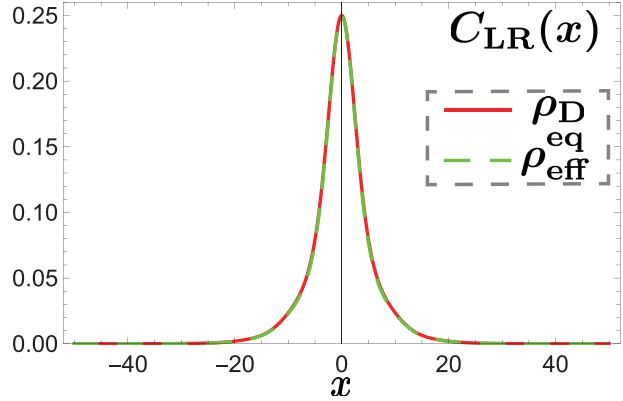}
\caption{Infinite-time average (red) and the equilibrium curve at the effective temperature $\beta_{\mathrm{eff}}=2/(\rho_0 g)$ of the cross-correlation function.
Parameters are set to $\rho_0=1$, $2m=1$, $g=0.5$, and $L=10^4$.
}
\label{fig:TL_eff}
\end{figure}

Since the cross-correlation function is written in terms of $\hat{\theta}_{\mathrm{s}}$, it is fully determined by $\rho_{\mathrm{GGE}}^{\mathrm{s}}$.
An explicit calculation shows that $\lambda_k^{\mathrm{s}}=-\ln[\tanh^2(r_{\mathrm{s}})]$.
We can define an effective inverse temperature governing a long-length scale of the spin component by
\beq
\beta_{\mathrm{eff}}:=\lim_{k\rightarrow 0}\frac{\lambda_k^{\mathrm{s}}}{\omega_k}=\frac{2}{\rho_0g}.
\eeq
As shown in Fig.~\ref{fig:TL_eff}, the canonical ensemble at the temperature $\beta_{\mathrm{eff}}^{-1}$ also reproduces $C_{\mathrm{LR}}(x)$ in the stationary state very well.
Thus, the analysis of the TL model predicts that the interference contrast is well fitted by the equilibrium curve at the inverse temperature $\beta_{\mathrm{eff}}=2/(\rho_0g)$~\cite{Kitagawa2011}.
This prediction has indeed been confirmed experimentally~\cite{Smith2013}.
This fact strongly suggests that the experimentally observed prethermalization is nothing but the EP in the TL model~\cite{Kaminishi_arXiv2017}.

\subsubsection{EP in the Lieb-Liniger model}
\label{sec:EP_LL}

We consider the same quench problem in the Lieb-Liniger model.
The pre-quench Hamiltonian is given by $\hat{H}_{\mathrm{ini}}\left[(\hat{\psi}_{\mathrm{L}}+\hat{\psi}_{\mathrm{R}})/\sqrt{2}\right]$ and the post-quench Hamiltonian is given by $\hat{H}_{\mathrm{fin}}=\hat{H}_{\mathrm{LL}}[\hat{\psi}_{\mathrm{L}}]+\hat{H}_{\mathrm{LL}}[\hat{\psi}_{\mathrm{R}}]$.
Although the TL model is a low-energy effective theory of the Lieb-Liniger model, as we see below, the EP in the Lieb-Liniger model is \textit{different} from that in the TL model~\cite{Kaminishi_arXiv2017}.

The key ingredients of the EP are the initial entanglement between the split subsystems and the energy degeneracies in $\hat{H}_{\mathrm{fin}}$.
The TL Hamiltonian has a large number of degeneracies, but many of them are lifted in the Lieb-Liniger model due to nonlinear interactions among bosons.
In the TL Hamiltonian after coherent splitting $\hat{H}_{\mathrm{TL}}[\hat{n}_{\mathrm{L}},\hat{\theta}_{\mathrm{L}}]+\hat{H}_{\mathrm{TL}}[\hat{n}_{\mathrm{R}},\hat{\theta}_{\mathrm{R}}]$, there are many degeneracies which stem from the continuous symmetry of
\beq
\begin{pmatrix}
\hat{b}_k^{\mathrm{L}} \\ \hat{b}_k^{\mathrm{R}}
\end{pmatrix}
\rightarrow
\begin{pmatrix}
\hat{b}_k^{\mathrm{L}'} \\ \hat{b}_k^{\mathrm{R}'}
\end{pmatrix}
=\begin{pmatrix}
\cos\varphi & \sin\varphi \\ -\sin\varphi & \cos\varphi
\end{pmatrix}
\begin{pmatrix}
\hat{b}_k^{\mathrm{L}} \\ \hat{b}_k^{\mathrm{R}}
\end{pmatrix}
\eeq
for $0\leq\varphi<2\pi$.
On the other hand, the Lieb-Liniger model has no such symmetry and no corresponding energy degeneracy.
The different types of energy degeneracies lead to different types of the EPs in the TL and the Lieb-Liniger models.

In the Lieb-Liniger model, the energy degeneracies due to translation symmetry ($\hat{\psi}(x)\rightarrow\hat{\psi}(x+a)$ for $a\in\mathbb{R}$) and space inversion symmetry ($\hat{\psi}(x)\rightarrow\hat{\psi}(-x)$) cause the EP~\cite{Kaminishi2015}.
While the continuous translation symmetry is associated with a local conservation law, i.e., the conservation of the total momentum, the inversion symmetry is a genuinely nonlocal and many-body symmetry.

To examine this aspect in more detail, we look into energy eigenstates of the Lieb-Liniger model.
They are obtained by the Bethe ansatz~\cite{Lieb1963}, and each energy eigenstate is characterized by the set of quantum numbers $(I_1,I_2,\dots,I_N)$, where $I_1<I_2<\dots<I_N$ and each $I_j$ is an integer when $N$ is odd and a half-odd number when $N$ is even.
For a set of quantum numbers $\{I_j\}_{j=1}^N$, we obtain a set of quasi-momenta (or rapidities) $\{k_j\}_{j=1}^N$ by solving the following Bethe ansatz equation:
\beq
k_j=\frac{2\pi}{L}I_j-\frac{2}{L}\sum_{l=1}^N\arctan\left(\frac{k_j-k_l}{c}\right),
\eeq
where $c:=2mg$.
The total momentum $P$ and the total energy $E$ in an eigenstate $|\{I_j\}_{j=1}^N\>$ are obtained by
\beq
P=\sum_{j=1}^Nk_j, \quad E=\frac{1}{2m}\sum_{j=1}^Nk_j^2.
\eeq
The energy degeneracies in the Lieb-Liniger model discussed above correspond to the degeneracies between the states $|\{I_j\}_{j=1}^N\>$ and $|\{-I_j\}_{j=1}^N\>$, where the corresponding sets of quasi-momenta are $\{k_j\}_{j=1}^N$ and $\{-k_j\}_{j=1}^N$, respectively.

Typically, the states $|\{I_j\}_{j=1}^N\>$ and $|\{-I_j\}_{j=1}^N\>$ are macroscopically distinct, and hence matrix elements $\<\{-I_j\}_{j=1}^N|\hat{O}|\{I_j\}_{j=1}^N\>$ of a local operator $\hat{O}$ would be very small in a macroscopic system and the degeneracy between them would play no role.
Kaminishi et al.~\cite{Kaminishi2015} have studied the effect of this type of energy degeneracies in a stationary state after the coherent splitting, and found that this type of energy degeneracies induce the EP at least for relatively small system sizes $L=N\sim 10$.

\begin{figure}[tb]
\centering
\includegraphics[width=14cm]{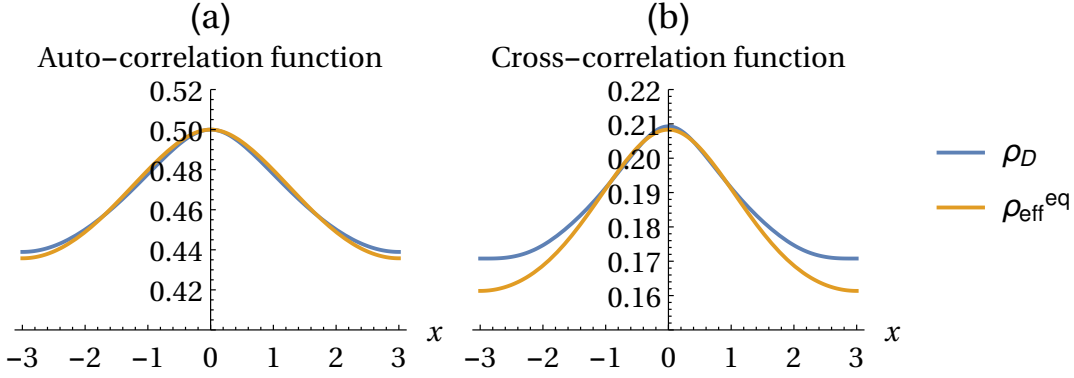}
\caption{Numerical results for (a) the auto-correlation function and (b) the cross-correlation function.
The parameters are set as $\rho_0=1$, $c=1$, and $L=6$.
Blue curves are the infinite-time averages (the expectation values in the diagonal ensemble $\rho_{\mathrm{D}}$) and yellow curves are the equilibrium curves at an effective temperature.
The effective temperature is determined by fitting the two curves in (a).
The two curves in (b) cannot be fitted well at any effective temperature.}
\label{fig:EP_LL}
\end{figure}

If we calculate the auto-correlation function $C_{\mathrm{L}}(x)$ in a stationary state, it can be fitted very well by an equilibrium curve at a certain temperature as shown in Fig.~\ref{fig:EP_LL} (a).
On the other hand, if we calculate the cross-correlation function $C_{\mathrm{LR}}(x)$, it \textit{cannot} be fitted by an equilibrium curve at some temperature as shown in Fig.~\ref{fig:EP_LL} (b).
This is in contrast to the EP in the TL model discussed in Sec.~\ref{sec:EP_TL}, in which $C_{\mathrm{LR}}(x)$ can be fitted by an equilibrium curve at $\beta_{\mathrm{eff}}=2/(\rho_0g)$.

In this way, the EP in the Lieb-Liniger model is qualitatively different from that in the TL model.
A good agreement between experiments~\cite{Gring2012,Smith2013} and a theoretical calculation in the TL model~\cite{Kitagawa2011} indicates that the EP in the Lieb-Liniger model found in Ref.~\cite{Kaminishi2015} is a phenomenon different from the prethermalization that was experimentally observed in Ref.~\cite{Gring2012}.

Experimental progress in cold-atomic systems has enabled us to implement well-controlled few-body systems~\cite{Serwane2011}, and the EP in the Lieb-Liniger model would be observable in such a few-body system.
Since prethermalization and thermalization can happen in small isolated quantum systems, it would be a fascinating problem to explore novel dynamical phenomena occurring only in relatively small quantum systems.

\section{Future prospects and concluding remarks}
\label{sec:conclusion}

In this article we have reviewed theoretical approaches to the problems of thermalization and prethermalization.
We have seen in Sec.~\ref{sec:quantum_equilibrium} that thermal equilibrium is characterized by its typicality; the thermodynamic typicality leads to the definition of macroscopic thermal equilibrium (MATE), while the canonical typicality leads to that of microscopic thermal equilibrium (MITE).
An approach to thermal equilibrium can thus be understood as a tendency towards more likely states starting from a less likely state.

Despite this simple intuitive picture, a complete understanding of thermalization is still elusive.
In Sec.~\ref{sec:thermalization}, we have seen that equilibration is a rather generic behavior of a large isolated quantum system, but thermalization is explained by a nontrivial property of the Hamiltonian, i.e., the eigenstate thermalization hypothesis (ETH) explained in Sec.~\ref{sec:ETH}, or a nontrivial property of the initial state, i.e., a sufficiently large effective dimension explained in Sec.~\ref{sec:macro_thermalization} and Sec.~\ref{sec:micro_sufficient}. 
Can we prove the ETH for a given interacting many-body system?
What is the relation between the ETH and the quantum chaos?
Can we obtain a good lower bound of the effective dimension after a quantum quench?
What time scales are to be expected for equilibration?
How do weak interactions with the environment, which induce decoherence, modify the results?
Is the many-body localization (MBL) possible in translation-invariant quantum systems?
Those fundamental questions remain open and currently under active research.

Nonequilibrium dynamics of many-body quantum systems presents several important problems.
A separation of relevant time scales in nonequilibrium dynamics induces prethermalization.
Although the decay of a prethermal state is described by the Pauli master equation in many cases, there are several exceptions.
One such exception has been discussed in Sec.~\ref{sec:exp_decay}, in which a prethermal state persists for an exponentially long time.
In this case, the mechanism of the exponentially long lifetime is well understood.
Another exception has been introduced in Sec.~\ref{sec:KCM}, in which dynamical constraints give rise  to slow dynamics.
Slow dynamics in some isolated quantum systems is similar to that observed in some classical (spin or molecular) glasses, but we do not completely understand what is responsible for the glassy dynamics in isolated quantum systems, and  we do not know how the glassy relaxation from a prethermal state is described.
It is thus an open problem to fully understand nonequilibrium dynamics associated with prethermalization.

Although we have focused on the dynamics of local quantities in this review, the dynamics of nonlocal quantities has also been actively studied.
In particular, much attention has recently been paid to the entanglement dynamics~\cite{Calabrese2005, Fagotti2008, Lauchli2008, Kim2013, Buyskikh2016, Nahum2017, Alba2017}.
Spreading of entanglement (``scrambling'') makes a sharp contrast to the transport of conserved quantities.
For example, in a many-body localized system, there is no transport of the energy throughout the system, while the entanglement grows without bound, although at a very slow rate (logarithmically in time)~\cite{Nandkishore_review2015,Altman_review2015}.
Recently, the validity of the ETH for nonlocal quantities has also been investigated through the calculation of the entanglement entropy in a single energy eigenstate~\cite{Deutsch2010,Beugeling2015_global,Garrison_arXiv2015,Vidmar2017,Fujita_arXiv2017,Vidmar-Rigol2017}.

How to characterize a stationary state in a small isolated system is also an interesting problem which is not covered in this review.
As we have emphasized, an isolated quantum system can still thermalize to some extent even if the system size is relatively small.
However, even if a given Hamiltonian satisfies the MITE-ETH in the thermodynamic limit, local quantities in a stationary state of a small system can deviate largely from their microcanonical averages~\cite{Biroli2010}.
In particular, it is known that the convergence to the behavior expected from the MITE-ETH is very slow in a nearly integrable system, and it behaves like an integrable system that does not thermalize at all when the system size is not sufficiently large.
That is why it is challenging to numerically calculate the entire thermalization process through prethermalization in a nearly integrable system.
In a small system, highly nonlocal many-body conserved quantities may be relevant to characterize the stationary state.
For example, we have seen in Sec.~\ref{sec:EP_LL} that the entanglement prethermalization in the Lieb-Liniger model is explained by a nonlocal conserved quantity associated with energy degeneracies due to the translation symmetry and the space-inversion symmetry.
In general, we do not know how we can construct a statistical ensemble that contains only a small number of parameters and describes the stationary state in such a case.

While we have not covered exciting experimental developments that motivate us to explore fundamental questions of theoretical physics, there is no doubt that the interplay between theory and experiment has been and will continue to bring fruitful cross-fertilization for investigating this and related fundamental problems.
In the problems of thermalization and prethermalization, experiments in well-controlled and well-isolated ultra-cold atoms have inspired theoretical research. 
From a viewpoint of ultra-cold atomic physics, studies on (pre)thermalization in feedback-controlled many-body systems, nonequilibrium dynamics in driven-dissipative systems, and the effect of successive quantum measurements on (pre)thermalization will be among the subjects that merit further study.

\section*{Acknowledgment}
T. M., T. N. I., and E. K. were supported by JSPS KAKENHI Grant Nos. 15K17718, 16H06718, and 16J03140, respectively.
M. U. acknowledges support by KAKENHI Grant No. JP26287088 from the Japan Society for the Promotion of Science, a Grant-in-Aid for Scientific Research on Innovative Areas ``Topological Materials Science'' (KAKENHI Grant No. JP15H05855), and the Photon Frontier Network Program from MEXT of Japan.

\bibliography{review_ref.bib}

\end{document}